\documentclass[useAMS,usenatbib]{mn2e}

\usepackage{graphicx}
\usepackage{txfonts}
\usepackage{natbib}
\usepackage{rotating}
\usepackage{float}
\usepackage{array}
\usepackage{mathenv}
\usepackage{multirow}
\usepackage{url}
\usepackage{longtable}

  \renewenvironment{thebibliography}[1]{%
    \begin{oldthebibliography}{#1}%
      \setlength{\parskip}{0ex}%
      \setlength{\itemsep}{0ex}%
  }%
  {%
    \end{oldthebibliography}%
  }

\newcommand{\edit}{}
\newcommand{\redit}{}
\begin{document}

  \title[An X-ray survey of the 2Jy sample I]{An X-ray survey of the 2Jy sample. I: is there an accretion mode dichotomy in radio-loud AGN?}


   \author[B. Mingo et. al.]{B. Mingo$^{1,2}$\thanks{E-mail:bmingo@extragalactic.info}, M. J. Hardcastle$^{1}$, J. H. Croston$^{3}$, D. Dicken$^{4}$, D. A. Evans$^{5}$, \newauthor R. Morganti$^{6,7}$, and C. Tadhunter$^{8}$ \\
   $^{1}$School of Physics, Astronomy \& Mathematics, University of Hertfordshire, College Lane, Hatfield AL10 9AB, UK\\
   		$^{2}$Department of Physics and Astronomy, University of Leicester, University Road, Leicester LE1 7RH, UK\\
		$^{3}$School of Physics and Astronomy, University of Southampton, Southampton SO17 1SJ, UK\\
		$^{4}$Institut d'Astrophysique Spatiale, Universit\'e Paris Sud, 91405 Orsay, France\\
		$^{5}$Harvard-Smithsonian Center for Astrophysics, 60 Garden Street, Cambridge, MA 02138, USA\\
		$^{6}$ASTRON, the Netherlands Institute for Radio Astronomy, Postbus 2, 7990 AA, Dwingeloo, The Netherlands\\
		$^{7}$Kapteyn Astronomical Institute, University of Groningen, P.O. Box 800, 9700 AV Groningen, The Netherlands\\
		$^{8}$Department of Physics and Astronomy, University of Sheffield, Hounsfield Road, Sheffield S3 7RH, UK }

   \date{Received ; accepted}

\maketitle

\begin{abstract}

\edit{We carry out a systematic study of the X-ray emission from the active nuclei of the $0.02<z<0.7$ 2Jy sample, using \textit{Chandra} and \textit{XMM-Newton} observations. We combine our results with those from mid-IR, optical emission line and radio observations, and add them to those of the 3CRR sources. We show that the low-excitation objects in our samples \redit{show signs} of radiatively inefficient accretion. We study the effect of the jet-related emission on the various luminosities, confirming that it is the main source of soft X-ray emission for our sources. We also find strong correlations between the accretion-related luminosities, and identify several sources whose optical classification is incompatible with their accretion properties. We derive the bolometric and jet kinetic luminosities for the samples and find a difference in the total Eddington rate between the low and high-excitation populations, with the former peaking at $\sim1$ per cent and the latter at $\sim20$ per cent Eddington. Our results are consistent with a simple Eddington switch when the effects of environment on radio luminosity and black hole mass calculations are considered. The apparent independence of jet kinetic power and radiative luminosity in the high-excitation population in our plots supports a model in which jet production and radiatively efficient accretion are not strongly correlated in high-excitation objects, though they have a common underlying mechanism.}

\end{abstract}


%

\section{Introduction}\label{Intro}

Our knowledge of active galactic nuclei (AGN), their observational properties and underlying mechanisms has vastly increased over the last few decades. We now know that these objects are powered through gas accretion onto some of the most supermassive black holes that sit in the centres of most galaxies \citep[e.g.][]{Magorrian1998}. Radio-loud objects are particularly important to our understanding of AGN, since, despite the fact that they constitute only a small fraction of the overall population, it is during this phase that the impact of the AGN on their surrounding environment (through the production of jets and large-scale outflows and shocks) can be most directly be observed and measured \citep[e.g.][]{Kraft2003,Cattaneo2009,Croston2011}. Moreover, radio galaxies make up over 30 per cent of the massive galaxy population, and it is likely that all massive galaxies go through a radio-loud phase, as the activity is expected to be cyclical \citep[e.g.][]{Best2005,Saikia2010}.

It is now commonly accepted that the dominant fuelling mechanism for radio-quiet objects is the accretion of cold gas onto the black hole from a radiatively efficient, geometrically thin, optically thick accretion disk \citep{Shakura1973}. However, this may not be the case for radio-loud objects. \citet{Hine1979} noticed the existence of a population of radio-loud objects which lacked the high-excitation optical emission lines traditionally associated with AGN. These so-called low-excitation or weak-line radio galaxies (LERGs or WLRGs) cannot be unified with the rest of the AGN population \edit{(high excitation galaxies in general, or HEGs, and radio galaxies in particular, or HERGs)}, since their differences are not merely observational or caused by orientation or obscuration. It has been argued that \edit{LERGs} accrete hot gas \citep[see e.g.][]{Hardcastle2007b,Janssen2012} in a radiatively inefficient manner, through optically thin, geometrically thick \edit{accretion flows \citep[RIAF, see e.g][]{Narayan1995,Quataert2003}}. These objects thus lack the traditional accretion structures (disk and torus) commonly associated with active nuclei \citep[see e.g.][]{VDWolk2010,FdezOntiveros2012,Mason2012}, and seem to be channeling most of the gravitational energy into the jets, rather than radiative output. This makes them very faint and hard to detect with any non-radio selected surveys.

Current models \citep[e.g.][]{Bower2006,Croton2006} suggest that the radiatively efficient process may be dominant at high redshifts, \redit{and to be related to the scaling relation between black hole mass and host galaxy's bulge mass} \citep[e.g.][]{Silk1998,Heckman2004}. \edit{Radiatively efficient accretion} may also be the mode involved in the apparent correlation (and delay) between episodes of star-formation and AGN activity in the host galaxies \citep[e.g][]{Hopkins2012,Ramos2013}. Radiatively inefficient accretion is believed to be more common at low redshifts \citep{Hardcastle2007b}, and to play a crucial role in the balance between gas cooling and heating, both in the host galaxy and in cluster environments \citep{McNamara2007,Antognini2012}. These two types of accretion are often called `quasar mode' and `radio mode', which is somewhat misleading, given that there are radiatively efficient AGN with jets and radio lobes. \redit{This change of a predominant accretion mode with redshift is applicable primarily to the largest galaxies and most massive supermassive black holes (SMBH), since smaller systems evolve differently.}

As pointed out by \edit{e.g. \citet{Laing1994,Blundell2001,Rector2001,Chiaberge2002,Hardcastle2009}}, it is important to note that the high/low-excitation division does not directly correlate with the FRI-FRII categories established by \citet{FR1974}, as is often thought. While most low-excitation objects seem to be FRI, there is a population of bona-fide FRII LERGs, as well as numerous examples of FRI HERGs \citep[e.g][]{Laing1994}. \edit{This lack of a clear division} is most likely caused by the complex underlying relation between fuelling, jet generation and environmental interaction. There seems to be a evidence for a difference in the Eddington rate between both populations \citep[see e.g.][]{Hardcastle2007b,Lin2010,Ho2009,Evans2011,Plotkin2012,Best2012,Russell2012,Mason2012}, with LERGs typically accreting at much lower rates ($<0.1$ Eddington) than HERGs. Estimating the jet kinetic power is also complicated, given that the radio luminosity of a source depends on the environmental density \citep{Hardcastle2013,Ineson2013} and given the apparent difference in the particle content and/or energy distribution for typical FRI and FRII \edit{jets and lobes} \citep[see e.g.][]{Croston2008b,Godfrey2013}.

\edit{In terms of their optical classification, HERGs are further split into quasars (QSOs), broad-line radio galaxies (BLRGs), and narrow-line radio galaxies (NLRGs), in consistency with the unified models, and in parallel with their radio-quiet counterparts (respectivelly, radio-quiet quasars, type 1 and type 2 radio-quiet AGN). We will use the optical classification for HERGs throughout this work.}

In this paper we analyse the X-ray emission from the 2Jy sample of radio galaxies \citep{WP1985}, with an approach based on that of \citet{Hardcastle2006,Hardcastle2009} used on the 3CRR galaxies. X-ray emission is less ambiguous than other wavelengths for an analysis of a sample such as the 2Jy, which contains a variety of populations, in that\edit{, at these high luminosities, and in the nuclear regions we are considering,} it is unequivocally linked to AGN activity. To fully understand the characteristics of this AGN activity, however, a multiwavelength approach is needed.

From works like those of \citet{Hardcastle2006,Hardcastle2009}, we do know that LERGs follow the correlation of narrow-line galaxies (NLRGs) between soft X-ray and radio emission \citep{Hardcastle1999}, reinforcing the hypothesis that in radio-loud objects this X-ray component originates in the jet. One of the crucial points we aim to investigate in this paper is the dissimilarity between the NLRG and LERG populations.

Our aim is to study the correlations between the luminosities of the sources at different wavelengths, to \edit{link the emission produced in regions at various distances from the central black hole: from the disk and corona to the torus, the jet and the lobes. In doing so, we will investigate how accretion translates into radiative and kinetic output across the whole radio-loud population.} 

\edit{While many of the sources in the 3CRR catalogue have been observed in great detail, the multiwavelength coverage is not uniform, and the sample is not statistically complete in the X-rays, being more complete for redshifts $<$0.5. The observations of the 2Jy sample, however, were taken with the explicit purpose of providing \redit{comparable measurements} for all the objects in the sample. This consistency provides us with the opportunity to test whether the conclusions reached by Hardcastle et al. can be extrapolated to all radio-loud AGN or are related to the biased redshift distribution of the 3CRR sources.}

\edit{Although it is well known that some of the physical mechanisms involved in radio-loud emission in AGN are similar to those found in X-ray binaries \citep[see e.g. the review by][]{Kording2006}, some caution must be applied, since there are also dissimilarities in the timescales and fuelling processes involved. In this work we will focus only on AGN, and the possible impact of our results may have on understanding their observational properties, calssification, accretion mode and the influence on their hosts.} 

For this paper we have used a concordance cosmology with $H_{0}=70$ km s$^{-1}$ Mpc$^{-1}$, $\Omega_{m}=0.3$ and $\Omega_{\Lambda}=0.7$.

%

\section{Data and Analysis}\label{Data}

\subsection{The Sample}\label{Sample}

The 2Jy sample \citep{WP1985,Tadhunter1993} is a sample of southern radio galaxies with flux greater than 2 Jy at 2.7 GHz\footnote{For the most up-to-date version of the catalogue and ancillary data, see \url{http://2Jy.extragalactic.info/}}. The subsample we study has consistent, uniform multiwavelength coverage (see Section \ref{Multidata} for details) and, since we only include the steep-spectrum sources, it \edit{contains only genuinely powerful radio galaxies, while avoiding most of the effects caused by the strong relativistic beaming found in flat-spectrum sources. Other than excluding \redit{beamed} sources, the radio selection, unlike those done in optical, IR or X-ray wavelengths, selects no preferential orientation.}

We analyse a statistically complete subsample of the 2Jy steep-spectrum sources defined by \citet{Dicken2008}, containing 45 objects with with $\delta <+10^{\circ}$ and redshifts $0.05<z<0.7$. Particle acceleration in the jet causes the radio spectrum to flatten, thus the steep-spectrum ($\alpha>0.5$, \redit{where we use the negative sign convention for $\alpha$}) selection of \citet{Dicken2008} excludes core and jet dominated sources. \edit{Flat-spectrum sources are typically blazars, whose nuclear emission is completely dominated by the jet, and, although they are a small fraction of the total population, they appear brighter due to the jet contribution. By excluding these sources we eliminate a possible source of bias.} Unlike \citet{Dicken2008}, we have not included the flat-spectrum, core-dominated sources 3C 273 and PKS 0521--36 for comparison. The subsample studied here has consistent, uniform multiwavelength coverage\edit{, and, being statistically complete, includes all the sources within the flux, sky area, spectral types, and redshift ranges defined}.

From a radio classification point of view, the sample is dominated by powerful sources, with 6 objects being Fanaroff-Riley type I (FR I), 7 compact sources (CSS), and 32 Fanaroff-Riley type II (FR II) \citep{Morganti1993,Morganti1999}. As for emission line classification, 10 sources are LERGs, 19 are NLRGs, 12 are BLRGs and 3 are QSOs \citep{Tadhunter1993,Tadhunter1998}.


\edit{We have included in our analysis the 3CRR sources with $z<1$ studied by \citet{Hardcastle2006,Hardcastle2009}. The 3CRR catalogue of \citet{Laing1983} includes all the extragalactic radio sources with a flux greater than 10.9 mJy at 178 MHz and $\delta>+10^{\circ}$. By combining the 3CRR and 2Jy catalogues we are effectively selecting a large sample of the most radio-luminous galaxies in the Universe. To further improve the overall statistics, we also include in this work \redit{8 new observations} of 3CRR sources not covered by \citet{Hardcastle2006,Hardcastle2009} (see Appendix \ref{3C_appendix} for details).}

\edit{The 2Jy sample does not \edit{spatially} overlap with the 3CRR catalogue, due to the different location of the sources (the 3CRR catalogue covers sources in the Northern hemisphere, the 2Jy sources are in the Southern hemisphere). Some of the brightest sources are included in the original 3C catalogue, as is the case for e.g. the BLRG 3C 18 (PKS 0038+09). Although we have excluded core-dominated sources (to minimise the effects of beaming), the 2Jy sample was selected at a higher frequency than the 3CRR sample. This higher frequency selection implies that, overall, more beamed sources are selected in the 2Jy sample than for the 3CRR, which is a possible caveat to the assumption that no preferential AGN orientation is selected. Some of the implications of this fact are discussed in Section \ref{Correlations}.}

\edit{Although the 3CRR catalogue contains a much larger number of sources than the 2Jy sample, it is not statistically complete in the X-rays, and has better coverage at lower redshifts. The observations of the 2Jy sample are also more homogeneous. While it may seem that studying a reduced number of sources from the 2Jy sample does not add much to the existing correlations, the characteristics of the sample and observations allow us to validate our previous results on the 3CRR catalogue, eliminating the low-redshift and inhomogeneous coverage biases. The 2Jy sample also contributes a large number of NLRGs and LERGs to the overall statistics, which are particularly important to test our scientific goals. The combination of both samples provides a very powerful tool to explore the entire population of radio-powerful AGN.}

Throughout this paper we have kept the existing optical line classifications for the objects in both the 2Jy and in the 3CRR samples, for consistency, but we point out when evidence suggests that \edit{the optical} classification does not accurately characterise a specific object. For the overall populations low-excitation and high-excitation can be used as synonyms for radiatively inefficient and efficient AGN, respectively, but it is important to keep in mind that this does not hold true for some objects. The LERG/HERG classification is observational, based on optical line ratios, and in some cases it is not a good diagnostic for the true nature of the accretion process involved (a radiatively efficient object will be classified as a LERG if its high-excitation lines are not detected, while a radiatively inefficient source may be classified as a HERG if high-excitation lines are observed, even if they are produced by a mechanism that is not related to the AGN, e.g. photoionization by stellar activity).

\subsection{X-ray Data}\label{Xdata}
There are 46 sources in our sample, with $0.05<z<0.7$. All have X-ray observations save for PKS 0117--15 (3C 38), which, unfortunately, was not observed by \textit{XMM-Newton}, and is thus excluded from our analysis. \edit{Our sample, therefore, contains 45 2Jy objects.} The list of galaxies in the sample and the \textit{Chandra} and \textit{XMM-Newton} observations is shown in Table \ref{objects_table}. \edit{Many of the observations were taken specifically for this project, \textit{Chandra} observations were requested for the low-$z$ sources to map any extended emission (jets, hotspots, lobes and any emission from a hot IGM for sources in dense environments). For the sources with $z>0.2$, where extended structures cannot be resolved, we requested \textit{XMM} observations instead, to maximise the signal to noise ratio of the AGN spectra, so as to allow spectral separation of the unresolved components. The new observations of the 2Jy sample used in this work are indicated in Table \ref{objects_table}. The list of new observations of 3CRR sources is given in Appendix \ref{3C_appendix}.}

\edit{By limiting the redshift range to $z>0.05$ we exclude both low power sources and those whose extended emission may not be fully covered by \textit{Chandra}. The extended emission (jets and lobes) in these low-z sources will be studied in detail in our second paper.}

We analysed \textit{Chandra} observations for the low-$z$ sources in our sample. When using archival data we only considered ACIS-S and ACIS-I observations without gratings, and discarded calibration or very short observations that did not significantly contribute to the statistics. When more than one spectrum was extracted for a source, we carried out simultaneous fits. We reduced the data using CIAO 4.3 and the latest CALDB. We included the correction for VFAINT mode to minimise the issues with the background for all the sources with a count rate below 0.01 counts s$^{-1}$ and observed in VFAINT mode. For sources with rates above this threshold and below 0.1 counts s$^{-1}$ the difference made by this correction is barely noticeable. For the brightest sources the software is not able to properly account for the high count rate, considering some of these events as background, and resulting in dark ``rings" appearing in the images, and the loss of a substantial number of counts. 

\edit{We extracted spectra for all the sources, using extraction regions consistent with those of \citet{Hardcastle2009}: a 2.5 pixel (1 px = 0.492 arcsec) radius circular region centered in the object as source, and an immediately external annulus, with an outer radius of 4 pixels, for the background, to minimise the contamination from any thermal components in the circumnuclear regions. For very bright sources we had to use larger regions to include most of the point-spread function (PSF)}, namely a 20 pixel radius circle for the source, and a 20 to 30 pixel circular annulus for the background. In the cases where pileup was present (PKS 0038+09, 0442--28, 0625--35, 0945--27, 1733--56, 1814--63, 2135--14), we corrected the \edit{auxiliary response file (ARF)} as described by \citet{Hardcastle2006} and \citet{Mingo2011}. We generated an energy versus flux table from an initial model fit, and fed it to ChaRT \edit{\citep[the Chandra Ray Tracer, ]{ChaRT}}, a tool that generates a PSF from a given model. Next, we fed the results to the tool MARX\edit{\footnote{See http://space.mit.edu/CXC/MARX/docs.html}}, which produces an image of the simulated PSF. We then generated a new events file from our original data and an annular extraction region, identical to the one we used to generate our spectra, but excluding the central few pixels. We used a code to fit a 5th-degree polynomial to the ratio of this events file and the whole simulated events file as a function of energy. \edit{The} code reads in the ARF generated by CIAO and scales the effective area at each energy, using the polynomial fit, to effectively correct for the missing effective area due to the exclusion of the central pixels. The code then writes a new ARF which can be used to correct for the effects of excluding the central pixels.

For the sources at $0.2<z<0.7$ we used \textit{XMM-Newton} observations. We extracted MOS and PN spectra for all of them, using SAS 11.0 and the latest calibration files. We used spatially coincident extraction regions for the three instruments whenever possible, using 30-arcsec source regions and off-source 90-arcsec background regions for the fainter sources, and 60-arcsec and 120-arcsec source and background regions, respectively, for the bright ones. Only a few observations were affected by flaring severe enough to require filtering. The most problematic case was PKS 1547--79, a faint source observed during a period of high flaring. We filtered the most severely affected parts of the observation.

Four low-$z$ sources (PKS 0404+03, 1814--63, 2135--14, 2221--02) have \textit{XMM} observations that we did not use, since the \textit{Chandra} spectra adequately characterised the AGN spectrum and had no contamination from any circumnuclear gas. For PKS 2314+03, however, we used both the \textit{Chandra} and \textit{XMM} observations, given that its spectrum is quite peculiar.

We rebinned all the spectra to 20 counts per bin (after background subtraction) to make them compatible with $\chi^{2}$ statistics.

\begin{table}\scriptsize
\caption{\edit{Objects in the 2Jy sample observed with \textit{Chandra} (ACIS-S except for PKS 0625--53 and PKS 2135--14, which were taken with the ACIS-I) and \textit{XMM-Newton} (MOS and PN). FRI and FRII stand for Fanaroff-Riley class I and II respectively, CSS stands for compact steep-spectrum. LERG, NLRG and BLRG stand, respectively, for low excitation, narrow-line and broad-line radio galaxy; Q stands for Quasar. New observations taken for this survey are indicated with an asterisk after the observation ID.}}\label{objects_table}
\centering
\setlength{\tabcolsep}{1.6pt}
\setlength{\extrarowheight}{3pt}
\begin{tabular}{ccccccc}\hline
PKS&FR Class&Type&$z$&Instrument&Obsid&Exp (ks)\\\hline
0023-26&CSS&NLRG&0.322&\textit{XMM}&0671870601*&19.55\\
0034-01&FRII&LERG&0.073&\textit{Chandra}&02176&28.18\\
0035-02&FRII&BLRG&0.220&\textit{Chandra}&09292&8.04\\
0038+09&FRII&BLRG&0.188&\textit{Chandra}&09293&8.05\\
0039-44&FRII&NLRG&0.346&\textit{XMM}&0651280901*&20.57\\
0043-42&FRII&LERG&0.116&\textit{Chandra}&10319*&18.62\\
0105-16&FRII&NLRG&0.400&\textit{XMM}&0651281001*&21.27\\
0213-13&FRII&NLRG&0.147&\textit{Chandra}&10320*&20.15\\
0235-19&FRII&BLRG&0.620&\textit{XMM}&0651281701*&13.67\\
0252-71&CSS&NLRG&0.566&\textit{XMM}&0651281601*&19.17\\
0347+05&FRII&LERG&0.339&\textit{XMM}&0651280801*&16.47\\
0349-27&FRII&NLRG&0.066&\textit{Chandra}&11497*&20.14\\
0404+03&FRII&NLRG&0.089&\textit{Chandra}&09299&8.18\\
0409-75&FRII&NLRG&0.693&\textit{XMM}&0651281901*&13.67\\
0442-28&FRII&NLRG&0.147&\textit{Chandra}&11498*&20.04\\
0620-52&FRI&LERG&0.051&\textit{Chandra}&11499*&20.05\\
0625-35&FRI&LERG&0.055&\textit{Chandra}&11500*&20.05\\
0625-53&FRII&LERG&0.054&\textit{Chandra}&04943&18.69\\
0806-10&FRII&NLRG&0.110&\textit{Chandra}&11501*&20.04\\
0859-25&FRII&NLRG&0.305&\textit{XMM}&0651282201*&13.85\\
\multirow{2}{*}{0915-11}&\multirow{2}{*}{FRI}&\multirow{2}{*}{LERG}&\multirow{2}{*}{0.054}&\textit{Chandra}&04969&98.2\\
&&&&\textit{Chandra}&04970&100.13\\
\multirow{2}{*}{0945+07}&\multirow{2}{*}{FRII}&\multirow{2}{*}{BLRG}&\multirow{2}{*}{0.086}&\textit{Chandra}&06842&30.17\\
&&&&\textit{Chandra}&07265&20.11\\
\multirow{2}{*}{1136-13}&\multirow{2}{*}{FRII}&\multirow{2}{*}{Q}&\multirow{2}{*}{0.554}&\textit{Chandra}&02138&9.82\\
&&&&\textit{Chandra}&03973&77.37\\
1151-34&CSS&Q&0.258&\textit{XMM}&0671870201*&18.67\\
1306-09&CSS&NLRG&0.464&\textit{XMM}&0671871201*&22.67\\
1355-41&FRII&Q&0.313&\textit{XMM}&0671870501*&14.97\\
1547-79&FRII&BLRG&0.483&\textit{XMM}&0651281401*&13.25\\
1559+02&FRII&NLRG&0.104&\textit{Chandra}&06841&40.18\\
1602+01&FRII&BLRG&0.462&\textit{XMM}&0651281201*&13.67\\
\multirow{2}{*}{1648+05}&\multirow{2}{*}{FRI}&\multirow{2}{*}{LERG}&\multirow{2}{*}{0.154}&\textit{Chandra}&05796&48.17\\
&&&&\textit{Chandra}&06257&50.17\\
1733-56&FRII&BLRG&0.098&\textit{Chandra}&11502*&20.12\\
1814-63&CSS&NLRG&0.063&\textit{Chandra}&11503*&20.13\\
1839-48&FRI&LERG&0.112&\textit{Chandra}&10321*&20.04\\
1932-46&FRII&BLRG&0.231&\textit{XMM}&0651280201*&13.18\\
1934-63&CSS&NLRG&0.183&\textit{Chandra}&11504*&20.05\\
1938-15&FRII&BLRG&0.452&\textit{XMM}&0651281101*&18.17\\
1949+02&FRII&NLRG&0.059&\textit{Chandra}&02968&50.13\\
1954-55&FRI&LERG&0.060&\textit{Chandra}&11505*&20.92\\
2135-14&FRII&Q&0.200&\textit{Chandra}&01626&15.13\\
2135-20&CSS&BLRG&0.635&\textit{XMM}&0651281801*&17.57\\
2211-17&FRII&LERG&0.153&\textit{Chandra}&11506*&20.04\\
2221-02&FRII&BLRG&0.057&\textit{Chandra}&07869&46.20\\
2250-41&FRII&NLRG&0.310&\textit{XMM}&0651280501*&13.67\\
\multirow{2}{*}{2314+03}&\multirow{2}{*}{FRII}&\multirow{2}{*}{NLRG}&\multirow{2}{*}{0.220}&\textit{XMM}&0651280101*&21.67\\
&&&&\textit{Chandra}&12734&8.05\\
2356-61&FRII&NLRG&0.096&\textit{Chandra}&11507*&20.05\\\hline
\end{tabular}
\end{table}

\subsection{Spectral Fitting}\label{Fitting}

For spectral fitting we used XSPEC version 12.5 and followed the same approach as \citet{Hardcastle2006,Hardcastle2009}, as follows. We considered the energy range between 0.4 and 7 keV for the \textit{Chandra} spectra, and 0.3 to 8 keV for the \textit{XMM} spectra. \edit{For the sources observed by \textit{XMM}, the PN, MOS1 and MOS2 spectra were fitted simultaneously. The same approach was taken for those sources with more than one \textit{Chandra} ovservation (PKS 0915-15, PKS 0945+07, PKS 1136-13, PKS 1648+05) and for PKS 2314+03, which was observed by both \textit{Chandra} and \textit{XMM} (see Table \ref{objects_table}).} 

\edit{The typical X-ray spectrum of a radio-loud AGN can be approximated with a phenomenological model consisting of three main components. The accretion-related emission is well modelled with a power law that contributes mostly at energies between 2 and 10 keV, as predicted by accretion models \citep[see e.g.][]{Haardt1991}, and is also found in radio-quiet objects, although the slope of the power law changes. The soft excess in radio-loud objects, however, is not dominated by reflection of the accretion-related emission onto the disk, but is related to the jet \citep[see e.g.][]{Hardcastle1999,Hardcastle2006,Hardcastle2009}. This soft emission often dominates below 1 keV, and is also well modelled with a power law. When the torus obscures part of the emission, an intrinsic absorption component must be added to the model as well. Some objects also show fluorescence Fe K$\alpha$ lines around 6.4 keV. When no obscuration is present (in broad-line objects), distinguishing both power law components is not possible. Given that the jet-related emission in broad-line sources may be further complicated by relativistic beaming, and for consistency with the work of \citet{Hardcastle2006,Hardcastle2009}, we have considered both power law components as one, when dealing with these sources. We are aware that this overestimates the luminosities (in the sense that the same luminosity may be ascribed to more than one component), and take this fact into account in our plots and correlation analysis.}

\edit{We approached the fitting process in a systematic manner, by fitting all the sources to a set of three possible models. We first fitted each spectrum to a model consisting to a single power law with fixed Galactic absorption (wabs), for which we used the weighted average extinction values of \citet{Dickey1990}; we call this component `unabsorbed' throughout this work, after \citet{Hardcastle2006,Hardcastle2009}. Secondly, we fitted the same model, adding an intrinsic absorption column (zwabs); we refer to this component as `accretion-related'. We then fitted a combination of both models, and assessed which of the three provided a best fit to the data. When the photon index of either power law could not be constrained, we fixed the values to $\Gamma$=2.0 and $\Gamma$=1.7, for the unabsorbed and accretion related component respectively \citep[these values are consistent with what is found in most radio-loud AGN, and follow the choices of][]{Hardcastle2006,Hardcastle2009}. When residuals were still present at high energies we added a redshifted Gaussian profile for the Fe K-$\alpha$ line (zgauss), as required by the data. In the cases where a single power law provided a best fit to the data, we calculated an upper limit on the luminosity of the other component by fixing the parameters of the existing model, and adding the missing component with a fixed photon index. We added a fixed intrinsic absorption column $N_{H}=10^{23}$ cm$^{-2}$ in the case of the accretion-related power law, a value consistent with what is seen in sources with detected, heavily absorbed components, and in agreement with that chosen by \citet{Evans2006} and \citet{Hardcastle2006,Hardcastle2009}. }


\redit{While, for consistency, we have used the foreground N$_H$ values of \citet{Dickey1990} for all the objects in the sample, we note that the Galactic extinction column may be underestimated for PKS 0404+03. \textit{Herschel}/SPIRE observations show unusually bright Galactic cirrus dust emission in this area (Dicken et al. 2014, in prep.).}
 
\edit{We derived the luminosity for the unabsorbed component, $L_{X_{u}}$, from the normalization of the unabsorbed power law, and used XSPEC to calculate the 2-10 keV unabsorbed luminosity (corresponding to the accretion-related component, corrected for intrinsic absorption), $L_{X_{a}}$. These energy ranges were chosen because they also allow direct comparison with the existing literature, and are consistent with our previous work.}

We are aware of the fact that the brightest sources are likely to have measurable variations in their luminosity over time, although the most variable sources are excluded by the steep-spectrum selection. \edit{Variability} is an intrinsic uncertainty characteristic of X-ray AGN studies, unavoidable unless follow-up observations are carried out for each source. We acknowledge that X-ray variability is a systematic effect that introduces scatter in our plots, and estimate the impact of variability and other systematics in Sections \ref{Correlations} and \ref{Eddington}.

Some of the sources in our sample observed by \textit{XMM} show signs of inhabiting rich environments, as shown in the optical by \citet{RamosAlmeida2010}, and Ramos Almeida et al. (2013, MNRAS, in press). Our extraction regions may not be able to fully account for this, hence some contamination of the soft X-ray component can be expected. PKS 0023-26 and PKS 0409-75 (together with PKS 0347+05, which has additional complications, as pointed out in Section \ref{0347+05}) are the sources where contamination from a thermal component may be most relevant, given that they are relatively faint in the [OIII] and mid-IR bands. We tested a model in which one of the power law components is replaced by a thermal one (apec) in these sources, and obtained worse fits than with the non-thermal model. We also attempted to quantify the amount of thermal emission by adding a thermal model on top of the two power laws, but the results were inconsistent due to the degeneracy between model components. Given that PKS 0023-26 is not clearly outlying in our plots, we assume that the dominant contribution to the soft X-ray emission is related to the AGN, rather than thermal emission. \redit{The case is less clear for PKS 0409-75, whose soft X-ray component is very bright, causing it to be an outlier. Beaming is not likely to be the cause of this excess, since the radio core is undetected at 20 GHz \citet{Dicken2008}, but it is possible that there is a contribution of inverse-Compton emission from the lobes, which are not resolved by \textit{XMM}. In both PKS 0023-26 and PKS 0409-75, an in-depth study of the ICM X-ray emission is needed to fully quantify its contribution to the AGN X-ray luminosity.}

\edit{The results of the spectral fits are displayed in Table \ref{fit_param_table}. The sources where a Fe K-$\alpha$ line was detected are listed in Table \ref{Fe_table}. Details for each individual source, and references to previous work, are given in Appendix \ref{notes}.
 For consistency, we have checked our results, both on the derived luminosities and the extended emission (which we will analyse in detail in our second paper) against those obtained by \citet{Siebert1996}, based on data from \textit{ROSAT}, and find them in good agreement.}

\begin{table*}\scriptsize
\caption[Best fit parameters for all the objects in the 2Jy sample]{\edit{Best fit parameters for all the objects in the sample. errors are calculated at $90$ per cent confidence. Where no errors are indicated the parameters were fixed to that value. Values preceded by a `$<$' contribution of a hypothetical second component. The subinidices 1 and 2 refer to the unabsorbed and accretion-related components, respectively. The net (background-subtracted) counts are given per observation, thus for sources observed by \textit{Chandra} only one value is given (two for the sources with two observations); for those observed by \textit{XMM-Newton} the three values correspond to the PN, MOS1, and MOS2 spectra, respectively. See Table \ref{objects_table} for details on the individual observations.}}\label{fit_param_table}
\centering
\setlength{\tabcolsep}{2.0pt}
\setlength{\extrarowheight}{3pt}
\begin{tabular}{cccccccccc}\hline
PKS&z&Foreground $N_H$&Intrinsic $N_H$&$\Gamma_{u}$&Norm 1&$\Gamma_{a}$&Norm 2&Net counts&$\chi^{2}$/DOF\\
&&$\times 10^{20}$ cm$^{-2}$&$\times 10^{22}$ cm$^{-2}$&&keV$^{-1}$cm$^{-2}$s$^{-1}$ @1keV&&keV$^{-1}$cm$^{-2}$s$^{-1}$ @1keV&&\\\hline
0023-26&0.322&1.76&$0.16^{+0.11}_{-0.10}$&$2.00$&$<1.06\times10^{-6}$&$1.84^{+0.51}_{-0.19}$&$1.73^{+0.55}_{-0.80}\times10^{-5}$&336/107/105&20.69/23.00\\
0034-01&0.073&2.89&$10.07^{+4.84}_{-3.04}$&$1.24^{+0.36}_{-0.34}$&$1.17^{+0.14}_{-0.15}\times10^{-5}$&$1.70$&$1.01^{+0.29}_{-0.28}\times10^{-4}$&490&10.02/20.00\\
0035-02&0.220&2.85&$3.34^{+1.36}_{-0.97}$&$2.00$&$1.44^{+0.14}_{-0.13}\times10^{-4}$&$1.70$&$2.59^{+0.41}_{-0.43}\times10^{-4}$&1091&31.12/49.00\\
0038+09&0.188&5.45&$10.00$&$0.97^{+0.07}_{-0.07}$&$1.05^{+0.07}_{-0.07}\times10^{-3}$&$1.70$&$<4.54\times10^{-4}$&1769&84.44/82.00\\
0039-44&0.346&2.56&$12.40^{+1.82}_{-1.76}$&$2.87^{+0.42}_{-0.18}$&$6.60^{+1.52}_{-1.52}\times10^{-6}$&$1.39^{+0.09}_{-0.17}$&$1.58^{+0.40}_{-0.51}\times10^{-4}$&1232/446/423&94.30/92.00\\
0043-42&0.116&2.70&$13.91^{+5.53}_{-4.08}$&$2.00$&$2.06^{+0.85}_{-0.86}\times10^{-6}$&$1.70$&$1.51^{+0.47}_{-0.40}\times10^{-4}$&203&6.32/5.00\\
0105-16&0.400&1.67&$16.13^{+7.34}_{-5.43}$&$1.59^{+0.18}_{-0.20}$&$2.65^{+0.21}_{-0.22}\times10^{-5}$&$1.50^{+0.41}_{-0.39}$&$1.84^{+2.01}_{-0.40}\times10^{-4}$&1687/708/661&125.16/137.00\\
0213-13&0.147&1.89&$18.40^{+4.66}_{-3.40}$&$1.69^{+1.05}_{-1.59}$&$4.62^{+1.23}_{-1.81}\times10^{-6}$&$1.77^{+0.18}_{-0.22}$&$1.31^{+0.64}_{-0.51}\times10^{-3}$&1150&45.85/50.00\\
0235-19&0.620&2.70&$0.00$&$1.43^{+0.25}_{-0.24}$&$7.30^{+1.07}_{-1.09}\times10^{-6}$&$1.43^{+0.25}_{-0.24}$&$7.30^{+1.07}_{-1.09}\times10^{-6}$&146/40/62&10.44/10.00\\
0252-71&0.566&3.66&$14.90^{+12.67}_{-7.09}$&$2.13^{+0.40}_{-0.33}$&$8.15^{+1.50}_{-2.14}\times10^{-6}$&$1.70$&$3.87^{+1.54}_{-1.04}\times10^{-5}$&373/103/105&20.40/24.00\\
0347+05&0.339&13.20&$74.85^{+39.25}_{-20.63}$&$1.95^{+0.25}_{-0.25}$&$1.54^{+0.17}_{-0.17}\times10^{-5}$&$1.70$&$1.57^{+1.42}_{-0.79}\times10^{-4}$&352/104/124&16.72/24.00\\
0349-27&0.066&1.00&$6.40^{+2.00}_{-1.52}$&$2.00$&$<2.67\times10^{-6}$&$1.70$&$2.55^{+0.49}_{-0.55}\times10^{-4}$&469&15.98/20.00\\
0404+03&0.089&12.10&$49.13^{+19.32}_{-15.61}$&$2.00$&$<1.05\times10^{-5}$&$1.70$&$2.78^{+1.72}_{-1.14}\times10^{-3}$&226&12.28/8.00\\
0409-75&0.693&8.71&$10.00$&$2.02^{+0.06}_{-0.07}$&$8.71^{+0.33}_{-0.35}\times10^{-5}$&$1.70$&$<4.64\times10^{-6}$&638/527/533&137.62/107.00\\
0442-28&0.147&2.32&$0.85^{+0.33}_{-0.31}$&$2.00$&$5.49^{+5.45}_{-5.30}\times10^{-5}$&$1.13^{+0.24}_{-0.17}$&$1.18^{+0.44}_{-0.30}\times10^{-3}$&2992&119.53/134.00\\
0620-52&0.051&5.32&$10.00$&$2.40^{+0.10}_{-0.10}$&$9.51^{+0.52}_{-0.52}\times10^{-5}$&$1.70$&$<1.27\times10^{-5}$&1070&39.24/47.00\\
0625-35&0.055&7.51&$5.99^{+2.56}_{-1.64}$&$2.00$&$1.11^{+0.09}_{-0.09}\times10^{-3}$&$1.70$&$2.74^{+0.57}_{-0.45}\times10^{-3}$&3940&221.05/173.00\\
0625-53&0.054&5.51&$10.00$&$2.00$&$<1.15\times10^{-5}$&$1.70$&$<8.51\times10^{-10}$&20&1.00/1.00\\
0806-10&0.110&7.65&$21.19^{+8.22}_{-6.18}$&$3.00^{+1.04}_{-1.37}$&$1.08^{+0.34}_{-0.43}\times10^{-5}$&$1.70$&$4.63^{+2.33}_{-1.52}\times10^{-4}$&449&18.17/18.00\\
0859-25&0.305&10.80&$38.25^{+21.16}_{-23.79}$&$1.61^{+0.41}_{-0.44}$&$9.13^{+1.53}_{-0.84}\times10^{-6}$&$1.67^{+0.39}_{-0.75}$&$1.81^{+31.00}_{-1.13}\times10^{-4}$&392/146/122&24.73/24.00\\
0915-11&0.054&4.94&$2.39^{+1.90}_{-1.35}$&$2.00$&$<2.11\times10^{-6}$&$1.35^{+0.60}_{-0.63}$&$2.49^{+4.58}_{-1.55}\times10^{-5}$&709/547&71.33/57.00\\
0945+07&0.086&3.01&$1.44^{+0.20}_{-0.18}$&$3.01^{+0.62}_{-0.57}$&$8.30^{+2.62}_{-2.52}\times10^{-5}$&$0.73^{+0.09}_{-0.09}$&$1.24^{+0.18}_{-0.14}\times10^{-3}$&5890/3778&468.26/434.00\\
1136-13&0.554&3.59&$6.06^{+4.33}_{-2.77}$&$2.00^{+0.09}_{-0.08}$&$3.22^{+0.08}_{-0.14}\times10^{-4}$&$1.48^{+0.44}_{-0.29}$&$1.38^{+0.22}_{-0.53}\times10^{-4}$&2970/17514&705.38/619.00\\
1151-34&0.258&7.70&$52.76^{+52.42}_{-28.83}$&$1.86^{+0.07}_{-0.07}$&$8.02^{+0.26}_{-0.28}\times10^{-5}$&$1.70$&$1.28^{+2.40}_{-0.90}\times10^{-4}$&2190/754/829&160.27/163.00\\
1306-09&0.464&3.03&$0.11^{+0.04}_{-0.04}$&$2.00$&$<2.24\times10^{-6}$&$1.77^{+0.09}_{-0.09}$&$7.25^{+0.62}_{-0.54}\times10^{-5}$&2317/823/806&163.12/169.00\\
1355-41&0.313&5.61&$0.27^{+0.24}_{+0.18}$&$2.00$&$1.11^{+0.39}_{-0.38}\times10^{-5}$&$1.70$&$2.84^{+8.48}_{-2.45}\times10^{-4}$&33250/11524/11095&843.39/722.00\\
1547-79&0.483&9.69&$99.28^{+594.53}_{-46.51}$&$2.00$&$1.60^{+0.18}_{-0.70}\times10^{-5}$&$1.70$&$2.51^{+1.86}_{-1.38}\times10^{-4}$&252/126/83&17.89/18.00\\
1559+02&0.104&6.42&$6.02^{+3.85}_{-2.89}$&$3.37^{+0.30}_{-0.25}$&$2.12^{+0.22}_{-0.26}\times10^{-5}$&$1.70$&$3.26^{+1.32}_{-0.97}\times10^{-5}$&635&19.22/23.00\\
1602+01&0.462&6.59&$0.00$&$1.68^{+0.03}_{-0.03}$&$2.79^{+0.06}_{-0.06}\times10^{-4}$&$1.68^{+0.03}_{-0.03}$&$2.79^{+0.06}_{-0.06}\times10^{-4}$&5052/2141/2063&396.19/362.00\\
1648+05&0.154&6.40&$10.00$&$0.80^{+1.34}_{-1.61}$&$5.01^{+1.85}_{-1.91}\times10^{-6}$&$1.70$&$<1.45\times10^{-3}$&31/80/&6.53/4.00\\
1733-56&0.098&8.89&$10.00$&$1.54^{+0.05}_{-0.05}$&$1.06^{+0.04}_{-0.04}\times10^{-3}$&$1.70$&$<1.67\times10^{-4}$&2991&142.22/133.00\\
1814-63&0.063&7.76&$2.00^{+0.28}_{-0.28}$&$2.00$&$2.03^{+1.35}_{-1.40}\times10^{-5}$&$1.26^{+0.18}_{-0.17}$&$1.90^{+0.23}_{-0.25}\times10^{-3}$&2795&119.13/126.00\\
1839-48&0.112&5.70&$10.00$&$1.35^{+0.23}_{-0.22}$&$1.17^{+0.21}_{-0.21}\times10^{-5}$&$1.70$&$<8.82\times10^{-6}$&183&6.98/8.00\\
1932-46&0.231&5.01&$0.00$&$1.82^{+0.08}_{-0.07}$&$6.20^{+0.29}_{-0.27}\times10^{-5}$&$1.82^{+0.08}_{-0.07}$&$6.20^{+0.29}_{-0.27}\times10^{-5}$&927/366/369&52.79/74.00\\
1934-63&0.183&6.15&$10.00$&$1.36^{+0.18}_{-0.18}$&$2.44^{+0.28}_{-0.28}\times10^{-5}$&$1.70$&$<1.88\times10^{-5}$&348&14.91/15.00\\
1938-15&0.452&9.66&$0.37^{+0.15}_{-0.12}$&$2.00$&$4.28^{+0.77}_{-1.89}\times10^{-5}$&$1.51^{+0.06}_{-0.08}$&$8.92^{+1.00}_{-0.63}\times10^{-5}$&2549/959/986&189.19/194.00\\
1949+02&0.059&14.80&$42.69^{+8.26}_{-3.06}$&$2.05^{+0.30}_{-0.29}$&$1.50^{+0.17}_{-0.17}\times10^{-5}$&$1.41^{+0.10}_{-0.14}$&$1.24^{+0.39}_{-0.49}\times10^{-3}$&1847&78.24/81.00\\
1954-55&0.060&4.61&$10.00$&$0.97^{+0.37}_{-0.38}$&$4.00^{+1.43}_{-1.38}\times10^{-6}$&$1.70$&$<7.34\times10^{-6}$&82&1.78/2.00\\
2135-14&0.200&4.73&$18.38^{+11.73}_{-7.97}$&$1.90^{+0.17}_{-0.12}$&$8.26^{+0.39}_{-0.39}\times10^{-4}$&$1.70$&$1.79^{+0.68}_{-0.41}\times10^{-3}$&2225&120.06/96.00\\
2135-20&0.635&3.38&$64.52^{+44.68}_{-24.56}$&$2.00$&$4.75^{+0.97}_{-1.01}\times10^{-6}$&$1.70$&$4.63^{+4.04}_{-2.60}\times10^{-5}$&167/31/46&12.24/14.00\\
2211-17&0.153&2.51&$10.00$&$2.00$&$<4.74\times10^{-6}$&$1.70$&$<4.87\times10^{-9}$&16&1.00/1.00\\
2221-02&0.057&5.01&$19.69^{+3.12}_{-1.97}$&$0.82^{+0.26}_{-0.29}$&$5.17^{+0.47}_{-0.50}\times10^{-5}$&$1.70$&$2.21^{+0.27}_{-0.17}\times10^{-3}$&3305&169.28/144.00\\
2250-41&0.310&1.48&$10.00$&$1.93^{+1.69}_{-1.16}$&$4.62^{+2.55}_{-2.77}\times10^{-6}$&$1.70$&$4.08^{+1640.00}_{-4.08}\times10^{-8}$&190/61/21&13.49/8.00\\
2314+03&0.220&5.22&$9.58^{+7.88}_{-4.17}$&$2.16^{+0.21}_{-0.19}$&$1.55^{+0.13}_{-0.17}\times10^{-5}$&$1.70$&$3.23^{+1.47}_{-0.79}\times10^{-5}$&586/195/209&67.19/46.00\\
2356-61&0.096&2.34&$14.68^{+1.50}_{-1.29}$&$3.08^{+0.57}_{-0.88}$&$8.43^{+2.02}_{-2.15}\times10^{-6}$&$1.70$&$9.82^{+0.98}_{-0.92}\times10^{-4}$&1107&47.84/47.00\\
&&&&&&&&\\\hline
\end{tabular}
\end{table*}

\subsection{Other Data}\label{Multidata}

\edit{As outlined in Section \ref{Intro}, multiwavelength data for the 2Jy sample were taken in a systematic manner, so that all the objects would have comparable measurements. This also allows us to establish a direct comparison with the existing data and analysis on the 3CRR sources \citep{Hardcastle2006,Hardcastle2009}.}

We used the VLA and ATCA data at 5 GHz (both for overall and core luminosities) from \citet{Morganti1993,Morganti1999}. Since only some of the 2Jy sources are covered by the Parkes catalog \citep{Parkes1990}, we calculated the spectral index from 408 MHz and 1.4 GHz observations (also from the Parkes catalog) and extrapolated the results to 178 MHz. We used this same spectral index to extrapolate the 151 MHz fluxes, needed to calculate the jet kinetic power (see Section \ref{Jet}). The low-frequency fluxes for PKS 1934-63 are upper limits, since the source is self-absorbed in radio.

For the infrared, we used 24 $\mu$m data taken by Spitzer, from \citet{Dicken2008,Dicken2009}. All the targets in the 2Jy sample have deep Spitzer and Herschel observations at 24, 70, 100 and 160 $\mu$m, and $\sim90$ per cent (including all the targets in the steep-spectrum subsample) have Spitzer/IRS mid-IR spectra \citep{Dicken2012}. The 3C sources were observed at 15 $\mu$m (rest-frame), \edit{a band that is similar enough to Spitzer's 24 $\mu$m (after rest-frame correction) to allow direct comparison}. We studied the behaviour of a number of sources at both wavelengths, and estimated that the deviation in luminosity caused by the difference between 15 and 24 $\mu$m was well below 10 per cent in all cases.

For the optical line classification we used the complete, deep GEMINI G-MOS-S data from \citet{Tadhunter1993,Tadhunter1998}. \edit{K-band magnitudes of the host galaxies} were taken from \citet{Inskip2010} and K-corrected using the relations given by \citet{Glazebrook1995} and \citet{Mannucci2001}. The values presented in the Tables are K-corrected.

For the 3CRR sources we used the data from \citet{Hardcastle2006,Hardcastle2009}. In this case the 178 MHz fluxes were measured as part of the sample definition, but 1.4 GHz and 151 MHz fluxes had to be extrapolated from these measurements and the 178-750 MHz spectral indices\footnote{For the complete database see \url{http://3crr.extragalactic.info/}}. Details of the 3CRR data are given in Appendix \ref{3C_appendix}.

%

\section{The X-ray 2Jy sample}\label{Xray}

\begin{figure}
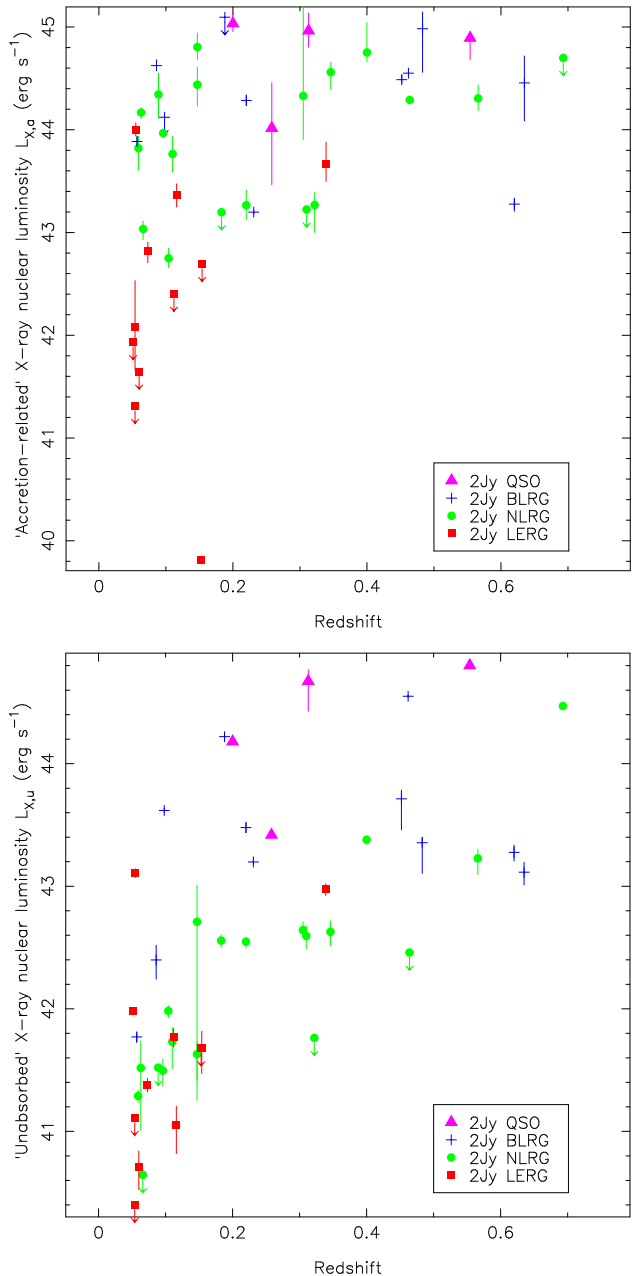

\begin{minipage}[t]{0.99\linewidth}
\centering
\includegraphics[width=0.99\textwidth]{z_XA_no3C}
\vspace{8pt}
\end{minipage}
\begin{minipage}[t]{0.99\linewidth}
\includegraphics[width=0.99\textwidth]{z_XJ_no3C}
\end{minipage}
\caption{X-ray luminosities for the 2Jy sources. Top: X-ray luminosity for the accretion-related component $L_{X_{a}}$ against redshift. Bottom: X-ray luminosity for the unabsorbed component $L_{X_{u}}$ against redshift. Red squares represent LERGs, green circles NLRGs, blue crosses BLRGs, and purple triangles QSOs. Arrows indicate upper limits.}\label{z_XA_no3C}
\end{figure}

In our analysis of the X-ray emission of the 2Jy objects we observe trends similar to those observed by \citet{Hardcastle2006,Hardcastle2009} for the 3CRR sources. The luminosity distribution of the sources versus redshift is as expected, \edit{with a large number of low-luminosity sources at low $z$, and mostly brighter objects detected at high $z$ (see Figure \ref{z_XA_no3C}). This effect is, at least in part, caused by the detection limits and sample selection criteria, but also by the well-known evolution of the AGN population with redshift.}

\begin{figure}
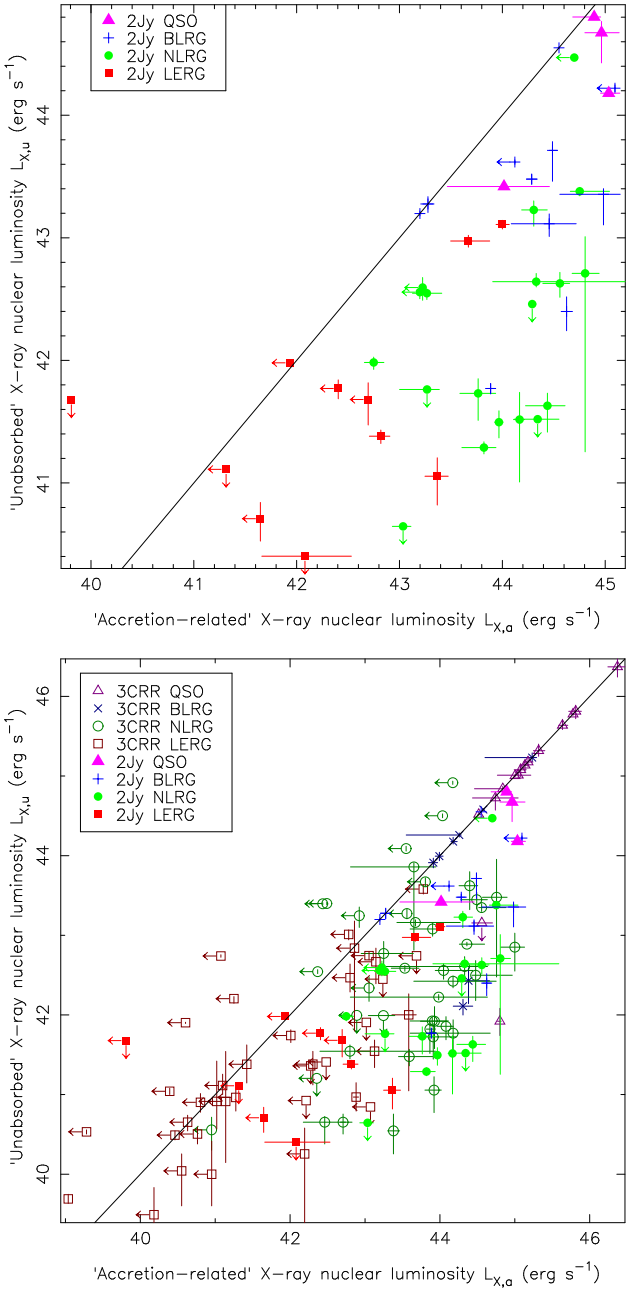

\begin{minipage}[t]{0.99\linewidth}
\centering
\includegraphics[width=0.99\textwidth]{XA_XJ_no3C}
\vspace{8pt}
\end{minipage}
\begin{minipage}[t]{0.99\linewidth}
\includegraphics[width=0.99\textwidth]{XA_XJ}
\end{minipage}
\caption{X-ray luminosity for the unabsorbed component $L_{X_{u}}$ against X-ray luminosity for the `accretion-related' component $L_{X_{a}}$. Top: only the 2Jy sources are plotted. Bottom: both the 2Jy and the 3CRR sources are plotted. Arrows indicate upper limits. Colours and symbols as in Figure \ref{z_XA_no3C}. A y=x line has been plotted as visual aid for the reader; this line does not represent a correlation.}\label{XA_XJ}
\end{figure}

It is important to keep in mind that the luminosities we derive for the X-ray components may suffer from contamination from each other. \edit{This effect is particularly evident in the broad-line and quasar-like objects. In these objects there is little or no intrinsic absorption to allow us to distinguish both components, thus we adopt the same value for $L_{X_{u}}$ and $L_{X_{a}}$. This effect can be seen in both panels of Figure \ref{z_XA_no3C}, and Figure \ref{XA_XJ}, where a few BLRGs and QSOs seem to have systematically higher luminosities than the rest of their populations.}

These plots show a distinct separation between the different emission-line populations. Low-excitation objects have much lower accretion-related X-ray emission than any of the other groups. This is consistent with the hypothesis in which LERGs lack the traditional radiatively efficient accretion features characteristic of the high-excitation population \citep[see e.g.][]{Hardcastle2007b}. The separation between narrow-line (NLRG) and broad-line (BLRG) objects is more striking in \edit{the bottom panel of Figure \ref{z_XA_no3C}} due both to the possible contamination by jet emission in broad-line objects, and to the influence of relativistic beaming, which `boosts' the soft X-ray emission in objects whose jets are viewed at small inclination angles.

The four LERGs that fall in the NLRG parameter space in \edit{the top panel of Figure \ref{z_XA_no3C} (having high, well-constrained $L_{X_{a}}$) may be, in fact, radiatively efficient objects}. 3C 15 (PKS 0034-01) is very luminous and has a relatively well constrained, obscured, hard component (see Section \ref{0034-01}). Although we do not detect unequivocal signs of a radiatively efficient accretion disk, in the form of an emission Fe K$\alpha$ line, this could be due to the low statistics, rather than the absence of the line itself. PKS 0043-42 does have a Fe K$\alpha$ line, and \citet{RamosAlmeida2011} find IR evidence for a torus (see also Section \ref{0043-42}). PKS 0625-35 (Section \ref{0625-35}) is extremely bright and is suspected to be a BL-Lac \citep{Wills2004}. In this case it is hard to tell whether there is any contamination from the jet emission on the accretion-related component, causing us to overestimate its luminosity, or whether this object is radiatively efficient in nature.

A special mention should be made of PKS 0347+05. This object \redit{was originally classified as a BLRG}, but recent evidence suggests that this is, in fact, a double system, with a LERG and a radio-quiet Seyfert 1 in close interaction (see Section \ref{0347+05}). We have decided to keep this object in our plots and classify it as a LERG based on its optical spectrum \citep{Tadhunter2012}, for consistency with the rest of our analysis, though it is a clear outlier in most of our plots.

\edit{The top panel of Figure \ref{XA_XJ}} shows the distribution of 2Jy sources according to the relation between their unabsorbed and accretion-related X-ray luminosities. Each population occupies a different area in the parameter space, with a certain degree of overlap between the brighter NLRGs and fainter BLRGs, as can be expected from unification models. For the same reason, there is some overlap between the fainter NLRGs and the brighter LERGs. However, it is evident from Figure \ref{XA_XJ} that LERGs have a much lower $L_{X_{a}}$/$L_{X_{u}}$ ratio than any of the other populations. \edit{The relative faintness of $L_{X_{a}}$ in LERGs} reinforces the conclusions from the previous paragraph about the nature of accretion in LERGs. Adding the 3CRR objects makes this even more evident, as can be seen in the equivalent plot by \citet{Hardcastle2009}. As in \edit{the top panel of Figure \ref{z_XA_no3C}}, the four `efficient' LERGs seem to fall in the parameter space occupied by NLRGs.

\begin{table}\scriptsize
\caption{Objects for which a Fe K$\alpha$ emission line was detected. \edit{errors are calculated at $90$ per cent confidence}. Where no errors are quoted the parameter had to be fixed for the overall model fit. \edit{For PKS 1151-34 the line energy had to be fixed after exploring the statistical space with the XSPEC command \textit{steppar}, since the program was not able to automatically find the best fit.}}\label{Fe_table}
\centering
\setlength{\tabcolsep}{1.6pt}
\setlength{\extrarowheight}{3pt}
\begin{tabular}{ccc}\hline
Source name&rest-frame energy&eq. width\\
&(keV)&(keV)\\\hline
0039-44&$6.32^{+0.68}_{-0.18}$&0.06\\
0043-42&$6.48^{+0.32}_{-0.05}$&0.88\\
0105-16&$6.22^{+0.78}_{-0.22}$&0.09\\
0409-75&$6.68^{+0.09}_{-0.15}$&0.44\\
0859-25&$6.51^{+0.47}_{-0.10}$&0.28\\
1151-34&$6.34$&0.10\\
1559+02&$6.44^{+0.05}_{-0.05}$&4.00\\
1814-63&$6.40^{+0.09}_{-0.07}$&0.15\\
1938-15&$6.51^{+0.07}_{-0.06}$&0.16\\
2221-02&$6.37^{+0.05}_{-0.05}$&0.17\\
2356-61&$6.30^{+0.08}_{-0.07}$&0.14\\\hline
\end{tabular}
\end{table}

\section{Correlations}\label{Correlations}

As described in Section \ref{Fitting}, from the analysis of the X-ray cores we derived the luminosity of the unabsorbed ($L_{X_{u}}$) and accretion-related components ($L_{X_{a}}$). For our analysis we compared these luminosities with those derived from the 178 MHz, 5 GHz (core), 24 $\mu$m and [OIII] fluxes, all of which are displayed in Table \ref{lumin_table}. As in the case of the 3CRR objects \citep{Hardcastle2009}, the 2Jy sample is a flux-limited sample, thus correlations are expected in luminosity-luminosity plots. We tested for partial correlation in the presence of redshift to account for this, following the method and code described by \citet{Akritas1996}, which takes into account upper limits in the data. \edit{In this method, $\tau$ is equivalent to Kendall's $\tau$, and $\sigma$ reprsents the dispersion of the data; we therefore consider the $\tau/\sigma$ ratio to assess the significance of the correlation.} The results of the partial correlation analysis are given in Table \ref{2Jy_correlations_table}. \edit{We have only added to the table results that add scientifically relevant information to those presented by \citet{Hardcastle2009}, rather than the full analysis.}

\begin{table*}\scriptsize
\caption{\edit{Luminosities for the sources in the 2Jy sample, following the format of \citet{Hardcastle2009}. The values are given as the logarithm of the luminosity in erg s$^{-1}$, upper limits are indicated with a `$<$' before the value. The columns represent, from left to right, the object name, classification, redshift, and luminosities at 178 MHz, 5 GHz (core), soft (jet-related) and hard (accretion-related) X-rays (followed by their respective 90 per cent confidence lower and upper bounds), mid-IR and [OIII].\edit{ The X-ray accretion-related luminosities have been corrected for intrinsic absorption.} We have converted the radio and IR luminosity densities into $\nu L_{\nu}$ to allow for direct comparison between the magnitudes in different bands. The errors for the radio, IR and [OIII] luminosities can be found in the original papers, listed in Section \ref{Multidata}. Where measurements could not be obtained their absence is indicated with a dash. The object types from Table \ref{objects_table} have been abbreviated as follows: E stands for LERG, N for NLRG, B for BLRG, Q for Quasar.}}\label{lumin_table}
\centering
\setlength{\tabcolsep}{2.5pt}
\setlength{\extrarowheight}{2pt}
\begin{tabular}{ccccccccccccc}\hline
PKS&Type&$z$&L$_{178}$&L$_{5}$&L$_{X_{u}}$&L$_{X_{u}}+$&L$_{X_{u}}-$&L$_{X_{a}}$&L$_{X_{a}}+$&L$_{X_{a}}$-&L$_{IR}$&L$_{[OIII]}$\\\hline
0023-26&N&0.322&43.16&-&$<$41.76&-&-&43.27&43.00&43.39&44.008&42.18\\
0034-01&E&0.073&41.60&41.25&41.38&41.32&41.43&42.82&42.71&42.91&43.079&40.49\\
0035-02&B&0.220&42.84&42.55&43.48&43.44&43.52&44.29&44.23&44.34&44.299&42.08\\
0038+09&B&0.188&42.55&41.54&44.22&44.19&44.25&$<$45.10&-&-&44.505&42.18\\
0039-44&N&0.346&43.13&40.65&42.63&42.51&42.72&44.56&44.39&44.66&45.219&43.04\\
0043-42&E&0.116&42.16&40.97&41.06&40.82&41.21&43.37&43.25&43.47&43.678&40.70\\
0105-16&N&0.400&43.49&40.61&43.38&43.34&43.41&44.75&44.66&45.04&44.835&42.40\\
0213-13&N&0.147&42.33&-&41.63&41.41&41.73&44.44&44.23&44.61&43.903&42.11\\
0235-19&B&0.620&43.88&-&43.28&43.21&43.34&43.28&43.21&43.34&45.350&43.28\\
0252-71&N&0.566&43.76&-&43.23&43.10&43.30&44.31&44.19&44.44&44.671&42.15\\
0347+05&B&0.339&42.93&40.33&42.97&42.92&43.02&43.67&43.50&43.88&44.224&40.96\\
0349-27&N&0.066&41.79&39.80&$<$40.65&-&-&43.03&42.93&43.11&43.056&41.08\\
0404+03&N&0.089&41.68&40.10&$<$41.52&-&-&44.34&44.11&44.55&43.878&41.46\\
0409-75&N&0.693&44.38&41.27&44.47&44.45&44.49&$<$44.70&-&-&44.599&42.11\\
0442-28&N&0.147&42.69&40.98&42.71&41.25&43.01&44.81&44.68&44.94&44.205&41.84\\
0620-52&E&0.051&41.29&40.89&41.98&41.96&42.00&$<$41.94&-&-&42.548&$<$39.41\\
0625-35&E&0.055&41.18&41.31&43.11&43.07&43.14&44.00&43.94&44.07&43.349&$<$40.48\\
0625-53&E&0.054&41.72&40.14&$<$41.11&-&-&$<$41.31&-&-&42.173&$<$40.04\\
0806-10&N&0.110&42.24&40.89&41.73&41.51&41.85&43.77&43.59&43.94&45.000&42.77\\
0859-25&N&0.305&43.26&42.08&42.64&42.60&42.71&44.33&43.91&45.59&44.542&41.98\\
0915-11&E&0.054&42.53&40.89&$<$40.40&-&-&42.08&41.66&42.53&42.920&40.46\\
0945+07&B&0.086&42.19&40.44&42.40&42.24&42.52&44.62&44.57&44.68&44.051&41.90\\
1136-13&Q&0.554&43.60&-&44.80&44.78&44.81&44.89&44.68&44.96&45.326&43.73\\
1151-34&Q&0.258&42.71&-&43.42&43.40&43.43&44.02&43.46&44.46&44.622&42.45\\
1306-09&N&0.464&43.14&-&$<$42.46&-&-&44.29&44.26&44.33&44.664&42.15\\
1355-41&Q&0.313&42.96&41.65&44.67&44.43&44.77&44.96&44.80&45.14&45.325&42.89\\
1547-79&B&0.483&43.46&40.95&43.36&43.11&43.40&44.98&44.56&45.15&44.941&43.43\\
1559+02&N&0.104&42.06&40.55&41.98&41.93&42.03&42.75&42.66&42.85&44.932&42.26\\
1602+01&B&0.462&43.70&42.25&44.55&44.54&44.56&44.55&44.54&44.56&44.884&42.81\\
1648+05&E&0.154&43.63&40.41&41.68&41.47&41.82&$<$42.69&-&-&43.174&40.65\\
1733-56&B&0.098&41.89&41.88&43.62&43.60&43.64&$<$44.12&-&-&43.952&41.81\\
1814-63&N&0.063&42.12&-&41.52&41.01&41.74&44.17&44.11&44.22&43.885&40.63\\
1839-48&E&0.112&41.97&41.36&41.77&41.69&41.84&$<$42.40&-&-&43.086&$<$39.36\\
1932-46&B&0.231&43.38&41.59&43.20&43.18&43.22&43.20&43.18&43.22&43.696&42.38\\
1934-63&N&0.183&$<$43.39&-&42.56&42.50&42.60&$<$43.20&-&-&44.302&42.08\\
1938-15&B&0.452&43.52&41.32&43.71&43.46&43.79&44.49&44.46&44.53&44.807&42.88\\
1949+02&N&0.059&41.55&39.58&41.29&41.24&41.34&43.82&43.61&43.94&44.290&41.86\\
1954-55&E&0.060&41.57&40.27&40.71&40.53&40.84&$<$41.65&-&-&42.429&$<$39.00\\
2135-14&Q&0.200&42.49&41.76&44.18&44.16&44.20&45.03&44.95&45.14&45.176&43.11\\
2135-20&B&0.635&43.71&-&43.12&43.01&43.20&44.46&44.09&44.72&45.020&43.15\\
2211-17&E&0.153&42.86&39.74&$<$41.68&-&-&$<$39.81&-&-&42.593&40.38\\
2221-02&B&0.057&41.74&40.46&41.77&41.73&41.81&43.89&43.85&43.93&44.315&42.23\\
2250-41&N&0.310&43.22&40.55&42.59&42.49&42.68&$<$43.22&-&-&44.654&42.70\\
2314+03&N&0.220&42.99&42.82&42.55&42.50&42.58&43.27&43.13&43.41&44.948&42.20\\
2356-61&N&0.096&42.64&40.57&41.50&41.37&41.59&43.97&43.92&44.01&44.075&41.95\\\hline
\end{tabular}
\end{table*}

While the relations between these luminosities can provide some insight into the physical processes going on in each source, it is important to keep in mind that there are several intrinsic effects that limit this insight, orientation, beaming, variability and environmental interference being perhaps the most relevant. These effects are also the most likely cause of scatter in the plots that we present in the following Sections. In this paper we therefore describe the correlations between these luminosities without reference to any particular model, merely attempting to establish the physical scenarios and measurement systematics that may cause these correlations to arise.

To allow direct comparison with the results of \citet{Hardcastle2009}, we have plotted both the 2Jy and the 3CRR objects in our Figures.\edit{ The bottom panel of Figure \ref{XA_XJ}} summarises the X-ray characteristics of both populations. In terms of sample size, we have multiwavelength luminosities for 45 2Jy objects and 135 3CRR sources (although in the latter the data are less complete, see the tables in Section \ref{3C_appendix}), more than doubling the number of objects studied by \citet{Hardcastle2009}.

The differences between the LERGs and HERGs observed in \edit{the top panel of Figure \ref{XA_XJ} are highlighted by the addition of the 3CRR objects (bottom panel)}, though it is also clearer that there is an overlap in the parameter space between BLRGs and NLRGs. M 87, 3C 326, and 3C 338, originally listed as NLRGs by \citet{Hardcastle2006,Hardcastle2009} have since been re-classified as LERGs \citep[][]{Buttiglione2009}. The LERG 3C 123 is probably more appropriately classified as a reddened NLRG, and the X-ray spectrum of 3C 200 is compatible with that of a radiatively efficient AGN, despite its LERG classification \citep[see Appendix A of][]{Hardcastle2006}.

\begin{figure}
\centering
\includegraphics[width=0.45\textwidth]{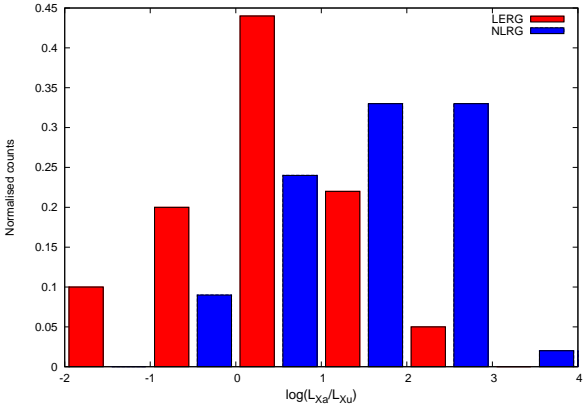}
\caption{Histogram of the $L_{X_{a}}/L_{X_{u}}$ for the 2Jy and 3CRR LERGs and NLRGs. Broad-line objects are excluded to avoid contamination. The ratios for LERGs are upper limits.}\label{L_ratio}
\end{figure}

\edit{Figure \ref{L_ratio} shows the ratio between $L_{X_{a}}$ and $L_{X_{u}}$ for the 2Jy and 3CRR LERGs and NLRGs. We have not included the broad-line objects in the plot because, even in the case where both components can be distinguished, contamination from each other and beaming may be an issue. It is quite clear in this plot that NLRGs have a systematically higher $L_{X_{a}}/L_{X_{u}}$, which is even more relevant when we consider the fact that for the vast majority of the LERGs we only have upper limits for $L_{X_{a}}$. This histogram already hints at the different nature of accretion and energy output in LERGs and HERGs. To fully separate the accretion-related contribution from the jet component, however, and to interpret these results, further analysis is needed. We address this issue in detail in Section \ref{Jet}.}

There are also some differences between the 2Jy and 3CRR populations, which can be partly attributed to the slightly different selection criteria used in both samples, and which may cause the 2Jy sample to have more beamed objects (as discussed in Section \ref{Sample}), as well as issues with sample completeness in the latter sample (the 3CRR sample is nearly complete in X-rays for low-$z$ objects, but not so for $z>0.5$). While we consider that these effects do not invalidate our results, it is essential to keep in mind that any selection criteria for an AGN sample introduce a certain bias. We will discuss other possible sources of bias in Section \ref{Eddington}.

\begin{table}\scriptsize
\caption[Partial correlations]{Results of partial correlation analysis described in Section \ref{Correlations}. The number of objects for each correlation is given in column 4, and it includes all the objects in the corresponding subsample given in column 3. The last column indicates the strength of the partial correlation between the quantities in columns 1 and 2 in the presence of redshift. We consider the correlation significant if $\tau/\sigma > 3$.}\label{2Jy_correlations_table}
\centering
\setlength{\tabcolsep}{2pt}
\setlength{\extrarowheight}{3pt}
\begin{tabular}{ccccccc}\hline
x&y&subsample&n&$\tau$&$\sigma$&$\tau/\sigma$\\\hline
$L_{178}$&$L_{Xu}$&2Jy+3CRR NLRG&106&0.214&0.045&4.726\\
&&2Jy NLRG&19&0.243&0.141&1.727\\
$L_{178}$&$L_{Xa}$&all&147&0.112&0.028&3.947\\
&&2Jy+3CRR HERG&99&0.113&0.038&2.969\\
&&2Jy+3CRR LERG&47&0.013&0.031&0.423\\
$L_{5}$&$L_{Xu}$&all&137&0.436&0.043&10.043\\
&&2Jy HERG+LERG&35&0.412&0.091&4.525\\
&&2Jy+3CRR, QSOs excluded&120&0.379&0.047&8.047\\
&&2Jy, QSOs excluded&33&0.395&0.102&3.886\\
$L_{5}$&$L_{Xa}$&all&137&0.252&0.046&5.531\\
&&2Jy+3CRR, QSOs excluded&120&0.143&0.046&3.090\\
&&2Jy+3CRR LERG&47&0.146&0.055&2.648\\
$L_{IR}$&$L_{Xu}$&all&117&0.338&0.054&6.297\\
&&2Jy+3CRR HERG&80&0.243&0.072&3.394\\
$L_{IR}$&$L_{Xa}$&all&117&0.476&0.046&10.440\\
&&2Jy+3CRR HERG&80&0.384&0.063&6.132\\
$L_{[OIII]}$&$L_{Xa}$&all&122&0.412&0.044&9.319\\
&&2Jy+3CRR HERG&86&0.323&0.057&5.665\\
$L_{178}$&$L_{IR}$&all&139&0.186&0.036&5.141\\
&&2Jy+3CRR HERG&102&0.195&0.047&4.172\\
&&2Jy+3CRR NLRG&59&0.168&0.060&2.782\\
&&2Jy+3CRR LERG&37&0.093&0.075&1.241\\
$L_{178}$&$L_{[OIII]}$&all&133&0.182&0.034&5.290\\
&&2Jy+3CRR HERG&96&0.188&0.044&4.242\\
&&2Jy+3CRR NLRG&53&0.138&0.056&2.474\\
&&2Jy+3CRR LERG&37&0.113&0.065&1.741\\
$L_{IR}$&$L_{[OIII]}$&all&111&0.586&0.064&9.126\\
&&2Jy HERG+LERG&45&0.660&0.101&6.504\\
&&2Jy+3CRR HERG&79&0.514&0.068&7.614\\
&&2Jy HERG&35&0.579&0.100&5.776\\
$L_{[OIII]}$&$Q$&2Jy+3CRR HERG&87&0.136&0.048&2.824\\
&&2Jy+3CRR NLRG&45&0.079&0.064&1.235\\
$L_{IR}$&$Q$&2Jy+3CRR HERG&87&0.154&0.050&3.063\\
&&2Jy+3CRR NLRG&45&0.143&0.079&1.813\\
$Q$&$L_{edd}$&all&102&0.274&0.096&2.851\\
&&2Jy+3CRR HERG&62&0.243&0.148&1.644\\
&&2Jy+3CRR NLRG&52&0.350&0.183&1.916\\
&&2Jy+3CRR LERG&40&0.213&0.109&1.955\\\hline
\end{tabular}
\end{table}

\subsection{X-ray/Radio correlations}\label{X_R}

\begin{figure}
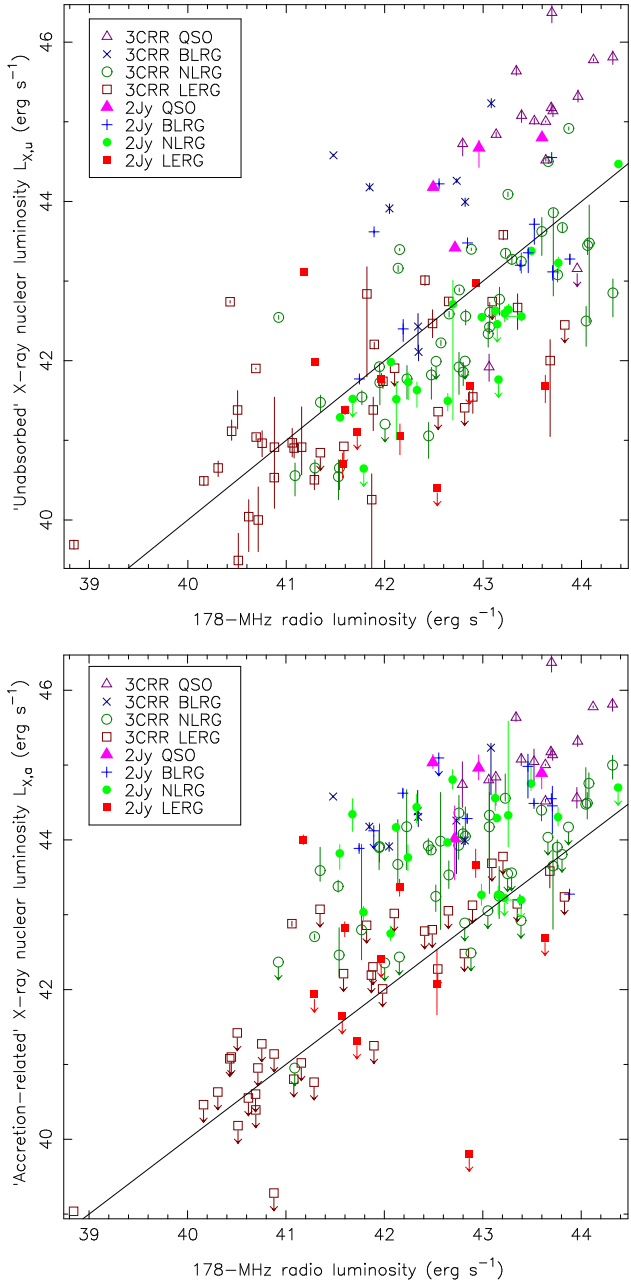

\begin{minipage}[t]{0.99\linewidth}
\centering
\includegraphics[width=0.99\textwidth]{R_XJ}
\vspace{8pt}
\end{minipage}
\begin{minipage}[t]{0.99\linewidth}
\includegraphics[width=0.99\textwidth]{R_XA}
\end{minipage}
\caption{Top: X-ray luminosity for the unabsorbed component $L_{X_{u}}$ as a function of 178 MHz total radio luminosity. Bottom: X-ray luminosity for the `accretion-related' component $L_{X_{a}}$ as a function of 178 MHz total radio luminosity. Both the 2Jy and the 3CRR sources are plotted. Arrows indicate upper limits. Colours and symbols as in Figure \ref{z_XA_no3C}. Line as in Figure \ref{XA_XJ}.}\label{R_XJ}
\end{figure}

\edit{The 178 MHz luminosity is not only an indicator of the time-averaged jet power, but also of the age of the source, and is related to the properties of the external environment \citep{Hardcastle2013}. By adding the 2Jy sources to the $L_{178}$/$L_{X_{u}}$ plot (top panel of Figure \ref{R_XJ}), a correlation between these quantities for the NLRGs is more readily apparent than it was for \citet{Hardcastle2009}, despite the scatter, and is significant in the partial correlation analysis (Table \ref{2Jy_correlations_table}). Although the 2Jy objects on their own do not show a significant correlation, the larger number of objects with respect to those of \citet{Hardcastle2009} enhances the significance of the correlation. Because of the fact that the 2Jy sample is statistically complete, this also allows us to rule out that the results previously obtained for the 3CRR sources are biased, as well as adding to the overall statistics.} 

\edit{The situation is not so clear for the BLRGs and QSOs, most likely due to the contamination from the accretion-related component. In the case of the LERGs the scatter is expected due to the fact that there are no selection effects on orientation. All of this suggests that there may be a weak physical link between the unabsorbed X-ray power (prior to beaming correction) and the overall radio power (related to the time-averaged AGN power).}

There is no apparent correlation between $L_{178}$/$L_{X_{u}}$ if only the 2Jy sources are considered (see Table \ref{2Jy_correlations_table}). This is most likely due to the large scatter in the jet-related quantities, and $L_{X_{u}}$ in particular, caused by the presence of beamed objects in the 2Jy sample, a consequence of the selection criteria, as well as the low number of sources. In fact, the value of $\tau$ in the $L_{178}$/$L_{X_{u}}$ correlation when only the 2Jy sources are considered is larger than it is for the combined 2Jy and 3CRR samples, but the scatter (indicated by $\sigma$) is much larger in the former case, resulting in $\tau/\sigma <3$. 

By contrast, and as already pointed out by \citet{Hardcastle2009}, there seems to be a strong correlation between $L_{178}$/$L_{X_{a}}$ for all the populations excluding the LERGs, which seem to lie mostly below the correlation (see bottom panel of Figure \ref{R_XJ} and Table \ref{2Jy_correlations_table}). The BLRGs and QSOs are not clearly outlying in this plot, despite the contamination from the jet-related X-ray component.

\begin{figure}
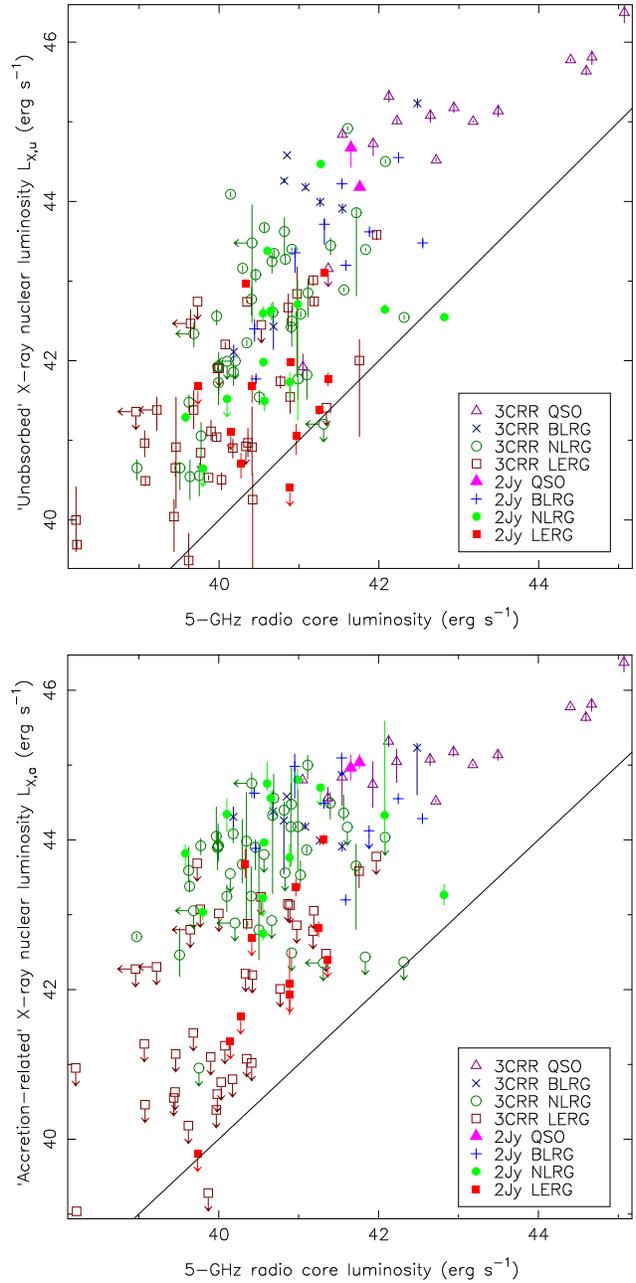

\begin{minipage}[t]{0.99\linewidth}
\centering
\includegraphics[width=0.99\textwidth]{RC_XJ}
\vspace{8pt}
\end{minipage}
\begin{minipage}[t]{0.99\linewidth}
\includegraphics[width=0.99\textwidth]{RC_XA}
\end{minipage}
\caption{Top: X-ray luminosity for the unabsorbed component $L_{X_{u}}$ as a function of 5 GHz radio core luminosity. Bottom: X-ray luminosity for the `accretion-related' component $L_{X_{a}}$ as a function of 5 GHz radio core luminosity. Both the 2Jy and the 3CRR sources are plotted. Arrows indicate upper limits. Colours and symbols as in Figure \ref{z_XA_no3C}. Line as in Figure \ref{XA_XJ}.}\label{RC_XJ}
\end{figure}

\edit{The top panel of Figure \ref{RC_XJ}} shows the relation between the 5 GHz core luminosity and the unabsorbed X-ray component. The correlation between these quantities is strong, despite the scatter, due to the fact that both quantities are subject to beaming. The fact that the LERGs lie in the same correlation as the NLRGs is evidence for the jet-related nature of the soft X-ray component in radio-loud sources \citep[see e.g.][and references within]{Worrall1987,Hardcastle2009}. \edit{The soft component observed in radio-quet AGN (either caused by reflection of the hard component on the accretion disk in the radiatively efficient AGN, or Comptonization in the radiatively inefficient sources) must still exist in radio-loud objects; in the latter, however, the jet-related emission dominates in the soft X-ray regime.} 

\edit{The bottom panel of Figure \ref{RC_XJ}} shows the relation between the 5 GHz core luminosity and the accretion-related X-ray component. In this plot it becomes apparent that the LERGs show a distinct behaviour, completely apart from the high-excitation population, and consistent with the hypothesis that these objects have a different accretion mechanism. The correlation between these two quantities is less strong than between $L_{5GHz,core}$ and $L_{X_{u}}$ (Table \ref{2Jy_correlations_table}), and all but disappears if the QSOs are removed. 

\edit{Correlations between both X-ray luminosities and the 5 GHz radio core luminosity are expected due to their mutual dependence on redshift. If the X-ray luminosity were simply related to the time-averaged AGN power, and independent from orientation and beaming, it would not be strongly correlated to the 5 GHz core luminosity (although there is a jet-disk connection relating both quantities, the scatter is larger than for purely jet-related components, weakening the correlation; see also Section \ref{accretion}). As argued by e.g. \citet{Hardcastle1999}, Doppler beaming can introduce up to three orders of magnitude of scatter in these correlations, given its strong influence on $L_{5 GHz,core}$. The correlation we observe between $L_{5GHz,core}$ and $L_{X_{u}}$, in particular, reinforces the hypothesis that the soft X-ray flux is related to jet emission in radio-loud sources.}

\subsection{X-ray/IR correlations}\label{X_I}

\edit{The main source of uncertainty in $L_{IR}$ comes from the dependence with the orientation of the dusty torus, which is believed to introduce a large uncertainty \citep[see e.g.][and references therein]{Hardcastle2009,Runnoe2012}. It is possible that some of the broad-line objects have some contamination from non-thermal (synchrotron) emission from the jet, although the dominant contribution to the mid-IR is dust-reprocessed emission from the torus. We discuss this point further later in this Section.}

\begin{figure}
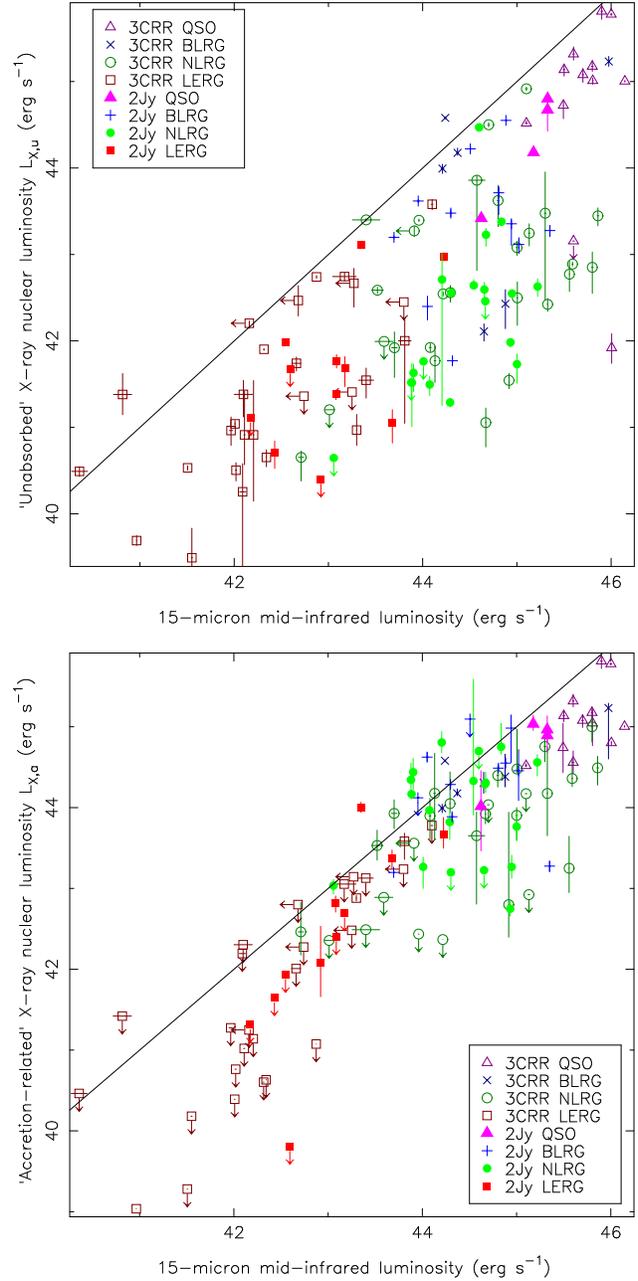

\begin{minipage}[t]{0.99\linewidth}
\centering
\includegraphics[width=0.99\textwidth]{I_XJ}
\vspace{8pt}
\end{minipage}
\begin{minipage}[t]{0.99\linewidth}
\includegraphics[width=0.99\textwidth]{I_XA}
\end{minipage}
\caption{Top: X-ray luminosity for the unabsorbed component $L_{X_{u}}$ as a function of total infrared (24 $\mu$m for the 2Jy sources, 15 $\mu$m for the 3CRR sources) luminosity. Bottom: X-ray luminosity for the `accretion-related' component $L_{X_{a}}$ as a function of total infrared (24 $\mu$m for the 2Jy sources, 15 $\mu$m for the 3CRR sources) luminosity. Both the 2Jy and the 3CRR sources are plotted. Arrows indicate upper limits. Colours and symbols as in Figure \ref{z_XA_no3C}. Line as in Figure \ref{XA_XJ}.}\label{I_XJ}
\end{figure}

Despite the large scatter, there is an evident overall correlation between $L_{IR}$ and $L_{X_{u}}$ \edit{(top panel of Figure \ref{I_XJ})}, which was already visible in the plots of \citet{Hardcastle2009} (see Table \ref{2Jy_correlations_table}). The 2Jy sources fill some of the gaps left by the 3CRR sources in the parameter space. The correlation disappears for individual populations, however. \edit{In the broad-line objects, it is possible that $L_{X_{u}}$ is affected by beaming.}

\redit{The correlation between $L_{IR}$ and $L_{X_{a}}$ is very strong \edit{(bottom panel of Figure \ref{I_XJ} and Table \ref{2Jy_correlations_table})}. The correlation is expected, since both luminosities are indicators of the overall power of the accretion disk. Some of the scatter in this correlation is likely to come from the fact that $L_{IR}$ is more dependent on orientation than $L_{X_{a}}$, and the way in which the latter is affected by obscuration (objects with a much larger $L_{IR}$ than $L_{X_{a}}$ are likely to be Compton-thick). The correlation between $L_{IR}$ and $L_{X_{a}}$ holds for radio-quiet objects at all orientations \citep[see e.g. the results of][on local Seyferts]{Gandhi2009,Asmus2011}, which suggests that non-thermal emission from the jet is either not affecting the quantities involved in the correlation, or is equally boosting both, as may be the case for some of the broad line objects with strong radio cores in our sample.}

Some of the NLRGs in our sample are quite heavily obscured, and we could only constrain an upper limit to their absorption column and accretion-related X-ray luminosity. These objects are probably Compton-thick, and lie to the lower right of the correlation in this plot. The most extreme example of \edit{such} behaviour is PKS 2250-41. PKS 1559+02 shows the largest departure from the correlation among the NLRGs, having a very small $L_{X_{a}}$ component when compared to $L_{IR}$, and is probably Compton-thick. The BLRG PKS 0235-19 is also very underluminous in X-rays, and a clear outlier in \edit{the bottom panel of Figure \ref{I_XJ}, which} is not expected for a broad-line object.

The behaviour of the LERGs in this figure is most significant, reinforcing the idea that LERGs cannot be explained as heavily obscured, `traditional', radiatively efficient AGN. LERGs are underluminous in X-rays, and lie below the correlation for HERGs. Adding an intrinsic absorption column $N_{H}=10^{24}$ cm$^{-2}$ is still insufficient to boost the X-ray luminosity of most of these objects enough to situate them on the correlation. The overlap between the populations happens mostly for objects whose emission-line classification is inconsistent with our best estimate of the accretion mode (the radiatively efficient LERGs mentioned in Section \ref{Xray}), and because of the large scatter caused by systematics.

The origin of the IR emission in `inefficient' LERGs should be questioned. We know from cases like M 87 that no accretion-related component is detected on small scales \citep[see Section 4.1 in][]{Hardcastle2009}, although IR emission is measured with \textit{Spitzer}. It is very likely that in these LERGs the IR emission is associated with the jet and the old stellar population, and is therefore not reliable as an estimator of accretion.

\subsection{X-ray/[OIII] correlations}\label{X_O}

\begin{figure}
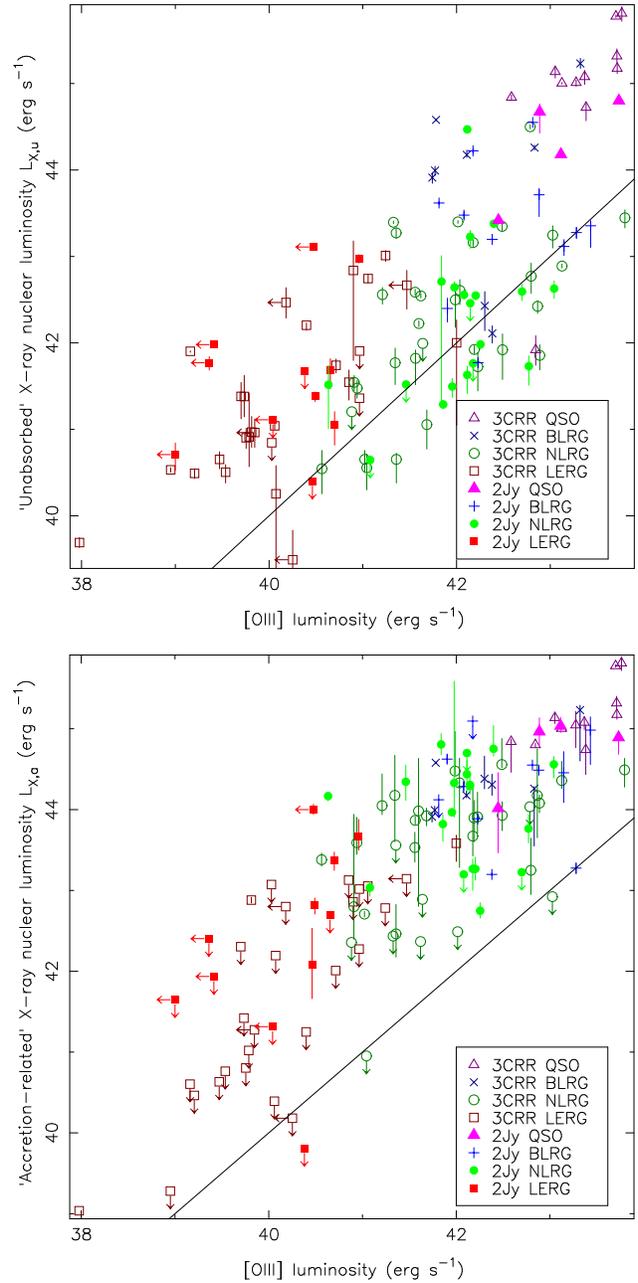

\begin{minipage}[t]{0.99\linewidth}
\centering
\includegraphics[width=0.99\textwidth]{O_XJ}
\vspace{8pt}
\end{minipage}
\begin{minipage}[t]{0.99\linewidth}
\includegraphics[width=0.99\textwidth]{O_XA}
\end{minipage}
\caption{Top: X-ray luminosity for the unabsorbed component $L_{X_{u}}$ against the [OIII] emission line luminosity. Bottom: X-ray luminosity for the `accretion-related' component $L_{X_{a}}$ against the [OIII] emission line luminosity. Both the 2Jy and the 3CRR sources are plotted. Arrows indicate upper limits. Colours and symbols as in Figure \ref{z_XA_no3C}. Line as in Figure \ref{XA_XJ}.}\label{O_XJ}
\end{figure}

The relation between the [OIII] and jet-related X-ray luminosity is shown in \edit{the top panel of Figure \ref{O_XJ}}. This plot is surprising in that it separates the populations quite clearly. \edit{This separation} is not expected a priori, since [OIII] traces the photoionizing power of the AGN, which is directly related to accretion, and not directly dependent on jet power, which is traced by $L_{X_{u}}$. \edit{The NLRG PKS 0409-75 is an outlier in the plot, having a much higher $L_{X_{u}}$ ($>10^{44}$ erg s$^{-1}$) than is expected from its $L_{[OIII]}$}. As detailed in Sections \ref{Fitting} and \ref{0409-75}, it is possible that the soft X-ray component in this source suffers from contamination from \redit{ inverse-Compton emission from the radio lobes}, since this object is in a dense environment.

\edit{The LERGs are underluminous in [OIII], as expected, and show a great deal of scatter due to the effect of the random orientation on their X-ray emission. Broad-line objects have boosted X-ray luminosities both due to beaming and to contamination from the accretion-related component, and lie towards the top right corner of the plot. The relative faintness in [OIII] of some objects can be explained by obscuration, as suggested by \citet{Jackson1990}. Obscuration, and the presence or absence of contamination from the accretion-related component in some broad-line objects, introduce scatter in this plot, and separate the BLRGs and QSOs from the NLRGs.}

As pointed out by \citet{Hardcastle2009}, there is a strong correlation between $L_{[OIII]}$ and $L_{X_{a}}$ (Table \ref{2Jy_correlations_table} and \edit{the bottom panel of Figure \ref{O_XJ})}, given that both quantities directly trace accretion \citep[see also][and Dicken et al. 2014, submitted]{Dicken2009}. As in the case of the correlation between $L_{IR}$ and $L_{X_{a}}$, the LERGs fall below the correlation expected for high-excitation objects (excepting the few `efficient' LERGs mentioned before). The scatter in this plot is much higher than that seen in \edit{the bottom panel of Figure \ref{I_XJ}}. Infrared emission is a better indicator of accretion than [OIII], since it is less contaminated by the jet and stellar processes, as well as easier to measure \edit{\citep[see also e.g. ][]{Dicken2009}.}

As for the case of \edit{the bottom panel of Figure \ref{I_XJ}}, PKS 1559+02 and PKS 2250-41 also fall below the correlation in \edit{the bottom panel of Figure \ref{O_XJ}}, reinforcing the hypothesis that these objects are Compton-thick. PKS 0235+05 is also an outlier in this plot, with a much lower $L_{X_{a}}$ than is expected for a BLRG.

\subsection{Radio/IR/[OIII] correlations}\label{R_IR}

\begin{figure}
\centering
\includegraphics[width=0.45\textwidth]{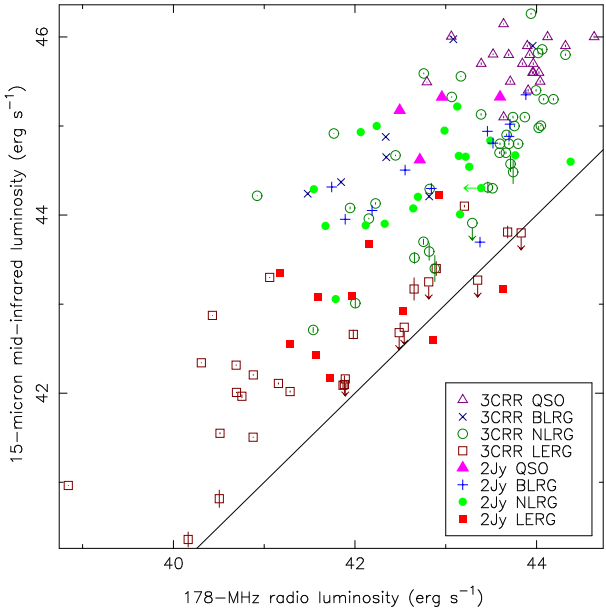}
\caption{Total infrared (24 $\mu$m for the 2Jy sources, 15 $\mu$m for the 3C sources) luminosity against the 178 MHz total radio luminosity. Arrows indicate upper limits. Colours and symbols as in Figure \ref{z_XA_no3C}. Line as in Figure \ref{XA_XJ}.}\label{R_I}
\end{figure}

\begin{figure}
\centering
\includegraphics[width=0.45\textwidth]{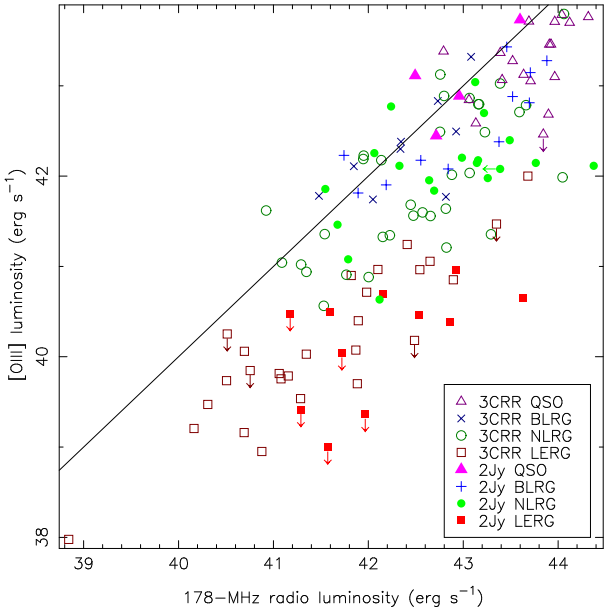}
\caption{[OIII] emission line luminosity against the 178 MHz total radio luminosity. Arrows indicate upper limits. Colours and symbols as in Figure \ref{z_XA_no3C}. Line as in Figure \ref{XA_XJ}.}\label{R_O}
\end{figure}

\citet{Hardcastle2009} found correlations between the overall radio luminosity and the infrared and [OIII] luminosities. We observe the same in our plots and correlation analysis (Figures \ref{R_I} and \ref{R_O}, and Table \ref{2Jy_correlations_table}), with the 2Jy sources filling some of the gaps in the parameter space. The LERGs have higher (relative) radio luminosities than the other populations, as expected. Beaming is likely to introduce scatter in the radio luminosity in both plots, while orientation is likely to influence the scatter in IR luminosities. For the 3CRR objects it can be seen that the broad-line objects have systematically higher [OIII] luminositites than narrow-line objects for the same luminosity \edit{\citep[see Figure \ref{R_O} and Figure 11 of ][]{Hardcastle2009}, but the situation is not so clear for the 2Jy sources alone, due to their redshift distribution.}

By contrast, and as observed by \citet{Hardcastle2009}, the radio core luminosity is not well correlated with either $L_{IR}$ or $L_{[OIII]}$. The QSOs have radio cores that are far more luminous than those of the other classes. All the populations, in fact, seem to be in different regions of the parameter space, with the broad-line objects having more luminous radio cores than the narrow-line objects for the same $L_{IR}$ and $L_{[OIII]}$ due to beaming, and LERGs being fainter in both plots, but also more radio-luminous, in proportion, than NLRGs.

\begin{figure}
\centering
\includegraphics[width=0.45\textwidth]{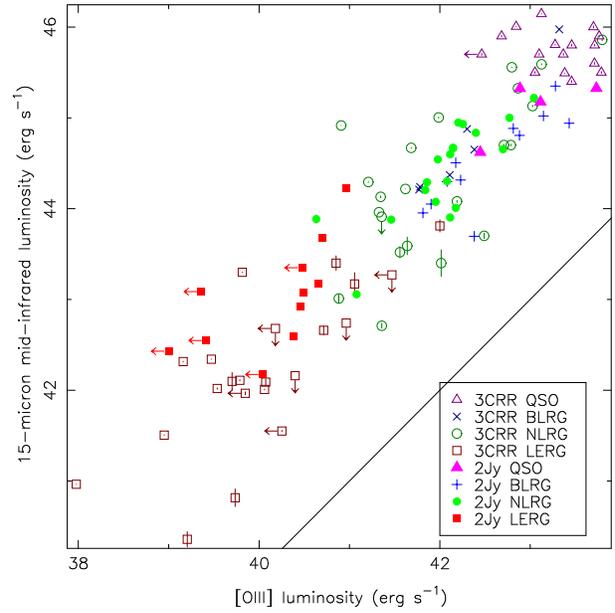}
\caption{Total infrared (24 $\mu$m for the 2Jy sources, 15 $\mu$m for the 3C sources) luminosity against the [OIII] emission line luminosity. Arrows indicate upper limits. Colours and symbols as in Figure \ref{z_XA_no3C}. Line as in Figure \ref{XA_XJ}.}\label{O_I}
\end{figure}

The correlation between $L_{IR}$ and $L_{[OIII]}$ is very strong (Figure \ref{O_I} and Table \ref{2Jy_correlations_table}), and made much clearer by the addition of the 2Jy objects.\redit{The recent results of Dicken et al. (2014, submitted) suggest that both quantities are affected to the same degree by orientation/extinction effects. Moreover, neither quantity is likely to be affected by beaming} (unless non-thermal contamination is substantial), which greatly reduces the scatter. Contamination from the jet is also likely to favour both quantities, mostly the IR emission, by the addition of a non-thermal component, \edit{but if shock-ionization is involved [OIII] emission may be boosted as well. Although the contributions from either mechanism are likely to be very different, and change for individual objects, they must be kept in mind. While expected, it is interesting to note that the scatter is much smaller when considering $L_{IR}$ and $L_{[OIII]}$, rather than the X-ray luminosities, where variability is much larger due to the shorter timescales involved.}

\section{Jet power and Eddington rates}\label{Eddington}

One of the hypotheses that has gained more strength in recent years over the mechanisms underlying accretion in LERGs postulates that there is an accretion rate switch between these objects and the high-excitation population at about $1-10$ per cent of the Eddington rate \citep[see e.g.][and references therein]{Best2012,Russell2012}. In this Section we aim to test this hypothesis, taking into account not just the radiative power from the AGN, but also the kinetic power of the jet, denoted $Q$ throughout this Section, after the definition of \citet{Willott1999}.

\subsection{Jet power estimations}\label{Jet}

\edit{To estimate the jet kinetic power we considered two possible correlations: that of \citet{Cavagnolo2010}, which relies on 1.4-GHz measurements, and of \citet{Willott1999}, which is derived from 151-MHz fluxes, with a correction factor $f=15$ \citep[see discussion in][]{Hardcastle2009}. \citet{Cavagnolo2010} derived their correlation from X-ray cavity measurements; this method, as pointed out by \citet{Russell2012}, is subject to uncertainties in the volume estimations and on how much of the \edit{accretion-derived} AGN power is actually transferred to the interstellar/intergalactic medium. Given that the objects in our samples are far more powerful than the ones considered by \citet{Cavagnolo2010}, it is possible that their correlation underestimates the jet powers in our case, but it is the best estimate based on actual data. \citet{Willott1999} derived their correlation from minimum energy synchrotron estimates and [OII] emission line measurements, which make the slope of the correlation somewhat uncertain, as well as introducing an additional uncertainty (in form of the factor $f$) in the normalization. }

As suggested by \citet{Croston2008b}, the particle content and energy distributions in FRI and FRII systems is probably very different \citep[but see also][]{Godfrey2013}, and we know there is a dependence of the jet luminosity with the environment \citep[jets are more luminous in denser environments, see e.g.][]{Hardcastle2013}, it is very likely that, a priori, a single correlation cannot be used across the entire population of radio-loud objects. However, \citeauthor{Godfrey2013} find that such a correlation does work, and conclude that environmental factors and spectral ageing `conspire' to reduce the radiative efficiency of FRII sources, effectively situating them on the same $Q_{jet}-L_{151}$ correlation as the low-power FRI galaxies. \edit{This effect} makes the use of these correlations qualitatively inaccurate, but quantitatively correct, within the assumptions, as approximations to the jet kinetic power.

We have repeated the luminosity versus jet power plots of \citet{Godfrey2013} for our sources, using both the \citet{Cavagnolo2010} and \citet{Willott1999} correlations, and we find them to agree very well, with slight divergences at the high and low ends of the distribution due to the different shapes of both correlations. For our analysis we have used the relation of \citet{Willott1999}, both for consistency with the analysis of \citet{Hardcastle2006,Hardcastle2009}, and because of the relatively higher reliability of low-frequency measurements. As a further check, we have compared the jet power we obtained for PKS 2211-17 with that obtained independently by \citet{Croston2011}, and have found them to agree within the uncertainties.

\edit{We thus derive the jet kinetic power, Q from the relation shown in eq. 12 of \citet{Willott1999}:}
\begin{equation}
Q=3 \times 10^{38}L_{151}^{6/7} W
\end{equation}\label{Willott} \edit{where $L_{151}$ is the luminosity at 151 MHz, in units of $10^{28}$ W Hz$^{-1}$ sr$^{-1}$.}

\subsection{Black hole masses, bolometric corrections and Eddington rates}\label{BH_bolo}

We calculated the black hole masses for the objects in our sample from the K$_{s}$-band magnitudes of \citet{Inskip2010} and a slight variation of the well-known correlation between these quantities and the black hole mass \citep{Graham2007}. We cross-tested the results with the black hole masses obtained from the r'-band magnitudes of \citet{RamosAlmeida2010} \citep[using the conversions to the B band and the corrections of][]{Fukugita1995} and the relations from \citet{Graham2007}, and found them to be mostly consistent, save for an overall effect that might be related to the different apertures used (the B-band derived masses tend to be smaller). 

15 of our objects are missing from the work of \citet{Inskip2010}. We obtained \textit{2MASS} magnitudes for some of them, so 11 sources do not have K-band measurements and are thus missing from the following tables and plots. Of these, 3 are QSOs, 4 BLRGs, 3 NLRGs and 1 LERG. Given that the black hole masses derived from K-band magnitudes for broad-line objects and QSOs are not reliable, we can assume that our sample is adequately covered. A further source of uncertainty for the $M_{BH}-L_{K}$ correlation originates from the fact that black hole masses in clusters are expected to be systematically higher \citep[see e.g.][]{Volonteri2012}. This is particularly important for LERGs inhabiting rich environments, a point we return to in the next Section.

When cross-checking \textit{UKIRT} and \textit{2MASS} observations for the 3CRR sources we found five objects where differences greater than 0.4 mag (after aperture and K corrections) were present between both instruments. After checking these discrepancies carefully, we have relied on \textit{2MASS} measurements whenever possible. It is important to keep in mind not only the limitations of the available data, but also the large degree of scatter present in the correlation of \citet{Graham2007}.

\begin{figure}
\centering
\includegraphics[width=0.45\textwidth]{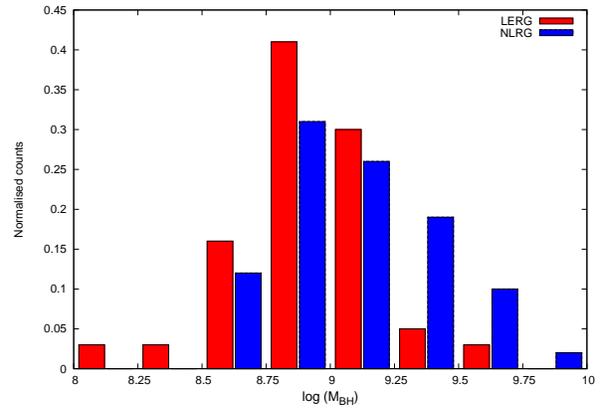}
\caption{Histogram of black hole masses for the 2Jy and 3CRR LERGs and NLRGs. Only narrow-line HERGs are included, to allow comparison between both samples and avoid issues with the unreliability of K-band derived black hole masses in broad-line objects.}\label{M_BH}
\end{figure}

\edit{The black hole masses for the 2Jy and 3CRR sources are given in Tables \ref{BH_table} and \ref{3C_BH_table}, respectively. We have plotted the histogram distribution of black hole masses in Figure \ref{M_BH}, to illustrate the range of masses covered and to investigate any systematic differences between LERGs and HERGs. We can see that the NLRGs tend to have slightly larger $M_{BH}$ than the LERGs, though there is no clear cut between the two populations. As mentioned earlier, this could be partly due to observational biases, and the fact that we are probably underestimating black hole masses for systems embedded in rich clusters, where most of the LERGs lie. The range of black hole masses could be contributing to the scatter in our plots.}

\edit{We derived the bolometric luminosity from the different bands, and studied their consistency. We used the correlations of \citet[][eq.21 ]{Marconi2004} for the X-ray 2-10 keV luminosity (L$_{X_{a}}$):
\begin{equation}
\log(L/L_{2-10keV})=1.54+0.24 \mathcal{L} +0.012 \mathcal{L}^{2} - 0.0015 \mathcal{L}^{3}
\end{equation} where $\mathcal{L}=(log(L)-12)$, and $L$ is the bolometric luminosity in units of $L_{\odot}$.} \edit{ We used the simple relation of \citet{Heckman2004} for the [OIII] luminosity ($L_{bol}=3500 L_{[OIII]}$) and the relation of \citet[][eq. 8]{Runnoe2012} for the IR luminosity at 24 $\mu$m:
\begin{equation}
\log(L_{iso})=(15.035 \pm 4.766) + (0.688 \pm 0.106) log(\lambda L_{\lambda})
\end{equation} where $L_{iso}$ assumes an isotropic bolometric luminosity (\citeauthor{Runnoe2012} recommend that a correction be made to account for orientation effects, so that $L_{bol}\sim 0.75 L_{iso}$, but we do not apply this correction). These bolometric luminosties obtained for the different bands are shown in Table \ref{BH_table}. }

It is worth noting that all these relations are a subject of debate. The $L_{X,2-10keV}$/$L_{bol}$ relation was initially postulated for bright quasars \citep{Elvis1994}, and although more complex relations like that of \citet{Marconi2004} agree with the initial results, they cannot be be fully applied to low-luminosity and low-excitation sources \citep[see e.g.][]{Ho2009}. The mid-IR luminosity seems to be a very reliable estimator of the bolometric luminosity of an AGN, despite issues with non-thermal contamination where a jet is present \citep[see e.g.][]{FdezOntiveros2012}, and a minor contribution from star formation. The main issue with this correlation lies in the dependence on orientation, which can introduce a bias of up to $40$ per cent \citep[see e.g.][]{Runnoe2012}. [OIII] has been widely used to assess the bolometric luminosity, given that the conversion factor between the two is just a constant, but it is not reliable when there are other sources of photoionization, it is known to underestimate the bolometric luminosity in low-excitation sources \citep[see e.g.][]{Netzer2009}, and is also orientation-dependent \citep{Jackson1990,Dicken2009}. 

\begin{table*}\scriptsize
\caption{K-band magnitudes, K-corrections \citep[calculated using the relations of][]{Glazebrook1995,Mannucci2001}, absolute magnitudes, black hole masses, Eddington luminosities, X-ray, [OIII] and infrared-derived Eddington ratios and jet Eddington ratios for the sources in the 2Jy sample. The K-band magnitudes from \citet{Inskip2010} are marked as I10 in the reference column, the magnitudes taken directly from the 2MASS catalogue are marked as 2M. The errors quoted for $L_{X,rad}/L_{X,Edd}$ are derived from both the errors in the X-ray powerlaw normalization and the errors in the intrinsic $N_H$, to show the maximum possible uncertainty. Where N$_H$ was fixed to $10^{23}$ cm$^{-2}$, the upper and lower values of the X-ray luminosity were calculated for $N_{H}=0$ and $N_{H}=10^{24}$ cm$^{-2}$ respectively. E stands for LERG, N for NLRG, B for BLRG, Q for Quasar.}\label{BH_table}
\centering
\setlength{\tabcolsep}{2.0pt}
\setlength{\extrarowheight}{3pt}
\begin{tabular}{ccccccccccccc}\hline
PKS&Type&Ref&$z$&mag K$_{s}$&K-corr&Mag K$_{s}$&$M_{BH}$&$L_{Edd}$&$L_{X,rad}$/$L_{X,Edd}$&$L_{[OIII],rad}$/$L_{[OIII],Edd}$&$L_{IR,rad}$/$L_{IR,Edd}$&$Q/L_{Edd}$\\\hline
&&&&&&&$\times10^{9}$ M$_{\odot}$&$\times10^{40}$ W&&&&\\\hline
0023-26&N&I10&0.322&15.036&-0.604&-26.70&1.67&2.17&$1.76^{+0.10}_{-0.11}\times10^{-3}$&$2.42\times10^{-2}$&$1.15\times10^{-2}$&$9.10\times10^{-2}$\\
0034-01&E&I10&0.073&12.569&-0.183&-25.21&0.53&0.69&$1.50^{+0.28}_{-0.28}\times10^{-3}$&$1.56\times10^{-3}$&$8.28\times10^{-3}$&$1.23\times10^{-2}$\\
0035-02&B&I10&0.220&14.107&-0.482&-26.47&1.40&1.81&$4.41^{+0.54}_{-0.47}\times10^{-2}$&$2.32\times10^{-2}$&$2.17\times10^{-2}$&$5.25\times10^{-2}$\\
0038+09&B&I10&0.188&14.299&-0.428&-25.94&0.93&1.21&$8.14^{+0.08}_{-0.56}\times10^{-1}$&$4.34\times10^{-2}$&$4.57\times10^{-2}$&$4.82\times10^{-2}$\\
0039-44&N&I10&0.346&15.411&-0.622&-26.53&1.46&1.89&$9.81^{+1.97}_{-1.84}\times10^{-2}$&$2.03\times10^{-1}$&$8.99\times10^{-2}$&$9.79\times10^{-2}$\\
0043-42&E&I10&0.116&12.999&-0.283&-25.94&0.94&1.22&$4.18^{+0.82}_{-0.65}\times10^{-3}$&$1.44\times10^{-3}$&$1.22\times10^{-2}$&$2.29\times10^{-2}$\\
0105-16&N&I10&0.400&15.419&-0.649&-26.91&1.96&2.55&$1.33^{+0.64}_{-0.39}\times10^{-1}$&$3.44\times10^{-2}$&$3.64\times10^{-2}$&$1.38\times10^{-1}$\\
0213-13&N&I10&0.147&13.502&-0.349&-26.07&1.03&1.33&$9.57^{+3.05}_{-2.85}\times10^{-2}$&$3.41\times10^{-2}$&$1.59\times10^{-2}$&$2.98\times10^{-2}$\\
0347+05&B&I10&0.339&14.286&-0.617&-27.59&3.28&4.27&$2.31^{+1.66}_{-0.66}\times10^{-2}$&$7.46\times10^{-4}$&$8.21\times10^{-3}$&$2.72\times10^{-2}$\\
0349-27&E&I10&0.066&12.853&-0.166&-24.68&0.36&0.46&$4.16^{+0.56}_{-0.55}\times10^{-3}$&$9.04\times10^{-3}$&$1.20\times10^{-2}$&$1.23\times10^{-2}$\\
0404+03&N&I10&0.089&13.417&-0.221&-24.85&0.41&0.53&$1.80^{+0.83}_{-0.62}\times10^{-1}$&$1.91\times10^{-2}$&$3.84\times10^{-2}$&$1.88\times10^{-2}$\\
0442-28&N&I10&0.147&13.160&-0.349&-26.41&1.33&1.73&$2.30^{+0.10}_{-0.09}\times10^{-1}$&$1.40\times10^{-2}$&$1.96\times10^{-2}$&$3.52\times10^{-2}$\\
0620-52&E&2M&0.051&9.801&-0.129&-27.11&2.27&2.95&$3.06^{+9.80}_{-3.06}\times10^{-5}$&$3.09\times10^{-5}$&$8.18\times10^{-4}$&$1.09\times10^{-3}$\\
0625-35&E&I10&0.055&10.724&-0.139&-26.36&1.29&1.68&$2.00^{+0.23}_{-0.18}\times10^{-2}$&$6.27\times10^{-4}$&$5.19\times10^{-3}$&$2.13\times10^{-3}$\\
0625-53&E&I10&0.054&10.042&-0.137&-27.00&2.09&2.72&$6.41^{+0.02}_{-6.41}\times10^{-6}$&$1.41\times10^{-4}$&$4.91\times10^{-4}$&$4.24\times10^{-3}$\\
0806-10&N&I10&0.110&12.137&-0.269&-26.67&1.62&2.11&$7.84^{+2.01}_{-1.69}\times10^{-3}$&$9.78\times10^{-2}$&$5.71\times10^{-2}$&$8.91\times10^{-3}$\\
0859-25&N&I10&0.305&14.758&-0.589&-26.83&1.83&2.38&$3.84^{+3.72}_{-2.05}\times10^{-2}$&$1.39\times10^{-2}$&$2.44\times10^{-2}$&$1.01\times10^{-1}$\\
0915-11&E&I10&0.054&10.868&-0.137&-26.18&1.12&1.45&$9.17^{+1.23}_{-0.62}\times10^{-5}$&$6.98\times10^{-4}$&$2.88\times10^{-3}$&$3.79\times10^{-2}$\\
0945+07&B&I10&0.086&12.376&-0.214&-25.81&0.84&1.10&$2.08^{+0.10}_{-0.07}\times10^{-1}$&$2.56\times10^{-2}$&$2.39\times10^{-2}$&$1.54\times10^{-2}$\\
1151-34&Q&2M&0.258&14.040&-0.537&-27.08&2.22&2.88&$1.23^{+1.53}_{-0.15}\times10^{-2}$&$3.40\times10^{-2}$&$2.29\times10^{-2}$&$2.77\times10^{-2}$\\
1306-09&N&I10&0.464&15.120&-0.666&-27.61&3.33&4.33&$1.88^{+0.02}_{-0.04}\times10^{-2}$&$1.13\times10^{-2}$&$1.63\times10^{-2}$&$4.35\times10^{-2}$\\
1355-41&Q&I10&0.313&12.744&-0.597&-28.91&8.95&11.63&$5.60^{+1.69}_{-0.12}\times10^{-2}$&$2.32\times10^{-2}$&$1.73\times10^{-2}$&$1.10\times10^{-2}$\\
1547-79&B&I10&0.483&15.185&-0.669&-27.66&3.44&4.47&$1.51^{+44.55}_{-0.91}\times10^{-1}$&$2.11\times10^{-1}$&$2.45\times10^{-2}$&$7.76\times10^{-2}$\\
1559+02&N&I10&0.104&12.205&-0.256&-26.46&1.38&1.80&$4.75^{+0.14}_{-1.67}\times10^{-4}$&$3.50\times10^{-2}$&$5.90\times10^{-2}$&$1.34\times10^{-2}$\\
1648+05&E&2M&0.154&12.550&-0.363&-27.14&2.33&3.03&$2.42^{+554.08}_{-2.42}\times10^{-4}$&$5.21\times10^{-4}$&$2.18\times10^{-3}$&$1.61\times10^{-1}$\\
1733-56&B&I10&0.098&12.485&-0.242&-26.03&1.00&1.30&$3.75^{+1.53}_{-0.03}\times10^{-2}$&$1.76\times10^{-2}$&$1.74\times10^{-2}$&$1.54\times10^{-2}$\\
1814-63&N&I10&0.063&11.896&-0.159&-25.52&0.68&0.88&$6.34^{+0.22}_{-0.46}\times10^{-2}$&$1.71\times10^{-3}$&$2.23\times10^{-2}$&$2.64\times10^{-2}$\\
1839-48&E&2M&0.112&11.841&-0.274&-27.01&2.11&2.74&$1.19^{+1.35}_{-1.19}\times10^{-4}$&$2.94\times10^{-5}$&$2.14\times10^{-3}$&$6.66\times10^{-3}$\\
1932-46&B&I10&0.231&14.971&-0.499&-25.84&0.86&1.12&$6.90^{+0.09}_{-0.18}\times10^{-3}$&$7.49\times10^{-2}$&$1.35\times10^{-2}$&$2.72\times10^{-1}$\\
1934-63&N&I10&0.183&14.023&-0.419&-26.14&1.09&1.41&$2.20^{+12.07}_{-2.20}\times10^{-3}$&$2.98\times10^{-2}$&$2.86\times10^{-2}$&$2.73\times10^{-2}$\\
1949+02&N&I10&0.059&11.333&-0.149&-25.92&0.92&1.20&$1.63^{+0.81}_{-0.43}\times10^{-2}$&$2.10\times10^{-2}$&$3.32\times10^{-2}$&$6.30\times10^{-3}$\\
2135-14&Q&2M&0.200&12.404&-0.449&-28.00&4.47&5.81&$1.40^{+0.38}_{-1.40}\times10^{-1}$&$7.82\times10^{-2}$&$2.74\times10^{-2}$&$8.54\times10^{-3}$\\
2211-17&E&I10&0.153&13.422&-0.361&-26.25&1.18&1.54&$2.81^{+0.16}_{-2.81}\times10^{-7}$&$5.46\times10^{-4}$&$1.72\times10^{-3}$&$7.27\times10^{-2}$\\
2221-02&B&I10&0.057&11.448&-0.144&-25.73&0.79&1.03&$2.31^{+0.25}_{-0.22}\times10^{-2}$&$5.77\times10^{-2}$&$4.17\times10^{-2}$&$3.20\times10^{-3}$\\
2250-41&N&I10&0.310&15.508&-0.594&-26.12&1.07&1.40&$4.35^{+28.50}_{-4.35}\times10^{-4}$&$1.25\times10^{-1}$&$4.97\times10^{-2}$&$1.56\times10^{-1}$\\
2356-61&N&I10&0.096&12.559&-0.237&-25.90&0.91&1.18&$2.58^{+0.19}_{-0.22}\times10^{-2}$&$2.68\times10^{-2}$&$2.35\times10^{-2}$&$3.24\times10^{-2}$\\
&&&&&&&&&&&&\\\hline
\end{tabular}
\end{table*}

\begin{figure}
\begin{minipage}[t]{0.48\textwidth}
\includegraphics[clip=true, trim=1.6cm 0cm 1.6cm 0cm, width=0.9\textwidth]{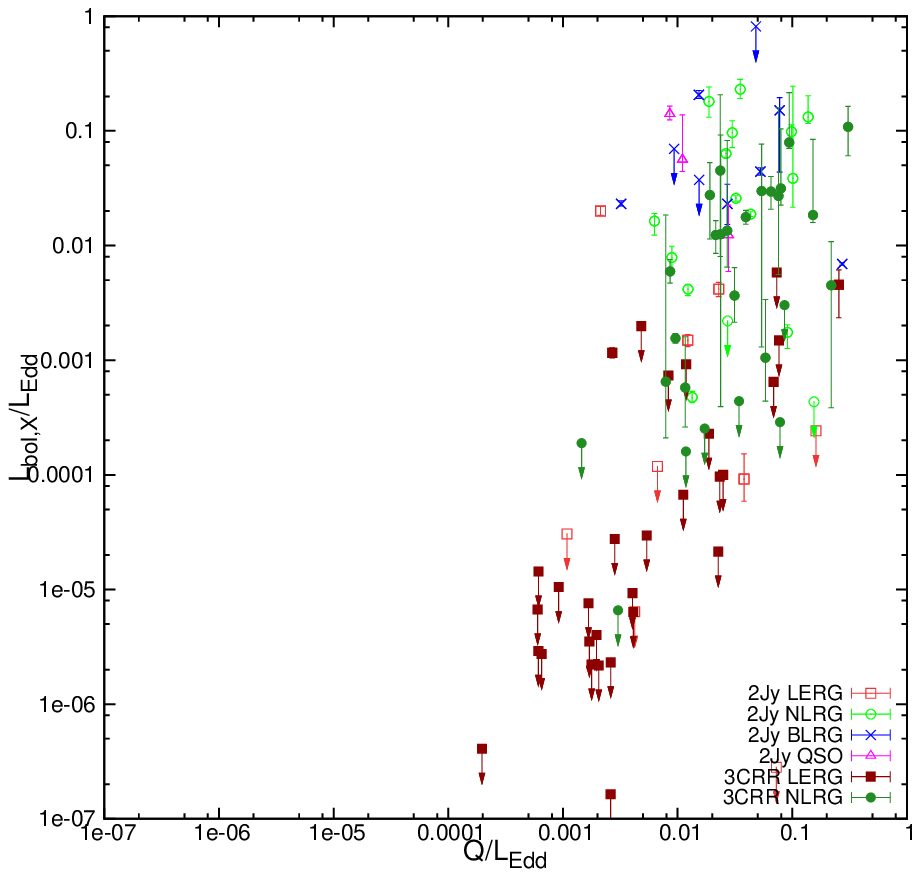}
\vspace{8pt}
\end{minipage}
\begin{minipage}[t]{0.48\textwidth}
\includegraphics[clip=true, trim=1.6cm 0cm 1.6cm 0cm, width=0.9\textwidth]{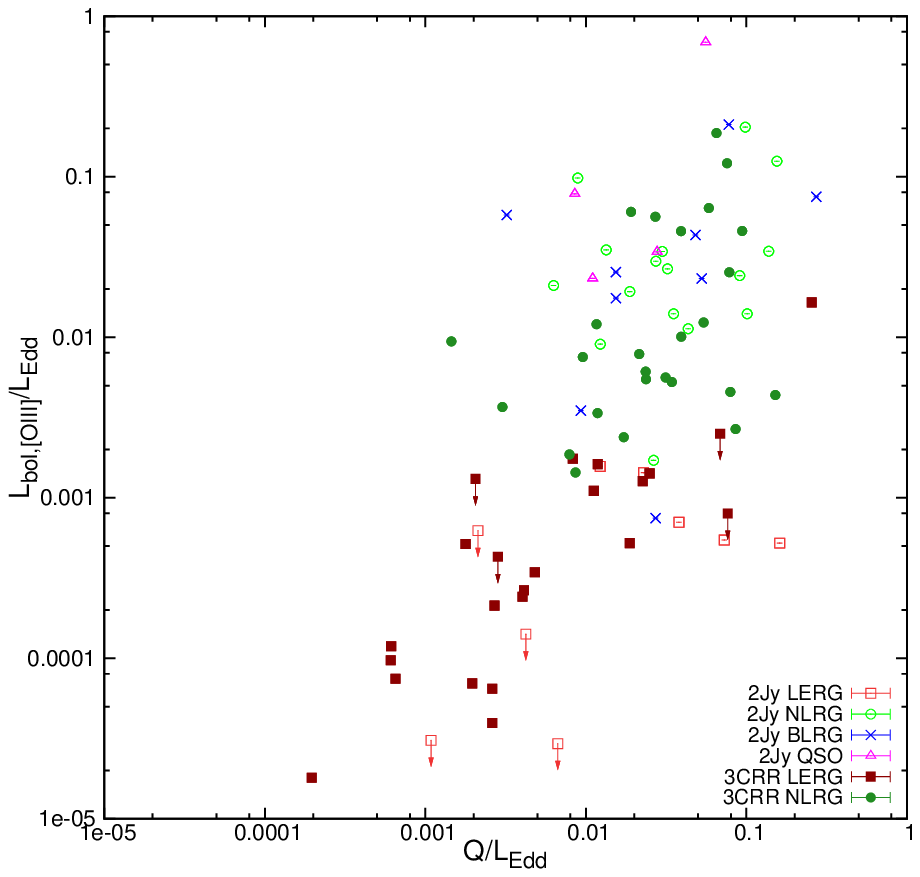}
\vspace{8pt}
\end{minipage}
\begin{minipage}[t]{0.48\textwidth}
\includegraphics[clip=true, trim=1.6cm 0cm 1.6cm 0cm, width=0.9\textwidth]{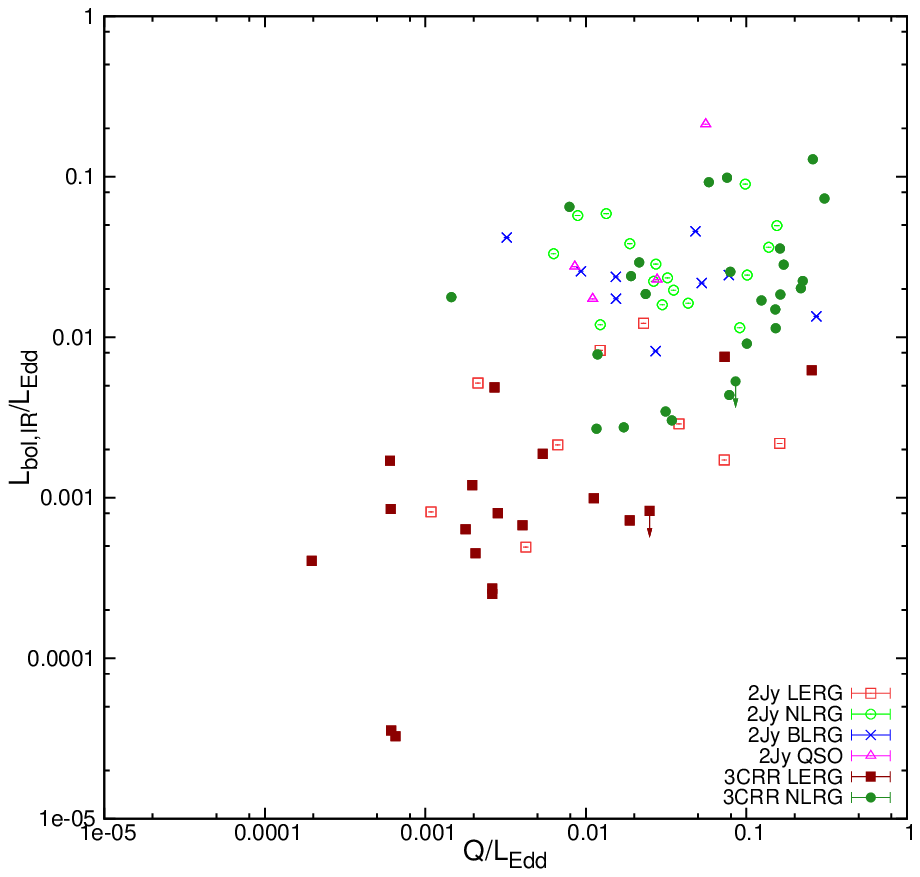}
\end{minipage}
\caption{$L_{bol}/L_{Edd}$ versus $Q/L_{Edd}$ for the 2Jy and the 3CRR sources. Top: $L_{bol,X}/L_{Edd}$ versus $Q/L_{Edd}$. Middle: $L_{bol,[OIII]}/L_{Edd}$ versus $Q/L_{Edd}$. Bottom: $L_{bol,IR}/L_{Edd}$ versus $Q/L_{Edd}$. Error bars reflect the uncertainties in the accretion-related luminosity, but not systematics such as the uncertainty in absorption or intrinsic variability. Arrows indicate upper limits}\label{Q_Edd}
\end{figure}

Jet power versus radiative luminosity plots can be enlightening in discerning the relative contributions of both components for each population. \edit{The top panel of Figure \ref{Q_Edd} shows $L_{bol,[X]}/L_{Edd}$ versus $Q/L_{Edd}$ for the 2Jy and the 3CRR sources, where Q is the jet power as defined by \citet{Willott1999}. The middle and bottom panels of Figure \ref{Q_Edd} show the same plot for [OIII] and IR derived bolometric luminosities, respectively. This latter panel of Figure \ref{Q_Edd} is the one with the best correlation (see Table \ref{2Jy_correlations_table})}. \edit{Reassuringly, in all the plots adding the contributions from the radiative output and the jet kinetic energy still results in sub-Eddington accretion, even in the brightest sources. }

\edit{The X-ray derived Eddington rates show the greatest degree of uncertainty on individual measurements, two orders of magnitude for some sources, and even higher for the (`inefficient') LERGs. The top panel of Figure \ref{Q_Edd} illustrates this fact clearly: the X-ray derived $L_{bol}/L_{Edd}$ spans two orders of magnitude more than that derived from the [OIII] and IR measurements (middle and bottom panels). This effect is most likely intrinsic to the nature of X-ray measurements of AGNs, where source variability, intrinsic absorption and beaming contribute to the scatter. LERGs seem to have systematically lower (by over three orders of magnitude in some cases) radiative Eddington rates in X-rays than they do when these rates are derived from IR or [OIII] measurements. Even assuming a much higher obscuration ($N_{H}=10^{24}$ cm$^{-2}$), their radiative Eddington rates would be far lower than those of the HERGs, which makes it unlikely that LERGs are simply Compton-thick HERGs.} 


\edit{Estimating $L/L_{Edd}$ is very challenging, particularly for radiatively inefficient sources, where models predict very little radiative emission. Can we, therefore, find a reliable probe for the accretion-related, radiative luminosity in LERGs? While IR measurements are most reliable to determine accretion in high-excitation sources, they appear to overestimate this component in LERGs. Most IR points in Figure \ref{Q_Edd} are detections, not upper limits, which is not consistent with the model predictions. As pointed out in Section \ref{X_O}, it is likely that in these objects the IR emission is associated with the jet and the old stellar population, rather than accretion. For the same reason, [OIII] measurements are also likely to be an overestimation, since shock-ionization by the jet can boost such emission. We conclude that for LERGs the Eddington rate is best derived from X-ray measurements, since we know that, once the possible contamination by $L_{X_{u}}$ is accounted for, any remaining radiative output must come from $L_{X_{a}}$.}

In all these plots a division between high and low-excitation sources is clearly visible. A trend between jet power and radiative luminosity can be observed for the LERGs. We can assume that a certain degree of contamination from jet emission is present in the radiative component in the three plots, and is probably causing this apparent trend.

Finally, we note that for the HERGs we do not see a decrease in jet power at high radiative luminosities, which indicates that, even if there is a switch between radiatively inefficient and efficient accretion (discussed below, Section \ref{switch}), jet generation is not switched off when radiatively efficient accretion takes over. There are several NLRGs, in fact, where the contribution from the jet kinetic luminosity is higher than that of the radiative luminosity \citep[see also][]{Punsly2011}.

\subsection{Radiative luminosity and jet power: is there a correlation?}\label{accretion}

\edit{While we know that the empirical relation between low-frequency radio emission and jet kinetic power shows a large scatter, and that environmental factors play a fundamental role in this relation, past work has suggested that there is a direct correlation between radiative luminosity and jet power in radio-loud AGN \citep[see e.g. Figure 1 in][, who find $Q=L_{NLR}^{0.9 \pm 0.2}$]{Rawlings1991}. According to this scenario, the radiative output of the AGN corresponds to a fraction of its accretion power (which holds true for radio-quiet sources), and so does the jet power, so that both magnitudes are correlated. However, our results in Figure \ref{Q_Edd} show that there is substantial scatter in this relationship even for the HERGs; when considering the HERG population only, there is no obvious correlation between the luminosity (in terms of Eddington) in the jet and in radiative output, and, while excluding the LERGs limits the dynamic range, they cannot be considered in the same terms, due to their different accretion properties. Therefore, we must ask: is there any evidence for a physical relation between $Q$ and $L_{rad}$, beyond the fact that they are both linked to accretion?}

We begin by noting that the interpretation of all these plots is complicated by the fact that that the radiative luminosity is essentially an instantaneous measurement, while the jet power is estimated from the large-scale radio lobes and so is a weighted average over the whole radio-loud lifetime of the source. This will certainly give rise to some of the scatter that we see, but it is not clear that it can account for the roughly one order of magnitude in dispersion about any relationship between radiative and kinetic luminosity.

\edit{We carried out tests for partial correlation between $Q$ and $L_{bol,[OIII]}$ for the HERGs, in the presence of their common dependence on redshift, and also looked at the relationship between $Q$ and $L_{bol,IR}$ (Table \ref{2Jy_correlations_table}). In the first case, which corresponds to the analysis of Rawlings \& Saunders, we cannot reject the null hypothesis at a $3\sigma$ level; in the second there is a weak correlation for the overall sample, which disappears when the broad-line objects are removed. Our sample size is larger than that of Rawlings \& Saunders: the crucial difference between our work and theirs is that we are not considering LERGs, which (because of their low jet powers and low emission-line/IR luminosities) would artificially strengthen any such correlation.}

If there is no real physical correlation between the jet power and emission-line power, why are there positive correlations between related quantities such as $L_{X_a}$ and $L_{178}$ (Table \ref{2Jy_correlations_table})? We propose that these quantities are largely the result of selection bias. Any object classified as a narrow-line radio galaxy or a quasar -- in other words, a classical AGN -- has a radiative luminosity that cannot fall much below $0.01L_{\rm Edd}$ (see further below, Section \ref{switch}) and cannot greatly exceed $L_{\rm Edd}$. By selecting the most luminous radio galaxies in the Universe, the 3CRR and 2Jy samples, we are selecting for objects that have the highest possible jet powers -- but these must also be limited by the accretion rate and so cannot greatly exceed $L_{\rm Edd}$. Thus the most radio-loud objects in the universe should always populate the top right of plots like \edit{Figure \ref{Q_Edd}}. However, crucially, this picture makes a prediction for less luminous samples of radio galaxies or radio-loud quasars that differs from that of the Rawlings \& Saunders
model. Classical AGN selected at lower radio powers are free to populate parts of the luminosity-luminosity plots to the left of the 3CRR/2Jy objects in \edit{Figure \ref{Q_Edd}}.

\begin{figure}
\includegraphics[clip=true, trim=1.6cm 0cm 1.6cm 0cm, width=0.45\textwidth]{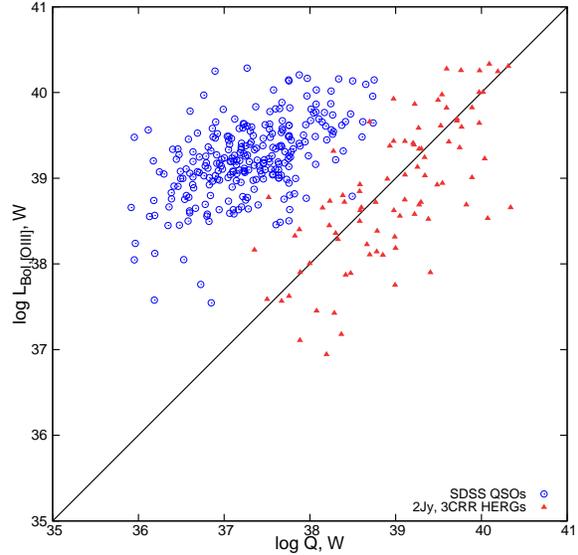}
\caption{$L_{bol,[OIII]}$ versus $Q$ for the 2Jy and the 3CRR high-excitation objects and the SDSS quasars from \citet{Punsly2011}. The line represents a 1:1 relation between both quantities.}\label{punsly}
\end{figure}

\edit{To test this picture we have plotted HERGs in our 2Jy and 3CRR samples next to the SDSS-selected quasars of \citet{Punsly2011}. We have tested this comparison sample for several reasons: (i) it is quite large, (ii) as it is not radio-selected it samples objects with a range of radio outputs (which are also lower than those of the 2Jy and 3CRR samples), (iii) and it contains [OIII] and $Q$ measures that we can directly compare to our own. Figure \ref{punsly} shows $L_{bol,[OIII]}$ versus $Q$ for the objects in the \citet{Punsly2011} sample. As predicted, the SDSS QSOs lie well to the left of the 3CRR objects: for a given radiative power, they generally have much lower jet powers than the 3CRR/2Jy objects. In the simple Rawlings \& Saunders model, these objects (with lower radio powers) would be expected to lie 2-3
orders of magnitude lower in radiative power as well. This reinforces the conclusions of \citet{Punsly2011}, who pointed out that there is no reason to expect
$L_{bol,[OIII]}$ and $Q$ to be correlated beyond the scaling with the central black hole. The picture also holds up for other samples for which jet power (or total radio power) has been correlated with $L_{[OIII]}$, including the 7C sources of \citep{Willott1999} and the SDSS QSOs from \citet{McLure2004}, which also lie systematically to the left of the line of equality in plots such as that in Figure \ref{punsly}. The limiting case is provided by studies of very low-luminosity radio-loud AGN such as that of \citet{Kaufmann2008}, where a very wide range of radio luminosities is necessarily sampled and where there is no apparent correlation between $Q$ and $L_{rad}$ at all. }

We can thus conclude that, for radiatively efficient accretion, the same mechanism that powers radiative emission also powers the jet. But while the fraction of accretion power that is converted to radiative luminosity lies in a relatively narrow range, that which is converted to jet power can vary much more widely, presumably through some \edit{yet to be determined controlling parameter such as black hole spin: through selection biases, this fraction of the total accretion power reaches a maximum of $\sim20$ per cent Eddington
in the 3CRR and 2Jy samples that are the subject of this paper.}

\subsection{An Eddington switch?}\label{switch}

\edit{We now explore the transition between HERGs and LERGs. Are both classes part of a continuous population? Is there a clear Eddington switch that makes an object efficient or inefficient, or or is the LERG/HERG difference controlled partly or wholly by other factors, as in the models of \citet{Hardcastle2007b}? And how are environmental and observational effects affecting the distribution?}

\begin{figure}
\includegraphics[clip=true, width=0.45\textwidth]{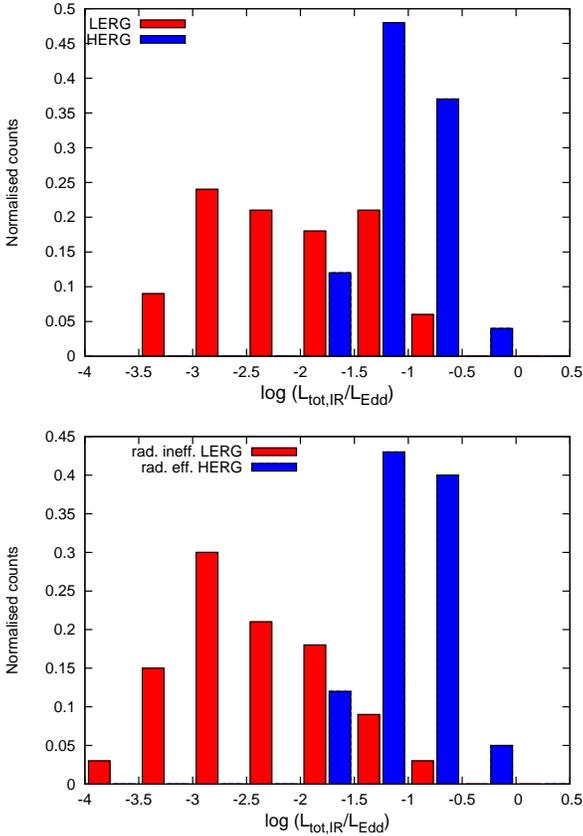}
\caption{Histograms of total Eddington rate [($L_{bol,IR}+Q)/L_{Edd}$] distribution for the 2Jy and the 3CRR sources. Broad-line objects are excluded from the HERGs to allow direct comparison between both samples. \textbf{Top:} all objects with available data are considered. \textbf{Bottom:} only radiatively inefficient LERGs and radiatively efficient HERGs are considered, and for the radiatively inefficient objects only $Q$ is considered for the total luminosity.}\label{Eddington_IR}
\end{figure}

We plotted histograms of the total Eddington luminosity $[(L_{rad}+Q_{jet})/L_{Edd}]$ for the three bands and the high/low-excitation populations. In all cases we found the distribution to be clearly bimodal, with HERGs having systematically higher Eddington rates (peaking at $\sim20$ per cent Eddington) than LERGs (peaking at $\sim1$ per cent Eddington). The narrowest distribution is that obtained from the IR data (Figure \ref{Eddington_IR}), but those derived from X-ray and [OIII] measurements have coincident peaks and outliers. 

Despite the fact that they have no influence on the result, we decided to remove the broad-line objects from the histograms to allow direct comparison between the 2Jy and 3CRR samples (we have no K-band measurements for 3CRR BLRGs and QSOs), and to remove the bias derived from black hole masses that are, at best, uncertain for these objects.

Before any conclusions can be drawn on the existence of an Eddington switch between LERGs and HERGs, it is important to consider the nature of outliers (i.e. high Eddington LERGs). The LERGs with high Eddington rates fall into two categories: `efficient' LERGs and cluster-embedded objects. To the former category belong PKS 0034-01 (3C 15), PKS 0043-42, PKS 0347+05, PKS 0625-35, 3C 123 and 3C 200 \citep[see][]{Hardcastle2006}, all of which show signs of radiatively efficient accretion \edit{(bright accretion-related emission in X-rays, bright mid-infrared emission, and, in some cases, a Fe K-$\alpha$ line, see also Appendix \ref{notes} for details)}. To the latter category belong PKS 2211-17, PKS 1648+05 (Hercules A), PKS 0915-11 (Hydra A), and 3C 438. All these objects (save perhaps for Hydra A, which has a peculiar spectrum) are bona-fide radiatively inefficient LERGs embedded in very dense clusters. It is possible that a boost of the jet luminosity due to the dense environment and an underestimation of the black hole mass \citep{Volonteri2012} are combining to produce this effect.

To test this effect we have redone the histogram assuming that the LERGs have no measurable radiative contribution from radiatively efficient accretion (that is, taking into account only $Q$ for these objects), and excluding all the sources for which the optical classification is inconsistent with our conclusions on the accretion mode (i.e. the `efficient' LERGs). We have also excluded 3C 319, since our preliminary results on new X-ray data show that this source may not be a radiatively inefficient AGN, but an efficient one that has recently switched off. The histogram is shown in the bottom panel of Figure \ref{Eddington_IR}. 

\edit{After removing the outliers and the radiative contribution for the LERGs, we find that all our LERGs have $L/L_{Edd}< 16$ per cent. 94 per cent of them have $L/L_{Edd}< 10$ per cent, with 91 per cent of them having $L/L_{Edd}< 3$ per cent. All the HERGs have $L/L_{Edd}>1$ per cent, with 88 per cent of the HERGs having $L/L_{Edd}> 3$ per cent, and 45 per cent of them having $L/L_{Edd}> 10$ per cent.}

Although the separation between the two populations is now clearer, there is still some overlap. The remaining LERGs with log$(Q/L_{Edd})>-1.5$ are the cluster-embedded objects mentioned above. While it is difficult to assess by how much the black hole mass is underestimated in these galaxies, some of the plots of \citet{Volonteri2012} show that \edit{these masses} could be off by over half an order of magnitude. \citet{Hardcastle2013} show that there is almost an order of magnitude scatter on the radio luminosity in their simulations for objects with the same jet powers, caused by the range of environmental densities tested. Therefore, the combination of these two effects could be enough to account for the high values of $Q$ in cluster-embedded objects, and the reason behind the overlap between the two populations. If this were the case, our results would be compatible with a simple switch.

%

\section{Summary and Conclusions}\label{Conclusions}

It is now clear that a classification that is based purely on morphological features, as that of \citet{FR1974}, emission-line properties, or orientation, \edit{as predicted by the simplest versions of models such as those described by \citet{Antonucci1993} and \citet{Urry1995},} cannot account for the underlying variety within the AGN population. As suggested by e.g. \citet{Hardcastle2007b,Lin2010,Antonucci2012,Best2012}, we need a classification that encompasses both the physical properties and the observational properties of AGNs. This is particularly important for the LERG/HERG case, since there is an underlying physical difference between the two overall populations, but there are cases in which the observational, optical line classification belies the true nature of the accretion mode. 

Although recent studies are beginning to take into account this intrinsic difference between the two populations, and progress is being made towards understanding the properties of LERGs, most samples still use restrictive selection criteria, employ only one or two energy bands to characterise the populations, contain objects which are misclassified, and use bolometric corrections that do not accurately describe the less powerful sources. In \edit{our} work we present consistent results that question the accuracy of some of these assumptions, and prove that further, more careful analysis is needed to understand the relationship between radiative output and jet production in the overall AGN population.

Throughout this work we have shown that the best way to reliably classify AGN populations is through a multiwavelength approach, which we use on our sample of 45 2Jy and 135 3CRR sources (more than double the size of that studied by \citet{Hardcastle2009}). We show that several objects classified as LERGs based on their optical spectra (PKS 0034-01, PKS 0043-42, PKS 0625-35, 3C 123, 3C 200 and more recently PKS 0347+05) are most likely radiatively efficient sources.

We find the same strong correlations between hard (2-10 keV) X-ray, mid-IR and [OIII] emission as \citet{Hardcastle2006,Hardcastle2009}, confirming that these quantities are all related to radiatively efficient accretion. We confirm the jet-related nature of the soft X-ray emission, as suggested by \citet{Hardcastle1999}. We also show that selection criteria must be taken into account when studying correlations between these quantities: relativistic beaming can introduce a large scatter in the plots, resulting in poorer partial correlations. We find that all the correlations of \citet{Hardcastle2009} become stronger by the addition of the 2Jy objects.

By comparing the accretion-related correlations, we show that mid-IR measurements are best to constrain the accretion properties of high-excitation objects, while for the low-excitation population X-rays are the best band to set an upper limit on radiatively efficient accretion, given that \edit{X-rays} are less subject to contamination from stellar processes and the presence of a jet (this is taken into account by the soft X-ray component, whose jet-related nature we confirm). Radio measurements are essential to establish the extent of radiatively inefficient accretion, and the amount of AGN power invested in the jet.

We emphasise the fact that bolometric corrections, $M_{BH}/L$ correlations and jet power estimations only give an overall indication of AGN behaviour, and may be inaccurate for individual sources, given the vast range of environments and nuclear powers involved. Further studies of individual SEDs and jet-environment interaction simulations are needed to establish how reliable these correlations are, in particular for the case of LERGs.

\edit{Despite these intrinsic limitations, we find very strong evidence of the radiatively inefficient nature of the LERGs, as well as confirmation for the fact that these objects accrete at very low Eddington rates ($<10$ per cent in all cases but one, with the distribution peaking at $\sim1$ per cent), as expected from the theoretical models \citep[e.g.][]{Narayan1995}. We find that the HERGs in our sample are narrowly distributed around 10 per cent Eddington rates, with roughly half of the objects having greater values. However, we find an overlap between both populations, which at first sight is not consistent with a simple switch at a given value of $L/L_{Edd}$. Even after discarding the objects whose classification belies the intrinsic accretion properties (i.e. plotting what we know are unequivocally radiatively efficient HERGs and radiatively inefficient LERGs), we find that LERGs embedded in very rich clusters have higher $L/L_{Edd}$. For these sources the central back hole masses may be underestimated and the lobe luminosity may be higher \citep{Hardcastle2013}. These two factors can account for the order of magnitude in $L/L_{Edd}$ that makes these objects overlap with the HERGs, in which case a simple switch between the two populations would be feasible.}

We do not see signs in our plots for radiatively efficient accretion completely taking over from jet production. In fact, we find several NLRGs in which the dominant energetic contribution from the AGN stems from the jet, rather than radiative luminosity. Selection on radio flux selects for the objects with the largest values of $Q$ at any given epoch. We find that jet kinetic power and radiative luminosity seem to have a common underlying mechanism, but are not correlated in radiatively efficient objects, confirming the conclusions of \citet{Punsly2011}. While a better understanding of the timescales and the addition of radio-quiet objects to the plots are necessary to fully understand whether these quantities are truly uncorrelated, our plots and correlation analysis seem to indicate that they are.

\edit{As part of our in-depth study of the 2Jy sample, in our second paper we will analyse the non-thermal, extended X-ray emission of the low-redshift sources observed by \textit{Chandra}, to characterise the properties of their jets, hotspots and lobes. This will be the first time a systematic study of this nature will have been carried out on a complete sample of radio galaxies, and it will allow us to gain further insight on the particle content and the effects of beaming across the entire radio-loud population. A third paper will study the environments of the 2Jy and 3CRR samples, focusing on the extended, thermal X-ray emission.}
%

\subsection*{Acknowledgements}

BM thanks the University of Hertfordshire for a PhD studentship. This work is based on observations obtained with \textit{XMM-Newton}, an ESA science mission with instruments and contributions directly funded by ESA Member States and NASA. It has also made use of new and archival data from \textit{Chandra} and software provided by the Chandra X-ray Center (CXC) in the application package CIAO. We thank Dr. B. Punsly for giving us access to his data for Figure \ref{punsly}. We thank the anonymous referee for the useful comments.
%

\bibliographystyle{mn2e_fixed}
\bibliography{2Jy}

\begin{thebibliography}{109}
\expandafter\ifx\csname natexlab\endcsname\relax\def\natexlab#1{#1}\fi

\bibitem[{{Akritas} \& {Siebert}(1996)}]{Akritas1996}
{Akritas} M.~G., {Siebert} J., 1996, \mnras, 278, 919

\bibitem[{{Antognini} {et~al}\mbox{.}(2012){Antognini}, {Bird}, \&
  {Martini}}]{Antognini2012}
{Antognini} J., {Bird} J., {Martini} P., 2012, \apj, 756, 116

\bibitem[{{Antonucci}(1993)}]{Antonucci1993}
{Antonucci} R., 1993, \araa, 31, 473

\bibitem[{{Antonucci}(2012)}]{Antonucci2012}
{Antonucci} R., 2012, Astronomical and Astrophysical Transactions, 27, 557

\bibitem[{{Asmus} {et~al}\mbox{.}(2011){Asmus}, {Gandhi}, {Smette},
  {H{\"o}nig}, \& {Duschl}}]{Asmus2011}
{Asmus} D., {Gandhi} P., {Smette} A., {H{\"o}nig} S.~F., {Duschl} W.~J., 2011,
  \aap, 536, A36

\bibitem[{{Balmaverde} {et~al}\mbox{.}(2008){Balmaverde}, {Baldi}, \&
  {Capetti}}]{Balmaverde2008}
{Balmaverde} B., {Baldi} R.~D., {Capetti} A., 2008, \aap, 486, 119

\bibitem[{{Best} \& {Heckman}(2012)}]{Best2012}
{Best} P.~N., {Heckman} T.~M., 2012, \mnras, 421, 1569

\bibitem[{{Best} {et~al}\mbox{.}(2005){Best}, {Kauffmann}, {Heckman},
  {Brinchmann}, {Charlot}, {Ivezi{\'c}}, \& {White}}]{Best2005}
{Best} P.~N., {Kauffmann} G., {Heckman} T.~M., {Brinchmann} J., {Charlot} S.,
  {Ivezi{\'c}} {\v Z}., {White} S.~D.~M., 2005, \mnras, 362, 25

\bibitem[{{Best} {et~al}\mbox{.}(1998){Best}, {Longair}, \&
  {Roettgering}}]{Best1998}
{Best} P.~N., {Longair} M.~S., {Roettgering} H.~J.~A., 1998, \mnras, 295, 549

\bibitem[{{Blundell} \& {Rawlings}(2001)}]{Blundell2001}
{Blundell} K.~M., {Rawlings} S., 2001, \apjl, 562, L5

\bibitem[{{Bower} {et~al}\mbox{.}(2006){Bower}, {Benson}, {Malbon}, {Helly},
  {Frenk}, {Baugh}, {Cole}, \& {Lacey}}]{Bower2006}
{Bower} R.~G., {Benson} A.~J., {Malbon} R., {Helly} J.~C., {Frenk} C.~S.,
  {Baugh} C.~M., {Cole} S., {Lacey} C.~G., 2006, \mnras, 370, 645

\bibitem[{{Buttiglione} {et~al}\mbox{.}(2009){Buttiglione}, {Capetti},
  {Celotti}, {Axon}, {Chiaberge}, {Macchetto}, \& {Sparks}}]{Buttiglione2009}
{Buttiglione} S., {Capetti} A., {Celotti} A., {Axon} D.~J., {Chiaberge} M.,
  {Macchetto} F.~D., {Sparks} W.~B., 2009, \aap, 495, 1033

\bibitem[{{Carter} {et~al}\mbox{.}(2003){Carter}, {Karovska}, {Jerius},
  {Glotfelty}, \& {Beikman}}]{ChaRT}
{Carter} C., {Karovska} M., {Jerius} D., {Glotfelty} K., {Beikman} S., 2003, in
  Astronomical Society of the Pacific Conference Series, Vol. 295, Astronomical
  Data Analysis Software and Systems XII, {H.~E.~Payne, R.~I.~Jedrzejewski, \&
  R.~N.~Hook}, ed., pp. 477--+

\bibitem[{{Cattaneo} {et~al}\mbox{.}(2009){Cattaneo}, {Faber}, {Binney},
  {Dekel}, {Kormendy}, {Mushotzky}, {Babul}, {Best}, {Br{\"u}ggen}, {Fabian},
  {Frenk}, {Khalatyan}, {Netzer}, {Mahdavi}, {Silk}, {Steinmetz}, \&
  {Wisotzki}}]{Cattaneo2009}
{Cattaneo} A. {et~al.}, 2009, \nat, 460, 213

\bibitem[{{Cavagnolo} {et~al}\mbox{.}(2010){Cavagnolo}, {McNamara}, {Nulsen},
  {Carilli}, {Jones}, \& {B{\^i}rzan}}]{Cavagnolo2010}
{Cavagnolo} K.~W., {McNamara} B.~R., {Nulsen} P.~E.~J., {Carilli} C.~L.,
  {Jones} C., {B{\^i}rzan} L., 2010, \apj, 720, 1066

\bibitem[{{Chiaberge} {et~al}\mbox{.}(2002){Chiaberge}, {Capetti}, \&
  {Celotti}}]{Chiaberge2002}
{Chiaberge} M., {Capetti} A., {Celotti} A., 2002, \aap, 394, 791

\bibitem[{{Croston} {et~al}\mbox{.}(2008){Croston}, {Hardcastle}, {Birkinshaw},
  {Worrall}, \& {Laing}}]{Croston2008b}
{Croston} J.~H., {Hardcastle} M.~J., {Birkinshaw} M., {Worrall} D.~M., {Laing}
  R.~A., 2008, \mnras, 386, 1709

\bibitem[{{Croston} {et~al}\mbox{.}(2011){Croston}, {Hardcastle}, {Mingo},
  {Evans}, {Dicken}, {Morganti}, \& {Tadhunter}}]{Croston2011}
{Croston} J.~H., {Hardcastle} M.~J., {Mingo} B., {Evans} D.~A., {Dicken} D.,
  {Morganti} R., {Tadhunter} C.~N., 2011, \apjl, 734, L28

\bibitem[{{Croton} {et~al}\mbox{.}(2006){Croton}, {Springel}, {White}, {De
  Lucia}, {Frenk}, {Gao}, {Jenkins}, {Kauffmann}, {Navarro}, \&
  {Yoshida}}]{Croton2006}
{Croton} D.~J. {et~al.}, 2006, \mnras, 365, 11

\bibitem[{{Danziger} {et~al}\mbox{.}(1984){Danziger}, {Fosbury}, {Boksenberg},
  {Goss}, \& {Bland}}]{Danziger1984}
{Danziger} I.~J., {Fosbury} R.~A.~E., {Boksenberg} A., {Goss} W.~M., {Bland}
  J., 1984, \mnras, 208, 589

\bibitem[{{de Vries} {et~al}\mbox{.}(1998){de Vries}, {O'Dea}, {Perlman},
  {Baum}, {Lehnert}, {Stocke}, {Rector}, \& {Elston}}]{deVries1998}
{de Vries} W.~H., {O'Dea} C.~P., {Perlman} E., {Baum} S.~A., {Lehnert} M.~D.,
  {Stocke} J., {Rector} T., {Elston} R., 1998, \apj, 503, 138

\bibitem[{{Dicken} {et~al}\mbox{.}(2009){Dicken}, {Tadhunter}, {Axon},
  {Morganti}, {Inskip}, {Holt}, {Gonz{\'a}lez Delgado}, \&
  {Groves}}]{Dicken2009}
{Dicken} D., {Tadhunter} C., {Axon} D., {Morganti} R., {Inskip} K.~J., {Holt}
  J., {Gonz{\'a}lez Delgado} R., {Groves} B., 2009, \apj, 694, 268

\bibitem[{{Dicken} {et~al}\mbox{.}(2012){Dicken}, {Tadhunter}, {Axon},
  {Morganti}, {Robinson}, {Kouwenhoven}, {Spoon}, {Kharb}, {Inskip}, {Holt},
  {Ramos Almeida}, \& {Nesvadba}}]{Dicken2012}
{Dicken} D. {et~al.}, 2012, \apj, 745, 172

\bibitem[{{Dicken} {et~al}\mbox{.}(2008){Dicken}, {Tadhunter}, {Morganti},
  {Buchanan}, {Oosterloo}, \& {Axon}}]{Dicken2008}
{Dicken} D., {Tadhunter} C., {Morganti} R., {Buchanan} C., {Oosterloo} T.,
  {Axon} D., 2008, \apj, 678, 712

\bibitem[{{Dickey} \& {Lockman}(1990)}]{Dickey1990}
{Dickey} J.~M., {Lockman} F.~J., 1990, \araa, 28, 215

\bibitem[{{Elvis} {et~al}\mbox{.}(1994){Elvis}, {Wilkes}, {McDowell}, {Green},
  {Bechtold}, {Willner}, {Oey}, {Polomski}, \& {Cutri}}]{Elvis1994}
{Elvis} M. {et~al.}, 1994, \apjs, 95, 1

\bibitem[{{Evans} {et~al}\mbox{.}(2011){Evans}, {Summers}, {Hardcastle},
  {Kraft}, {Gandhi}, {Croston}, \& {Lee}}]{Evans2011}
{Evans} D.~A., {Summers} A.~C., {Hardcastle} M.~J., {Kraft} R.~P., {Gandhi} P.,
  {Croston} J.~H., {Lee} J.~C., 2011, \apjl, 741, L4

\bibitem[{{Evans} {et~al}\mbox{.}(2006){Evans}, {Worrall}, {Hardcastle},
  {Kraft}, \& {Birkinshaw}}]{Evans2006}
{Evans} D.~A., {Worrall} D.~M., {Hardcastle} M.~J., {Kraft} R.~P., {Birkinshaw}
  M., 2006, \apj, 642, 96

\bibitem[{{Fanaroff} \& {Riley}(1974)}]{FR1974}
{Fanaroff} B.~L., {Riley} J.~M., 1974, \mnras, 167, 31P

\bibitem[{{Fern{\'a}ndez-Ontiveros}
  {et~al}\mbox{.}(2012){Fern{\'a}ndez-Ontiveros}, {Prieto}, {Acosta-Pulido}, \&
  {Montes}}]{FdezOntiveros2012}
{Fern{\'a}ndez-Ontiveros} J.~A., {Prieto} M.~A., {Acosta-Pulido} J.~A.,
  {Montes} M., 2012, Journal of Physics Conference Series, 372, 012006

\bibitem[{{Fukugita} {et~al}\mbox{.}(1995){Fukugita}, {Shimasaku}, \&
  {Ichikawa}}]{Fukugita1995}
{Fukugita} M., {Shimasaku} K., {Ichikawa} T., 1995, \pasp, 107, 945

\bibitem[{{Gandhi} {et~al}\mbox{.}(2009){Gandhi}, {Horst}, {Smette},
  {H{\"o}nig}, {Comastri}, {Gilli}, {Vignali}, \& {Duschl}}]{Gandhi2009}
{Gandhi} P., {Horst} H., {Smette} A., {H{\"o}nig} S., {Comastri} A., {Gilli}
  R., {Vignali} C., {Duschl} W., 2009, \aap, 502, 457

\bibitem[{{Glazebrook} {et~al}\mbox{.}(1995){Glazebrook}, {Peacock}, {Miller},
  \& {Collins}}]{Glazebrook1995}
{Glazebrook} K., {Peacock} J.~A., {Miller} L., {Collins} C.~A., 1995, \mnras,
  275, 169

\bibitem[{{Godfrey} \& {Shabala}(2013)}]{Godfrey2013}
{Godfrey} L.~E.~H., {Shabala} S.~S., 2013, ArXiv e-prints

\bibitem[{{Graham}(2007)}]{Graham2007}
{Graham} A.~W., 2007, \mnras, 379, 711

\bibitem[{{Haardt} \& {Maraschi}(1991)}]{Haardt1991}
{Haardt} F., {Maraschi} L., 1991, \apjl, 380, L51

\bibitem[{{Hardcastle} {et~al}\mbox{.}(2007{\natexlab{a}}){Hardcastle},
  {Croston}, \& {Kraft}}]{HArdcastle2007}
{Hardcastle} M.~J., {Croston} J.~H., {Kraft} R.~P., 2007{\natexlab{a}}, \apj,
  669, 893

\bibitem[{{Hardcastle} {et~al}\mbox{.}(2006){Hardcastle}, {Evans}, \&
  {Croston}}]{Hardcastle2006}
{Hardcastle} M.~J., {Evans} D.~A., {Croston} J.~H., 2006, \mnras, 370, 1893

\bibitem[{{Hardcastle} {et~al}\mbox{.}(2007{\natexlab{b}}){Hardcastle},
  {Evans}, \& {Croston}}]{Hardcastle2007b}
{Hardcastle} M.~J., {Evans} D.~A., {Croston} J.~H., 2007{\natexlab{b}}, \mnras,
  376, 1849

\bibitem[{{Hardcastle} {et~al}\mbox{.}(2009){Hardcastle}, {Evans}, \&
  {Croston}}]{Hardcastle2009}
{Hardcastle} M.~J., {Evans} D.~A., {Croston} J.~H., 2009, \mnras, 396, 1929

\bibitem[{{Hardcastle} \& {Krause}(2013)}]{Hardcastle2013}
{Hardcastle} M.~J., {Krause} M.~G.~H., 2013, \mnras, 601

\bibitem[{{Hardcastle} \& {Worrall}(1999)}]{Hardcastle1999}
{Hardcastle} M.~J., {Worrall} D.~M., 1999, \mnras, 309, 969

\bibitem[{{Heckman} {et~al}\mbox{.}(2004){Heckman}, {Kauffmann}, {Brinchmann},
  {Charlot}, {Tremonti}, \& {White}}]{Heckman2004}
{Heckman} T.~M., {Kauffmann} G., {Brinchmann} J., {Charlot} S., {Tremonti} C.,
  {White} S.~D.~M., 2004, \apj, 613, 109

\bibitem[{{Hine} \& {Longair}(1979)}]{Hine1979}
{Hine} R.~G., {Longair} M.~S., 1979, \mnras, 188, 111

\bibitem[{{Ho}(2009)}]{Ho2009}
{Ho} L.~C., 2009, \apj, 699, 626

\bibitem[{{Holt} {et~al}\mbox{.}(2008){Holt}, {Tadhunter}, \&
  {Morganti}}]{Holt2008}
{Holt} J., {Tadhunter} C.~N., {Morganti} R., 2008, \mnras, 387, 639

\bibitem[{{Holt} {et~al}\mbox{.}(2009){Holt}, {Tadhunter}, \&
  {Morganti}}]{Holt2009}
{Holt} J., {Tadhunter} C.~N., {Morganti} R., 2009, \mnras, 400, 589

\bibitem[{{Hopkins}(2012)}]{Hopkins2012}
{Hopkins} P.~F., 2012, \mnras, 420, L8

\bibitem[{{Ineson} {et~al}\mbox{.}(2013){Ineson}, {Croston}, {Hardcastle},
  {Kraft}, {Evans}, \& {Jarvis}}]{Ineson2013}
{Ineson} J., {Croston} J.~H., {Hardcastle} M.~J., {Kraft} R.~P., {Evans} D.~A.,
  {Jarvis} M., 2013, ArXiv e-prints

\bibitem[{{Inskip} {et~al}\mbox{.}(2007){Inskip}, {Tadhunter}, {Dicken},
  {Holt}, {Villar-Mart{\'{\i}}n}, \& {Morganti}}]{Inskip2007}
{Inskip} K.~J., {Tadhunter} C.~N., {Dicken} D., {Holt} J.,
  {Villar-Mart{\'{\i}}n} M., {Morganti} R., 2007, \mnras, 382, 95

\bibitem[{{Inskip} {et~al}\mbox{.}(2010){Inskip}, {Tadhunter}, {Morganti},
  {Holt}, {Ramos Almeida}, \& {Dicken}}]{Inskip2010}
{Inskip} K.~J., {Tadhunter} C.~N., {Morganti} R., {Holt} J., {Ramos Almeida}
  C., {Dicken} D., 2010, \mnras, 407, 1739

\bibitem[{{Jackson} \& {Browne}(1990)}]{Jackson1990}
{Jackson} N., {Browne} I.~W.~A., 1990, \nat, 343, 43

\bibitem[{{Janssen} {et~al}\mbox{.}(2012){Janssen}, {R{\"o}ttgering}, {Best},
  \& {Brinchmann}}]{Janssen2012}
{Janssen} R.~M.~J., {R{\"o}ttgering} H.~J.~A., {Best} P.~N., {Brinchmann} J.,
  2012, \aap, 541, A62

\bibitem[{{Kauffmann} {et~al}\mbox{.}(2008){Kauffmann}, {Heckman}, \&
  {Best}}]{Kaufmann2008}
{Kauffmann} G., {Heckman} T.~M., {Best} P.~N., 2008, \mnras, 384, 953

\bibitem[{{K{\"o}rding} {et~al}\mbox{.}(2006){K{\"o}rding}, {Jester}, \&
  {Fender}}]{Kording2006}
{K{\"o}rding} E.~G., {Jester} S., {Fender} R., 2006, \mnras, 372, 1366

\bibitem[{{Kraft} {et~al}\mbox{.}(2005){Kraft}, {Hardcastle}, {Worrall}, \&
  {Murray}}]{Kraft2005}
{Kraft} R.~P., {Hardcastle} M.~J., {Worrall} D.~M., {Murray} S.~S., 2005, \apj,
  622, 149

\bibitem[{{Kraft} {et~al}\mbox{.}(2003){Kraft}, {V{\'a}zquez}, {Forman},
  {Jones}, {Murray}, {Hardcastle}, {Worrall}, \& {Churazov}}]{Kraft2003}
{Kraft} R.~P., {V{\'a}zquez} S.~E., {Forman} W.~R., {Jones} C., {Murray} S.~S.,
  {Hardcastle} M.~J., {Worrall} D.~M., {Churazov} E., 2003, \apj, 592, 129

\bibitem[{{Laing} {et~al}\mbox{.}(1994){Laing}, {Jenkins}, {Wall}, \&
  {Unger}}]{Laing1994}
{Laing} R.~A., {Jenkins} C.~R., {Wall} J.~V., {Unger} S.~W., 1994, in
  Astronomical Society of the Pacific Conference Series, Vol.~54, The Physics
  of Active Galaxies, {Bicknell} G.~V., {Dopita} M.~A., {Quinn} P.~J., eds., p.
  201

\bibitem[{{Laing} {et~al}\mbox{.}(1983){Laing}, {Riley}, \&
  {Longair}}]{Laing1983}
{Laing} R.~A., {Riley} J.~M., {Longair} M.~S., 1983, \mnras, 204, 151

\bibitem[{{Lane} {et~al}\mbox{.}(2004){Lane}, {Clarke}, {Taylor}, {Perley}, \&
  {Kassim}}]{Lane2004}
{Lane} W.~M., {Clarke} T.~E., {Taylor} G.~B., {Perley} R.~A., {Kassim} N.~E.,
  2004, \aj, 127, 48

\bibitem[{{Lilly} \& {Longair}(1984)}]{Lilly1984}
{Lilly} S.~J., {Longair} M.~S., 1984, \mnras, 211, 833

\bibitem[{{Lin} {et~al}\mbox{.}(2010){Lin}, {Shen}, {Strauss}, {Richards}, \&
  {Lunnan}}]{Lin2010}
{Lin} Y.-T., {Shen} Y., {Strauss} M.~A., {Richards} G.~T., {Lunnan} R., 2010,
  \apj, 723, 1119

\bibitem[{{Magorrian} {et~al}\mbox{.}(1998){Magorrian}, {Tremaine},
  {Richstone}, {Bender}, {Bower}, {Dressler}, {Faber}, {Gebhardt}, {Green},
  {Grillmair}, {Kormendy}, \& {Lauer}}]{Magorrian1998}
{Magorrian} J. {et~al.}, 1998, \aj, 115, 2285

\bibitem[{{Mannucci} {et~al}\mbox{.}(2001){Mannucci}, {Basile}, {Poggianti},
  {Cimatti}, {Daddi}, {Pozzetti}, \& {Vanzi}}]{Mannucci2001}
{Mannucci} F., {Basile} F., {Poggianti} B.~M., {Cimatti} A., {Daddi} E.,
  {Pozzetti} L., {Vanzi} L., 2001, \mnras, 326, 745

\bibitem[{{Marconi} {et~al}\mbox{.}(2004){Marconi}, {Risaliti}, {Gilli},
  {Hunt}, {Maiolino}, \& {Salvati}}]{Marconi2004}
{Marconi} A., {Risaliti} G., {Gilli} R., {Hunt} L.~K., {Maiolino} R., {Salvati}
  M., 2004, \mnras, 351, 169

\bibitem[{{Martel} {et~al}\mbox{.}(1999){Martel}, {Baum}, {Sparks}, {Wyckoff},
  {Biretta}, {Golombek}, {Macchetto}, {de Koff}, {McCarthy}, \&
  {Miley}}]{Martel1999}
{Martel} A.~R. {et~al.}, 1999, \apjs, 122, 81

\bibitem[{{Mason} {et~al}\mbox{.}(2012){Mason}, {Lopez-Rodriguez}, {Packham},
  {Alonso-Herrero}, {Levenson}, {Radomski}, {Ramos Almeida}, {Colina},
  {Elitzur}, {Aretxaga}, {Roche}, \& {Oi}}]{Mason2012}
{Mason} R.~E. {et~al.}, 2012, \aj, 144, 11

\bibitem[{{McLure} \& {Jarvis}(2004)}]{McLure2004}
{McLure} R.~J., {Jarvis} M.~J., 2004, \mnras, 353, L45

\bibitem[{{McNamara} \& {Nulsen}(2007)}]{McNamara2007}
{McNamara} B.~R., {Nulsen} P.~E.~J., 2007, \araa, 45, 117

\bibitem[{{Mingo} {et~al}\mbox{.}(2011){Mingo}, {Hardcastle}, {Croston},
  {Evans}, {Hota}, {Kharb}, \& {Kraft}}]{Mingo2011}
{Mingo} B., {Hardcastle} M.~J., {Croston} J.~H., {Evans} D.~A., {Hota} A.,
  {Kharb} P., {Kraft} R.~P., 2011, \apj, 731, 21

\bibitem[{{Morganti} {et~al}\mbox{.}(2011){Morganti}, {Holt}, {Tadhunter},
  {Ramos Almeida}, {Dicken}, {Inskip}, {Oosterloo}, \&
  {Tzioumis}}]{Morganti2011}
{Morganti} R., {Holt} J., {Tadhunter} C., {Ramos Almeida} C., {Dicken} D.,
  {Inskip} K., {Oosterloo} T., {Tzioumis} T., 2011, \aap, 535, A97

\bibitem[{{Morganti} {et~al}\mbox{.}(1993){Morganti}, {Killeen}, \&
  {Tadhunter}}]{Morganti1993}
{Morganti} R., {Killeen} N.~E.~B., {Tadhunter} C.~N., 1993, \mnras, 263, 1023

\bibitem[{{Morganti} {et~al}\mbox{.}(1999){Morganti}, {Oosterloo}, {Tadhunter},
  {Aiudi}, {Jones}, \& {Villar-Martin}}]{Morganti1999}
{Morganti} R., {Oosterloo} T., {Tadhunter} C.~N., {Aiudi} R., {Jones} P.,
  {Villar-Martin} M., 1999, \aaps, 140, 355

\bibitem[{{Narayan} \& {Yi}(1995)}]{Narayan1995}
{Narayan} R., {Yi} I., 1995, \apj, 452, 710

\bibitem[{{Netzer}(2009)}]{Netzer2009}
{Netzer} H., 2009, \mnras, 399, 1907

\bibitem[{{Ojha} {et~al}\mbox{.}(2004){Ojha}, {Fey}, {Johnston}, {Jauncey},
  {Tzioumis}, \& {Reynolds}}]{Ojha2004b}
{Ojha} R., {Fey} A.~L., {Johnston} K.~J., {Jauncey} D.~L., {Tzioumis} A.~K.,
  {Reynolds} J.~E., 2004, \aj, 127, 1977

\bibitem[{{Plotkin} {et~al}\mbox{.}(2012){Plotkin}, {Anderson}, {Brandt},
  {Markoff}, {Shemmer}, \& {Wu}}]{Plotkin2012}
{Plotkin} R.~M., {Anderson} S.~F., {Brandt} W.~N., {Markoff} S., {Shemmer} O.,
  {Wu} J., 2012, \apjl, 745, L27

\bibitem[{{Prieto} {et~al}\mbox{.}(1993){Prieto}, {Walsh}, {Fosbury}, \& {di
  Serego Alighieri}}]{Prieto1993}
{Prieto} M.~A., {Walsh} J.~R., {Fosbury} R.~A.~E., {di Serego Alighieri} S.,
  1993, \mnras, 263, 10

\bibitem[{{Punsly} \& {Zhang}(2011)}]{Punsly2011}
{Punsly} B., {Zhang} S., 2011, \apjl, 735, L3

\bibitem[{{Quataert}(2003)}]{Quataert2003}
{Quataert} E., 2003, Astronomische Nachrichten Supplement, 324, 435

\bibitem[{{Ramos Almeida} {et~al}\mbox{.}(2011){Ramos Almeida}, {Dicken},
  {Tadhunter}, {Asensio Ramos}, {Inskip}, {Hardcastle}, \&
  {Mingo}}]{RamosAlmeida2011}
{Ramos Almeida} C., {Dicken} D., {Tadhunter} C., {Asensio Ramos} A., {Inskip}
  K.~J., {Hardcastle} M.~J., {Mingo} B., 2011, \mnras, 413, 2358

\bibitem[{{Ramos Almeida} {et~al}\mbox{.}(2013){Ramos Almeida},
  {Rodr{\'{\i}}guez Espinosa}, {Acosta-Pulido}, {Alonso-Herrero}, {P{\'e}rez
  Garc{\'{\i}}a}, \& {Rodr{\'{\i}}guez-Eugenio}}]{Ramos2013}
{Ramos Almeida} C., {Rodr{\'{\i}}guez Espinosa} J.~M., {Acosta-Pulido} J.~A.,
  {Alonso-Herrero} A., {P{\'e}rez Garc{\'{\i}}a} A.~M.,
  {Rodr{\'{\i}}guez-Eugenio} N., 2013, \mnras, 429, 3449

\bibitem[{{Ramos Almeida} {et~al}\mbox{.}(2010){Ramos Almeida}, {Tadhunter},
  {Inskip}, {Morganti}, {Holt}, \& {Dicken}}]{RamosAlmeida2010}
{Ramos Almeida} C., {Tadhunter} C.~N., {Inskip} K.~J., {Morganti} R., {Holt}
  J., {Dicken} D., 2010, \mnras, 1609

\bibitem[{{Rawlings} \& {Saunders}(1991)}]{Rawlings1991}
{Rawlings} S., {Saunders} R., 1991, \nat, 349, 138

\bibitem[{{Rector} \& {Stocke}(2001)}]{Rector2001}
{Rector} T.~A., {Stocke} J.~T., 2001, \aj, 122, 565

\bibitem[{{Rinn} {et~al}\mbox{.}(2005){Rinn}, {Sambruna}, \&
  {Gliozzi}}]{Rinn2005}
{Rinn} A.~S., {Sambruna} R.~M., {Gliozzi} M., 2005, \apj, 621, 167

\bibitem[{{Risaliti} {et~al}\mbox{.}(2003){Risaliti}, {Woltjer}, \&
  {Salvati}}]{Risaliti2003}
{Risaliti} G., {Woltjer} L., {Salvati} M., 2003, \aap, 401, 895

\bibitem[{{Runnoe} {et~al}\mbox{.}(2012){Runnoe}, {Brotherton}, \&
  {Shang}}]{Runnoe2012}
{Runnoe} J.~C., {Brotherton} M.~S., {Shang} Z., 2012, \mnras, 422, 478

\bibitem[{{Russell} {et~al}\mbox{.}(2012){Russell}, {McNamara}, {Edge},
  {Hogan}, {Main}, \& {Vantyghem}}]{Russell2012}
{Russell} H.~R., {McNamara} B.~R., {Edge} A.~C., {Hogan} M.~T., {Main} R.~A.,
  {Vantyghem} A.~N., 2012, ArXiv e-prints

\bibitem[{{Saikia} \& {Jamrozy}(2009)}]{Saikia2010}
{Saikia} D.~J., {Jamrozy} M., 2009, Bulletin of the Astronomical Society of
  India, 37, 63

\bibitem[{{Sambruna} {et~al}\mbox{.}(2000){Sambruna}, {Chartas}, {Eracleous},
  {Mushotzky}, \& {Nousek}}]{Sambruna2000}
{Sambruna} R.~M., {Chartas} G., {Eracleous} M., {Mushotzky} R.~F., {Nousek}
  J.~A., 2000, \apjl, 532, L91

\bibitem[{{Sambruna} {et~al}\mbox{.}(2006){Sambruna}, {Gliozzi}, {Tavecchio},
  {Maraschi}, \& {Foschini}}]{Sambruna2006}
{Sambruna} R.~M., {Gliozzi} M., {Tavecchio} F., {Maraschi} L., {Foschini} L.,
  2006, \apj, 652, 146

\bibitem[{{Shakura} \& {Sunyaev}(1973)}]{Shakura1973}
{Shakura} N.~I., {Sunyaev} R.~A., 1973, \aap, 24, 337

\bibitem[{{Siebert} {et~al}\mbox{.}(1996){Siebert}, {Brinkmann}, {Morganti},
  {Tadhunter}, {Danziger}, {Fosbury}, \& {di Serego Alighieri}}]{Siebert1996}
{Siebert} J., {Brinkmann} W., {Morganti} R., {Tadhunter} C.~N., {Danziger}
  I.~J., {Fosbury} R.~A.~E., {di Serego Alighieri} S., 1996, \mnras, 279, 1331

\bibitem[{{Silk} \& {Rees}(1998)}]{Silk1998}
{Silk} J., {Rees} M.~J., 1998, \aap, 331, L1

\bibitem[{{Simpson} {et~al}\mbox{.}(2000){Simpson}, {Ward}, \&
  {Wall}}]{Simpson2000}
{Simpson} C., {Ward} M., {Wall} J.~V., 2000, \mnras, 319, 963

\bibitem[{{Tadhunter} {et~al}\mbox{.}(2002){Tadhunter}, {Dickson}, {Morganti},
  {Robinson}, {Wills}, {Villar-Martin}, \& {Hughes}}]{Tadhunter2002}
{Tadhunter} C., {Dickson} R., {Morganti} R., {Robinson} T.~G., {Wills} K.,
  {Villar-Martin} M., {Hughes} M., 2002, \mnras, 330, 977

\bibitem[{{Tadhunter} {et~al}\mbox{.}(1993){Tadhunter}, {Morganti}, {di
  Serego-Alighieri}, {Fosbury}, \& {Danziger}}]{Tadhunter1993}
{Tadhunter} C.~N., {Morganti} R., {di Serego-Alighieri} S., {Fosbury} R.~A.~E.,
  {Danziger} I.~J., 1993, \mnras, 263, 999

\bibitem[{{Tadhunter} {et~al}\mbox{.}(1998){Tadhunter}, {Morganti}, {Robinson},
  {Dickson}, {Villar-Martin}, \& {Fosbury}}]{Tadhunter1998}
{Tadhunter} C.~N., {Morganti} R., {Robinson} A., {Dickson} R., {Villar-Martin}
  M., {Fosbury} R.~A.~E., 1998, \mnras, 298, 1035

\bibitem[{{Tadhunter} {et~al}\mbox{.}(2012){Tadhunter}, {Ramos Almeida},
  {Morganti}, {Holt}, {Rose}, {Dicken}, \& {Inskip}}]{Tadhunter2012}
{Tadhunter} C.~N., {Ramos Almeida} C., {Morganti} R., {Holt} J., {Rose} M.,
  {Dicken} D., {Inskip} K., 2012, \mnras, 427, 1603

\bibitem[{{Uchiyama} {et~al}\mbox{.}(2007){Uchiyama}, {Urry}, {Coppi}, {Van
  Duyne}, {Cheung}, {Sambruna}, {Takahashi}, {Tavecchio}, \&
  {Maraschi}}]{Uchiyama2007}
{Uchiyama} Y. {et~al.}, 2007, \apj, 661, 719

\bibitem[{{Urry} \& {Padovani}(1995)}]{Urry1995}
{Urry} C.~M., {Padovani} P., 1995, \pasp, 107, 803

\bibitem[{{van der Wolk} {et~al}\mbox{.}(2010){van der Wolk}, {Barthel},
  {Peletier}, \& {Pel}}]{VDWolk2010}
{van der Wolk} G., {Barthel} P.~D., {Peletier} R.~F., {Pel} J.~W., 2010, \aap,
  511, A64

\bibitem[{{Volonteri} \& {Ciotti}(2012)}]{Volonteri2012}
{Volonteri} M., {Ciotti} L., 2012, ArXiv e-prints

\bibitem[{{Wall} \& {Peacock}(1985)}]{WP1985}
{Wall} J.~V., {Peacock} J.~A., 1985, \mnras, 216, 173

\bibitem[{{Willott} {et~al}\mbox{.}(1999){Willott}, {Rawlings}, {Blundell}, \&
  {Lacy}}]{Willott1999}
{Willott} C.~J., {Rawlings} S., {Blundell} K.~M., {Lacy} M., 1999, \mnras, 309,
  1017

\bibitem[{{Wills} {et~al}\mbox{.}(2004){Wills}, {Morganti}, {Tadhunter},
  {Robinson}, \& {Villar-Martin}}]{Wills2004}
{Wills} K.~A., {Morganti} R., {Tadhunter} C.~N., {Robinson} T.~G.,
  {Villar-Martin} M., 2004, \mnras, 347, 771

\bibitem[{{Worrall} {et~al}\mbox{.}(1987){Worrall}, {Tananbaum}, {Giommi}, \&
  {Zamorani}}]{Worrall1987}
{Worrall} D.~M., {Tananbaum} H., {Giommi} P., {Zamorani} G., 1987, \apj, 313,
  596

\bibitem[{{Wright} \& {Otrupcek}(1990)}]{Parkes1990}
{Wright} A., {Otrupcek} R., 1990, in PKS Catalog (1990), p.~0

\end{thebibliography}

\appendix
\section{Notes on individual objects}\label{notes}

\subsection{PKS 0023--26}\label{0023-26}

PKS 0023--26 has a young stellar population \citep{Dicken2012}, and redshifted HI emission consistent with infalling gas \citep{Holt2008}. Its X-ray spectrum is quite \redit{atypical} for what is expected in NLRGs, with a dominating jet-related component and low intrinsic absorption. 

\subsection{PKS 0034--01 (3C 15)}\label{0034-01}

PKS 0034--01 has a radio morphology that is intermediate between that of an FR I and an FR II. The host galaxy has a dust lane \citep{Martel1999}. Although this object is classified as a LERG, in our plots it is near the luminosity break between LERGs and NLRGs ($L_{X,2-10keV}=6.6\times 10^{42}$ erg s$^{-1}$, see Table \ref{fit_param_table}). Its spectrum is relatively obscured ($N_{H}\sim10^{23}$ cm$^{-2}$), and requires two power law components. We do not detect a Fe K-$\alpha$ line, as we did for PKS 0043--42. It is unclear whether PKS 0034--01 is a ``true'' (albeit somewhat atypical) LERG, a low-luminosity NLRG, or an intermediate case. The absence of a torus \citep{VDWolk2010} seems to point towards the first possibility, though its poor environment makes it difficult to explain where the hot gas for a radiatively-inefficient accretion scenario might come from.

\subsection{PKS 0035--02 (3C 17)}\label{0035-02}

The optical spectrum of PKS 0035--02 shows double-peaked Balmer lines. Its X-ray spectrum shows two distinct components and some intrinsic absorption, which is not overly frequent in broad-line objects due to orientation. 

\subsection{PKS 0038+09 (3C 18)}\label{0038+09}

The X-ray spectrum of this \redit{BLRG} is bright (we had to correct it for pileup), and is well described with a single power law component, with no traces of intrinsic absorption, as is expected for most broad-line objects.

\subsection{PKS 0039--44}\label{0039-44}

The optical nucleus of this NLRG seems to be dusty, and it is believed to have two components \citep{RamosAlmeida2010}, which are not resolved in our \textit{XMM} images. Its X-ray spectrum is bright, with two distinct components, some intrinsic absorption and a prominent Fe K-$\alpha$ line.

\subsection{PKS 0043--42}\label{0043-42}

PKS 0043--42 has a very extended radio morphology, and no detectable radio core \citep{Morganti1999}. Although it is classified as a LERG, PKS 0043--42 is most likely a high-excitation object where the strong emission lines are simply not detected. \citet{RamosAlmeida2010} find distinct evidence for a clumpy torus in their \textit{Spitzer} data, and its X-ray spectrum shows clear signatures of radiatively efficient accretion, in the form of a bright hard component and a Fe K-$\alpha$ emission line. Its high luminosity situates this object in the parameter space occupied by the fainter NLRGs in our plots.

\subsection{PKS 0105--16 (3C 32)}\label{0105-16}

The spectrum we extracted from the \textit{XMM} images is quite typical for a NLRG, with two components, intrinsic absorption and a noticeable Fe K-$\alpha$ line.

\subsection{PKS 0213--13 (3C 62)}\label{0213-13}

The spectrum of PKS 0213--13 is dominated by the hard component, with the soft, jet-related component being very faint.

\subsection{PKS 0235--19}\label{0235-19}

The X-ray spectrum of \redit{the BLRG} PKS 0235--19 is not very bright, and is best modelled with a single powerlaw with foreground absorption. This faintness is unexpected, given that this source is very bright in the [OIII] and mid-IR bands, making it a clear outlier in our plots \edit{(see the bottom panels of Figures \ref{I_XJ} and \ref{O_XJ}).}

\subsection{PKS 0252--71}\label{0252-71}

PKS 0252--71 has a compact radio morphology. Its X-ray spectrum is quite faint, it features two distinct components (the jet-related one being brighter) and some absorption.

\subsection{PKS 0347+05}\label{0347+05}

This object was previously classified as a BLRG, but a recent study by \citet{Tadhunter2012} suggests that this is in fact a double system with a radio-loud object and a Seyfert 1 radio-quiet AGN. It was already known that this was an interacting system \citep{RamosAlmeida2010,Inskip2010}, but given these recent results is is very possible that we are measuring data from both objects, given that the galaxies are only 5 arcsec apart, and we are thus unable to resolve them. The optical spectra analysed by \citet{Tadhunter2012} suggest that the broad lines previously attributed to the radio source belong instead to the Seyfert, and the line ratios seem to indicate that the radio galaxy is a LERG. They suggest that this latter source is just a relic, having recently switched off, since they do not detect the radio core. The \textit{XMM} images show that the emission is centered between both sources, with more emission coming from the region associated with the radio source. The spectrum we analyse has a relatively bright soft component, and a much brighter, though heavily absorbed hard component, although the $N_{H}$ column is not very well constrained. This is not compatible with the spectrum of a Seyfert 1 galaxy; we therefore assume that the radio-loud AGN is still active \redit{\citep[in contrast to the suggestion of][]{Tadhunter2012}}, and is indeed the main contributor to the X-ray spectrum, which is more consistent with that of a NLRG. We have decided to use the optical LERG classification for this source, although that makes it an outlier in most of our plots, due to its brightness. The X-ray excess (as compared to other LERGs) could be attributed to the contribution from the Seyfert core, and the IR excess to the presence of star formation \citep{Dicken2012}, but in any case the true nature of this source remains uncertain. To further complicate the scenario, the rich ICM could also be contributing to the soft X-ray emission, although our data do not allow us to quantify this effect.

\subsection{PKS 0349--27}\label{0349-27}

This well-known FR II galaxy has some remarkable optical features, including \redit{a spectacular extended emission line nebulosity \citep{Danziger1984}}. The X-ray spectrum shows very little absorption and is completely dominated by the hard component; we were only able to obtain an upper limit on the jet-related power law (see \ref{fit_param_table}).

\subsection{PKS 0404+03 (3C 105)}\label{0404+03}

The host of PKS 0404+03 has been studied in detail in the optical and IR \citep[see][and references therein]{Inskip2010}, despite the presence of a nearby star and the high foreground $N_{H}$ column. The \textit{Chandra} spectrum is somewhat atypical, with high intrinsic absorption, very faint soft emission ($\sim20$ photon counts between 0.4 and 3 keV) and a bright accretion-related component. \redit{It is possible that the high intrinsic absorption we estimate is a consequence of an underestimation in the foreground extinction (see Section\ref{Fitting}).}

\subsection{PKS 0409--75}\label{0409-75}

This FR II has the highest redshift in our sample, and is one of the brightest radio sources in the Southern hemisphere \citep{Morganti1999}. It has a young stellar population \citep{Dicken2012} and it seems to have a double optical nucleus \citep{RamosAlmeida2010}. Its X-ray spectrum is also very bright, and atypical, with the jet-related component clearly dominant, no detectable intrinsic absorption and only an upper limit on the accretion-related component, which, however, has a detectable Fe K-$\alpha$ line. As pointed out in Section \ref{Fitting}, this object is an outlier in most of the plots, having a much brighter soft soft X-ray luminosity than would be expected from the correlations. Given that it lies in a relatively dense cluster environment, \redit{it is possible that some of the soft X-ray excess may be caused by inverse-Compton emission from the radio lobes, which are not resolved by \textit{XMM}. This would be consistent with the fact that PKS 0409--75 is also an outlier in \edit{the top panel of Figure \ref{RC_XJ}}}. A detailed study of the X-ray emission from the ICM is needed to assess its contribution to the soft X-ray luminosity of this NLRG.

\subsection{PKS 0442--28}\label{0442-28}

The spectrum of this source is very bright, with low intrinsic absorption (atypical for a NLRG) and a strong accretion-related component. There seems to be some excess emission around 5-6 keV, indicating the possible presence of a Fe K-$\alpha$ line, but adding a Gaussian component to the best fit model did not improve the statistics.

\subsection{PKS 0620--52}\label{0620-52}

This LERG has the lowest redshift in our sample, and shows evidence for a young stellar population \citep{Dicken2012}. Its spectrum is quite faint; we were able to detect and fit the soft component, but obtained only an upper limit on the accretion-related emission.

\subsection{PKS 0625--35}\label{0625-35}

This object is suspected to be a BL Lac \citep{Wills2004}. Although optically classified as a LERG, it is clear from our data that this is not a ``standard'' low-excitation object. The \textit{Chandra} image shows a large streak, and is piled up. The spectrum is very bright, with some intrinsic absorption and two power law components. In our plots PKS 0625--35 sits near the low-luminosity end of the NLRGs, its accretion-related luminosity being only below that of PKS 0043--42 and PKS 0034--01, which are both ``dubious'' LERGs. Beaming might account for the enhanced luminosity.

\subsection{PKS 0625--53}\label{0625-53}

This LERG is hosted by a dumbbell galaxy, which is also the brightest member in Abell 3391. It has an FR I radio morphology with a wide-angled tail \citep{Morganti1999} and a deflected jet. A strong nuclear component is not detected in the IR \citep{Inskip2010}, consistent with the classification as a low-excitation object. The \textit{Chandra} image shows a very faint nucleus; our spectrum only has one bin, which allows us to constrain an upper limit to the luminosity.

\subsection{PKS 0806--10 (3C 195)}\label{0806-10}

Our \textit{Chandra} spectrum is bright, with a strong accretion-related component and some intrinsic absorption. Although its accretion-related luminosity is somewhat smaller than expected (this is the most luminous object in the 2Jy sample at $z<2$ in the [OIII] and mid-IR bands), it falls within the overall correlations in \edit{the bottom panels of Figures \ref{I_XJ} and \ref{O_XJ}}, thus it is not likely to be Compton-thick.

\subsection{PKS 0859--25}\label{0859-25}

This NLRG seems to have a double nucleus \citep{RamosAlmeida2010}. Its \textit{XMM} spectrum is remarkable in that it shows a very prominent Fe K-$\alpha$ line.

\subsection{PKS 0915--11 (3C 218, Hydra A)}\label{0915-11}

Hydra A is a very well studied galaxy. It sits in the center of a rich cluster and is one of the most powerful local radio sources \citep[see e.g.][and references therein]{Lane2004}. The optical emission lines are very weak, and the K-band imaging does not show a nuclear point source \citep{Inskip2010}. It also shows evidence for recent star formation \citep{Dicken2012}, which is not common in cluster-centre galaxies, but can be attributed to a recent merger \citep[][report the presence of a dust lane]{RamosAlmeida2010}. The AGN is very faint in X-rays, and its spectrum has a rather peculiar shape \citep{Sambruna2000,Rinn2005}, possibly because of contamination from thermal emission that our region selection cannot fully correct for. This situates Hydra A slightly apart form the bulk of the LERG population in our diagrams, relatively close to the LERG/HERG divide. The intrinsic $N_{H}$ and the soft emission are rather well constrained, but the error in the normalization of the hard power law component is quite large, which is reflected in the large error bars in our plots. 

\subsection{PKS 0945+07 (3C 227)}\label{0945+07}

This is a well-known BLRG, with a very extended optical emission line region \citep{Prieto1993}. The \textit{Chandra} spectrum is very bright, and requires pileup correction and some care when selecting the extraction region (there is a faint streak in the image). It is well modelled with two power laws and low, but well constrained, intrinsic absorption \citep{HArdcastle2007}.

\subsection{PKS 1136--13}\label{1136-13}

This QSO has a very prominent jet which is visible in optical \citep{RamosAlmeida2010} and infrared \citep{Uchiyama2007}, and extremely bright in the \textit{Chandra} image, which also shows a prominent streak \citep{Sambruna2006}. The spectrum had to be corrected for pileup, and is modelled well with two components (the soft emission being dominant) and low intrinsic absorption.

\subsection{PKS 1151--34}\label{1151-34}

This QSO seems to be interacting with a nearby spiral galaxy \citep{RamosAlmeida2010}. Although the PAH features in the \textit{Spitzer} observations seem to indicate a young stellar population, this is not confirmed by the far-IR observations \citep{Dicken2012}. This source has double-peaked Balmer lines, and it is clearly radiatively efficient: the \textit{XMM} spectrum is rather bright, and well modelled with two power laws (the hard component being much brighter than the soft one), a surprisingly high absorption column (which is also not very well constrained, see Table \ref{fit_param_table}), and a Fe K-$\alpha$ emission line.

\subsection{PKS 1306--09}\label{1306-09}

PKS 1306--09 has a double optical nucleus \citep{Inskip2010,RamosAlmeida2010}. Its \textit{XMM} spectrum shows no signs of a jet-related component, and requires some intrinsic absorption.

\subsection{PKS 1355--41}\label{1355-41}

The \textit{XMM} spectrum requires two power law components and very low intrinsic absorption.

\subsection{PKS 1547--79}\label{1547-79}

PKS 1547--79 shows a double nucleus both in the optical \citep{RamosAlmeida2010} and IR images \citep{Inskip2010}. Its \textit{XMM} spectrum is rather peculiar, and not very bright, probably due to the high redshift. There may be signs of thermal contamination in the soft emission, and heavy intrinsic absorption is required for a good fit, but is very poorly constrained. This is very atypical for a BLRG, and possibly an effect of the poor spectral quality (the observation suffers from rather heavy flare contamination for about $70$ per cent of the exposure time), but careful flare-filtering and rebinning of the data resulted in no improvements in the fits.

\subsection{PKS 1559+02 (3C 327)}\label{1559+02}

The host galaxy of this NLRG is very massive, and seems to have a bifurcated dust lane \citep{Inskip2010,RamosAlmeida2010}, which crosses the nucleus. \citet{VDWolk2010} report a large infrared excess that extends beyond what is expected for a torus. The \textit{Chandra} image shows a very bright nucleus, which is close to the edge of the S3 chip. The spectrum is best fit with two components and low intrinsic absorption, and a Fe K-$\alpha$ emission line, which is not very well constrained \citep{HArdcastle2007}. As for PKS 0409--75 (Section \ref{0409-75}), it is remarkable that the Fe line is detected despite the faintness of the accretion-related component. As pointed out in Section \ref{Correlations}, it is very likely that this object is Compton-thick, given that its accretion-related X-ray luminosity is much fainter than what should be expected from its [OIII] and IR luminosities.

\subsection{PKS 1602+01 (3C 327.1)}\label{1602+01}

The host galaxy seems to have a double optical nucleus \citep{RamosAlmeida2010} and perhaps an extended emission line region \citep{Morganti1999}. The \textit{XMM} spectrum has two bright components, with no intrinsic absorption.

\subsection{PKS 1648+05 (3C 348, Hercules A)}\label{1648+05}

Hercules A is a cluster-embedded LERG with some unusual radio properties \citep{Morganti1993}. Dust features are detected in the optical images \citep{RamosAlmeida2010}. Its nuclear X-ray spectrum is very faint, with soft emission being the main contributor. We were only able to constrain an upper limit for the hard component.

\subsection{PKS 1733--56}\label{1733-56}

The host galaxy of PKS 1733--56 shows clear evidence of recent star formation \citep{Dicken2012}, and a disturbed optical morphology \citep{RamosAlmeida2010,Inskip2010}. The \textit{Chandra} spectrum is very bright, and had to be corrected for pileup. It is also quite typical of a BLRG, with low intrinsic absorption which does not allow us to distinguish clearly between both components. There is a faint excess $\sim6$ keV which could be related to a Fe K-$\alpha$ emission line, but adding an extra component does not improve the fit.

\subsection{PKS 1814--63}\label{1814-63}

PKS 1814--63 shows clear traces of an optical disk and a dust lane \citep{RamosAlmeida2010,Inskip2010}, which is atypical for a system with this radio luminosity \citep{Morganti2011}. It also shows evidence for starburst activity \citep{Dicken2012} and has an extended emission line region \citep{Holt2008,Holt2009}. Its \textit{Chandra} spectrum is bright and dominated by a relatively unobscured hard component, as typical for NLRGs. It also has a Fe K-$\alpha$ emission line.

\subsection{PKS 1839--48}\label{1839-48}

This FR I is another example of a cluster-embedded LERG \citep{RamosAlmeida2010}. \citet{VDWolk2010} report no detection of a dusty torus, which is consistent with the classification of this object as low-excitation. Its X-ray spectrum has a relatively bright soft component, but no traces of accretion-related emission, for which we were only able to constrain an upper limit. 

\subsection{PKS 1932--46}\label{1932-46}

The host of this BLRG shows signs of ongoing star formation \citep{Dicken2012}, has an extended emission line region \citep{Inskip2007} and its core seems to be relatively faint in the K band \citep{Inskip2010}, its IR luminosity is also rather low in our plots, while it is quite bright in [OIII]. The X-ray spectrum is not very bright, and is best modelled with a single, unobscured component, which does not allow us to distinguish between jet and accretion-related emission. This is consistent with the interpretation of \citet{Inskip2007}, who suggest that the nucleus has switched off, but such a short time ago that this information has not yet the extended narrow-line region

\subsection{PKS 1934--63}\label{1934-63}

This source has a compact double radio morphology \citep{Ojha2004b} and is optically very blue \citep{RamosAlmeida2010}. It also shows evidence for infalling gas \citep{Holt2008,Holt2009}. Its radio spectrum is prototypical for a gigahertz-peaked source, and is self-absorbed, thus we could only derive an upper limit to its 178  and 151 MHz fluxes. Its X-ray spectrum is dominated by the soft component, and we are not able to disentangle the obscuring column from the hard component, nor do we detect the Fe K-$\alpha$ reported by \citet{Risaliti2003} from their \textit{Beppo-SAX} observations. It is possible that this object is heavily obscured, although we do not see an excess in IR emission to support this.

\subsection{PKS 1938--15}\label{1938-15}

The spectrum of this BLRG has two components and a low intrinsic $N_{H}$ column. It has an excess compatible with a Fe K-$\alpha$ emission line; adding this component improves the fit slightly. 

\subsection{PKS 1949+02 (3C 403)}\label{1949+02}

PKS 1949+02 is a NLRG with an X-shaped radio morphology, and as such it has been studied in some detail \citep[see][and references therein]{RamosAlmeida2010}. Its X-ray spectrum has also been studied in detail \citep{Kraft2005,Balmaverde2008}, it is dominated by the hard component, rather obscured, and it has a very prominent Fe K-$\alpha$ emission line.

\subsection{PKS 1954--55}\label{1954-55}

The spectrum of this LERG is rather faint, and only a soft component is detected.

\subsection{PKS 2135--14}\label{2135-14}

The spectrum of this QSO is bright, and had to be corrected for pileup. It has two distinct components and some intrinsic obscuration. There is some excess above 5 keV which we have not been able to model. 

\subsection{PKS 2135--20}\label{2135-20}

The host of this BLRG shows evidence for star formation \citep{Dicken2012}, and is classified as a ULIRG. Although the quality of the spectrum is rather poor, given the low luminosity of the source (for a BLRG) and the high redshift, we detect two components, heavy (although not very well constrained) intrinsic absorption, and some excess that could be compatible with a Fe K-$\alpha$ emission line, although it is unclear due to our low statistics.

\subsection{PKS 2211--17 (3C 444)}\label{2211-17}

PKS 2211--17 is another example of a cluster-embedded LERG \citep{Croston2011}. Its nuclear spectrum is very faint, with only $\sim20$ counts in the 0.4-7 keV energy range. We could only derive upper limits for both X-ray components.

\subsection{PKS 2221--02 (3C 445)}\label{2221-02}

This object is a relatively well-known BLRG. It has a very bright nucleus in the K band \citep{Inskip2010} and  an extended emission line region \citep{Balmaverde2008}. The \textit{Chandra} spectrum is bright, but not heavily piled up. The hard component dominates, and we detect a rather prominent Fe K-$\alpha$ emission line.

\subsection{PKS 2250--41}\label{2250-41}

This source has a rather bright extended [OIII] line emission \citep{Tadhunter2002}. Its \textit{XMM} spectrum has a very faint accretion-related component, for which we were only able to derive an upper limit, although this is clearly a high-excitation object. As for PKS 1559+02, it is very likely that this object is Compton-thick.

\subsection{PKS 2314+03 (3C 459)}\label{2314+03}

The host galaxy of this NLRG is classified as an ultraluminous infrared galaxy (ULIRG) due to its intense star formation activity \citep{Dicken2012,Tadhunter2002}, and it also has a strong radio core \citep{Morganti1999}, which offsets it slightly from the rest of the NLRG population in our 5 GHz plots. The X-ray spectrum has two distinct components and some intrinsic absorption, with some excess in the soft energy range.

\subsection{PKS 2356--61}\label{2356-61}

This NLRG has a spectrum clearly dominated by the accretion-related component, with a noticeable Fe K-$\alpha$ emission line.

\section{3CRR tables}\label{3C_appendix}

The X-ray properties of the 3CRR sources are largely taken from \citet{Hardcastle2009}. They differ from the results presented in that paper in two ways: firstly, we make use of complete \textit{Chandra} observations of the 3CRR sample with $z<0.1$, which will be presented by Evans et al. (in prep.); secondly, we have used \textit{XMM-Newton} observations taken with the aim of completing the X-ray observations of the 3CRR sample at $z<0.5$, which are listed in Table \ref{new-3c-obs}. These were a uniform set of observations with a nominal on-source time of 15 ks, but some were badly affected by flaring. They were all analysed in the manner described by \citet{Hardcastle2009} and in the text.

\begin{table}
\caption{Observational details for the 3CRR sources with new \textit{XMM-Newton} data. Post-filtering livetimes are given for MOS1, MOS2 and PN.}\label{new-3c-obs}
\begin{tabular}{lrr}
\hline
Source&Observation ID&Livetimes (s)\\
\hline
3C\,19&0600450701&13143, 13848, 7786\\
3C\,42&0600450301&18181, 17841, 15165\\
3C\,46&0600450501&7716, 7677, 4343\\
3C\,67&0600450801&9319, 9832, 8719\\
4C14.27&0600450401&14390, 14351, 11203\\
3C\,314.1&0600450101&17949, 18645, 12092\\
3C\,319&0600450201&7470, 7062, 4979\\
3C\,341&0600450601&16659, 16624, 12995\\
\hline
\end{tabular}
\end{table}

In Table \ref{3C_BH_table} we give the K-band magnitudes and derived quantities for the 3CRR sources discussed in the text, and Table \ref{3C_lumin_table} gives a complete list of the 3CRR luminosities and emission-line classifications, an update of the table presented by \citet{Hardcastle2009}.

\begin{table*}\scriptsize
\caption{K-band magnitudes, K-corrections \citep[calculated using the relations of][]{Glazebrook1995,Mannucci2001}, absolute magnitudes, black hole masses, Eddington luminosities, X-ray, [OIII] and infrared-derived Eddington ratios and jet Eddington ratios for the sources in the 3CRR sample. The errors quoted for L$_{X,rad}$/L$_{X,Edd}$ are derived from the errors in the X-ray powerlaw normalization. Values preceded by a `$<$' indicate upper limits. E stands for LERG, N for NLRG, B for BLRG, Q for Quasar. The K magnitudes given correspond to the following references: L \citet{Lilly1984}, S \citet{Simpson2000}, V \citet{deVries1998}, B \citet{Best1998}. 2M stands for sources where the measurements were taken directly from 2MASS.}\label{3C_BH_table}
\centering
\setlength{\tabcolsep}{2.0pt}
\setlength{\extrarowheight}{1pt}
\begin{tabular}{ccccccccccccc}\hline
PKS&Type&Ref&$z$&mag K$_{s}$&K-corr&Mag K$_{s}$&$M_{BH}$&$L_{Edd}$&$L_{X,rad}$/$L_{X,Edd}$&$L_{[OIII],rad}$/$L_{[OIII],Edd}$&$L_{IR,rad}$/$L_{IR,Edd}$&$Q/L_{Edd}$\\\hline
&&&&&&&$\times10^{9}$ M$_{\odot}$&$\times10^{40}$ W&&&&\\\hline
4C12.03&E&L&0.156&13.130&-0.367&-26.60&1.54&2.00&$<9.20\times10^{-4}$&$1.62\times10^{-3}$&-&$1.18\times10^{-2}$\\
3C20&N&L&0.174&14.060&-0.403&-25.96&0.95&1.24&$3.16^{+7.29}_{-0.90}\times10^{-2}$&$4.58\times10^{-3}$&$2.57\times10^{-2}$&$7.98\times10^{-2}$\\
3C28&E&L&0.195&13.570&-0.441&-26.77&1.75&2.27&$<9.97\times10^{-5}$&$1.42\times10^{-3}$&$8.27\times10^{-4}$&$2.49\times10^{-2}$\\
3C31&E&2M&0.017&8.481&-0.043&-25.77&0.82&1.07&$<2.92\times10^{-6}$&$9.72\times10^{-5}$&$8.55\times10^{-4}$&$6.12\times10^{-4}$\\
3C33&N&S&0.060&11.720&-0.150&-25.54&0.69&0.90&$2.75^{+2.54}_{-1.62}\times10^{-2}$&$6.04\times10^{-2}$&$2.41\times10^{-2}$&$1.91\times10^{-2}$\\
3C35&E&L&0.068&11.770&-0.170&-25.80&0.84&1.09&$<1.98\times10^{-3}$&$3.43\times10^{-4}$&-&$4.82\times10^{-3}$\\
3C42&N&S&0.395&15.140&-0.648&-27.16&2.36&3.07&$2.99^{+4.67}_{-2.86}\times10^{-2}$&$1.24\times10^{-2}$&-&$5.42\times10^{-2}$\\
3C46&N&V&0.437&14.830&-0.660&-27.74&3.67&4.78&-&$4.59\times10^{-2}$&-&$3.92\times10^{-2}$\\
3C55&N&L&0.735&16.540&-0.763&-27.50&3.05&3.97&-&-&$1.29\times10^{-1}$&$2.59\times10^{-1}$\\
3C66B&E&2M&0.022&9.500&-0.055&-25.36&0.60&0.78&$<2.22\times10^{-6}$&$5.14\times10^{-4}$&$6.37\times10^{-4}$&$1.78\times10^{-3}$\\
3C76.1&E&2M&0.032&10.870&-0.083&-24.95&0.44&0.57&$<2.76\times10^{-5}$&$4.30\times10^{-4}$&$8.04\times10^{-4}$&$2.83\times10^{-3}$\\
3C79&N&S&0.256&14.420&-0.534&-26.67&1.62&2.11&$2.71^{+13.20}_{-2.15}\times10^{-2}$&$1.22\times10^{-1}$&$9.85\times10^{-2}$&$7.58\times10^{-2}$\\
3C83.1B&E&2M&0.026&10.850&-0.065&-24.41&0.29&0.38&$<2.95\times10^{-5}$&-&$1.88\times10^{-3}$&$5.38\times10^{-3}$\\
3C84&N&2M&0.018&8.126&-0.045&-26.26&1.19&1.55&$<1.90\times10^{-4}$&$9.41\times10^{-3}$&$1.79\times10^{-2}$&$1.46\times10^{-3}$\\
3C98&N&2M&0.031&10.930&-0.078&-24.74&0.38&0.49&$1.56^{+0.15}_{-0.15}\times10^{-3}$&$7.52\times10^{-3}$&-&$9.57\times10^{-3}$\\
3C123&E&L&0.218&13.960&-0.479&-26.67&1.63&2.12&$4.54^{+1.58}_{-2.20}\times10^{-3}$&$1.65\times10^{-2}$&$6.21\times10^{-3}$&$2.55\times10^{-1}$\\
3C153&N&S&0.277&14.220&-0.560&-27.09&2.24&2.91&$<4.39\times10^{-4}$&$5.25\times10^{-3}$&$3.04\times10^{-3}$&$3.43\times10^{-2}$\\
3C171&N&S&0.238&14.720&-0.510&-26.17&1.11&1.45&$2.96^{+1.05}_{-0.89}\times10^{-2}$&$1.87\times10^{-1}$&-&$6.53\times10^{-2}$\\
DA240&E&2M&0.036&10.724&-0.091&-25.30&0.58&0.75&$<1.17\times10^{-5}$&$2.66\times10^{-4}$&-&$4.81\times10^{-1}$\\
3C172&N&L&0.519&15.670&-0.675&-27.36&2.75&3.58&-&-&$9.14\times10^{-3}$&$8.61\times10^{-4}$\\
3C192&N&S&0.060&12.120&-0.151&-25.14&0.51&0.66&$5.78^{+10.30}_{-3.16}\times10^{-4}$&$1.21\times10^{-2}$&$2.69\times10^{-3}$&$1.16\times10^{-2}$\\
3C200&E&V&0.458&15.590&-0.665&-27.11&2.27&2.95&$<5.85\times10^{-3}$&-&$7.58\times10^{-3}$&$7.32\times10^{-2}$\\
3C223&N&S&0.137&13.770&-0.328&-25.59&0.72&0.93&$1.35^{+6.90}_{-0.70}\times10^{-2}$&$5.64\times10^{-2}$&-&$2.71\times10^{-2}$\\
3C228&N&L&0.552&16.250&-0.682&-26.95&2.02&2.62&$4.51^{+6.33}_{-4.13}\times10^{-3}$&-&$2.02\times10^{-2}$&$2.18\times10^{-1}$\\
3C236&E&L&0.099&12.220&-0.244&-26.29&1.22&1.59&$<7.36\times10^{-4}$&$1.75\times10^{-3}$&-&$8.32\times10^{-3}$\\
3C263.1&N&L&0.824&16.610&-0.818&-27.79&3.81&4.95&-&-&$2.24\times10^{-2}$&$2.24\times10^{-1}$\\
3C264&E&2M&0.021&9.489&-0.053&-25.26&0.56&0.73&$<4.01\times10^{-6}$&$6.97\times10^{-5}$&$1.20\times10^{-3}$&$1.96\times10^{-3}$\\
3C272.1&E&2M&0.003&6.222&-0.007&-23.46&0.14&0.18&$<4.09\times10^{-7}$&$1.80\times10^{-5}$&$4.04\times10^{-4}$&$1.96\times10^{-4}$\\
3C274&E&2M&0.004&5.812&-0.011&-25.38&0.61&0.79&$<1.64\times10^{-7}$&$3.95\times10^{-5}$&$2.52\times10^{-4}$&$2.61\times10^{-3}$\\
3C274.1&N&L&0.422&15.360&-0.657&-27.12&2.29&2.97&$<3.02\times10^{-3}$&$2.67\times10^{-3}$&$5.32\times10^{-3}$&$8.56\times10^{-2}$\\
3C280&N&B&0.996&16.800&-0.902&-28.19&5.17&6.72&$1.09^{+0.55}_{-0.48}\times10^{-1}$&-&$7.33\times10^{-2}$&$3.06\times10^{-1}$\\
3C284&N&L&0.239&13.990&-0.512&-26.91&1.96&2.54&$1.26^{+7.94}_{-1.22}\times10^{-2}$&$5.45\times10^{-3}$&-&$2.37\times10^{-2}$\\
3C285&N&L&0.079&12.440&-0.198&-25.53&0.68&0.89&$5.94^{+1.63}_{-1.23}\times10^{-3}$&$1.44\times10^{-3}$&-&$8.61\times10^{-3}$\\
3C288&E&L&0.246&13.420&-0.521&-27.56&3.21&4.17&$<9.65\times10^{-5}$&-&$1.14\times10^{-3}$&$2.34\times10^{-2}$\\
3C289&N&B&0.967&16.720&-0.891&-28.18&5.13&6.67&-&-&$3.57\times10^{-2}$&$1.62\times10^{-1}$\\
3C293&E&2M&0.045&10.841&-0.115&-25.77&0.82&1.07&$1.16^{+0.12}_{-0.11}\times10^{-3}$&$2.14\times10^{-4}$&$4.87\times10^{-3}$&$2.69\times10^{-3}$\\
3C295&N&L&0.461&14.330&-0.665&-28.39&5.99&7.79&$1.85^{+6.61}_{-0.26}\times10^{-2}$&$4.36\times10^{-3}$&$1.49\times10^{-2}$&$1.52\times10^{-1}$\\
3C296&E&2M&0.024&8.764&-0.061&-26.30&1.23&1.60&$<1.44\times10^{-5}$&$1.19\times10^{-4}$&$3.55\times10^{-5}$&$6.12\times10^{-4}$\\
3C300&N&L&0.272&15.110&-0.554&-26.16&1.10&1.43&$<2.88\times10^{-4}$&$2.54\times10^{-2}$&$4.37\times10^{-3}$&$7.83\times10^{-2}$\\
3C305&N&2M&0.042&10.643&-0.106&-25.75&0.81&1.05&$<6.56\times10^{-6}$&$3.67\times10^{-3}$&-&$3.01\times10^{-3}$\\
3C310&E&L&0.054&11.660&-0.137&-25.39&0.61&0.80&-&$2.29\times10^{-4}$&$7.24\times10^{-4}$&$1.87\times10^{-2}$\\
3C315&N&L&0.108&12.920&-0.266&-25.84&0.87&1.12&$<2.53\times10^{-4}$&$2.37\times10^{-3}$&$2.73\times10^{-3}$&$1.71\times10^{-2}$\\
3C319&E&L&0.192&14.910&-0.436&-25.15&0.51&0.66&$<1.49\times10^{-3}$&$7.99\times10^{-4}$&$2.54\times10^{-3}$&$7.65\times10^{-2}$\\
3C321&N&L&0.096&12.220&-0.237&-26.24&1.17&1.52&$6.48^{+172.00}_{-4.38}\times10^{-4}$&$1.86\times10^{-3}$&$6.48\times10^{-2}$&$7.88\times10^{-3}$\\
3C326&E&L&0.090&13.070&-0.222&-25.20&0.53&0.69&-&$2.14\times10^{-5}$&$9.48\times10^{-4}$&$2.26\times10^{-2}$\\
NGC6109&E&2M&0.030&10.325&-0.076&-25.27&0.56&0.73&$<3.51\times10^{-6}$&-&-&$1.70\times10^{-3}$\\
3C337&N&B&0.635&16.550&-0.709&-27.05&2.17&2.82&-&-&$1.14\times10^{-2}$&$1.52\times10^{-1}$\\
3C338&E&2M&0.030&9.170&-0.077&-26.50&1.43&1.86&$<2.32\times10^{-6}$&$6.47\times10^{-5}$&$2.72\times10^{-4}$&$2.61\times10^{-3}$\\
3C340&N&B&0.775&16.920&-0.788&-27.29&2.60&3.38&-&-&$2.84\times10^{-2}$&$1.71\times10^{-1}$\\
3C341&N&L&0.448&15.330&-0.663&-27.31&2.64&3.43&$1.05^{+2.32}_{-0.61}\times10^{-3}$&$6.38\times10^{-2}$&$9.22\times10^{-2}$&$5.81\times10^{-2}$\\
NGC6251&E&2M&0.024&9.026&-0.062&-26.14&1.08&1.41&$<6.72\times10^{-6}$&-&$1.70\times10^{-3}$&$6.01\times10^{-4}$\\
3C346&N&L&0.162&13.100&-0.379&-26.73&1.70&2.21&$<1.60\times10^{-4}$&$3.36\times10^{-3}$&$7.84\times10^{-3}$&$1.18\times10^{-2}$\\
3C349&N&L&0.205&14.470&-0.458&-26.00&0.98&1.27&$1.76^{+0.26}_{-0.24}\times10^{-2}$&$1.00\times10^{-2}$&-&$3.91\times10^{-2}$\\
3C352&N&B&0.806&16.720&-0.807&-27.61&3.32&4.32&-&-&$1.85\times10^{-2}$&$1.63\times10^{-1}$\\
3C386&E&2M&0.018&9.673&-0.045&-24.71&0.37&0.48&$<2.19\times10^{-6}$&$1.31\times10^{-3}$&$4.53\times10^{-4}$&$2.05\times10^{-3}$\\
3C388&E&L&0.091&11.960&-0.225&-26.34&1.26&1.64&$<6.68\times10^{-5}$&$1.10\times10^{-3}$&$9.90\times10^{-4}$&$1.12\times10^{-2}$\\
3C433&N&L&0.102&11.900&-0.250&-26.69&1.65&2.15&$1.24^{+0.42}_{-0.38}\times10^{-2}$&$7.86\times10^{-3}$&$2.94\times10^{-2}$&$2.15\times10^{-2}$\\
3C436&N&L&0.215&13.840&-0.474&-26.76&1.73&2.26&$3.67^{+2.75}_{-1.52}\times10^{-3}$&$5.62\times10^{-3}$&$3.45\times10^{-3}$&$3.15\times10^{-2}$\\
3C438&E&L&0.290&13.900&-0.574&-27.54&3.16&4.11&$<6.48\times10^{-4}$&$2.51\times10^{-3}$&$1.20\times10^{-3}$&$6.87\times10^{-2}$\\
3C441&N&B&0.708&16.200&-0.747&-27.72&3.62&4.71&-&-&$1.70\times10^{-2}$&$1.24\times10^{-1}$\\
3C442A&E&2M&0.027&9.860&-0.069&-25.57&0.70&0.92&$<7.58\times10^{-6}$&-&-&$1.67\times10^{-3}$\\
3C449&E&2M&0.017&9.070&-0.044&-25.31&0.58&0.75&$<2.73\times10^{-6}$&$7.45\times10^{-5}$&$3.27\times10^{-5}$&$6.51\times10^{-4}$\\
3C452&N&L&0.081&12.030&-0.202&-26.00&0.98&1.27&$4.50^{+16.20}_{-3.70}\times10^{-2}$&$6.09\times10^{-3}$&$1.85\times10^{-2}$&$2.35\times10^{-2}$\\
NGC7385&E&2M&0.024&9.540&-0.062&-25.62&0.73&0.95&$<1.05\times10^{-5}$&-&-&$9.18\times10^{-4}$\\
3C457&N&L&0.428&15.720&-0.658&-26.80&1.79&2.33&$7.93^{+13.70}_{-0.89}\times10^{-2}$&$4.61\times10^{-2}$&-&$9.42\times10^{-2}$\\
3C465&E&2M&0.029&10.070&-0.075&-25.52&0.68&0.88&$<9.31\times10^{-6}$&$2.42\times10^{-4}$&$6.75\times10^{-4}$&$4.04\times10^{-3}$\\\hline
\end{tabular}
\end{table*}

\onecolumn

\scriptsize
\begin{longtable}{ccccccccccccccc}
\caption{Luminosities for the sources in the 3CRR sample, following the format of \citet{Hardcastle2006,Hardcastle2009} (see also Table \ref{lumin_table}). The values are given as the logarithm of the luminosity in erg s$^{-1}$, upper limits are indicated with a `$<$' sign before the value. We have converted the radio and IR luminosity densities into $\nu L_{\nu}$ to allow for direct comparison between the magnitudes in different bands. Where measurements could not be obtained their absence is indicated with a dash. E stands for LERG, N for NLRG, B for BLRG, Q for Quasar.}\label{3C_lumin_table}\\
\hline
PKS&Type&z&L$_{178}$&L$_{5}$&L$_{X_{u}}$&$~~~~~$&$~~~~~$&L$_{X_{a}}$&$~~~~~$&$~~~~~$&L$_{IR}$&$~~~~~$&L$_{[OIII]}$&L$_{[OII]}$\\\hline
\endfirsthead
\multicolumn{15}{c}%
{\tablename\ \thetable\ -- \textit{Continued from previous page}}\\
\hline
PKS&Type&z&L$_{178}$&L$_{5}$&L$_{X_{u}}$&$~~~~~$&$~~~~~$&L$_{X_{a}}$&$~~~~~$&$~~~~~$&L$_{IR}$&$~~~~~$&L$_{[OIII]}$&L$_{[OII]}$\\\hline
\endhead
\hline \multicolumn{15}{r}{\textit{Continued on next page}}\\
\endfoot
\hline
\endlastfoot
4C12.03&E&0.156&42.10&40.00&$<$41.91&-&-&$<$43.02&-&-&-&-&40.97&-\\
3C6.1&N&0.840&43.87&41.61&44.92&44.89&44.94&$<$44.17&-&-&45.100&0.010&-&42.15\\
3C16&E&0.405&43.09&39.73&$<$42.74&-&-&$<$43.69&-&-&-&-&-&41.81\\
3C19&N&0.482&43.25&40.14&44.09&44.06&44.12&$<$43.55&-&-&-&-&-&-\\
3C20&N&0.174&42.82&39.97&42.56&42.45&42.64&44.05&43.94&44.44&44.293&0.004&41.21&40.73\\
3C22&B&0.938&43.96&41.95&-&-&-&-&-&-&45.900&0.010&-&43.16\\
3C28&E&0.195&42.54&$<$38.96&$<$41.36&-&-&$<$42.27&-&-&$<$42.740&-&40.96&41.81\\
3C31&E&0.017&40.31&39.45&40.65&40.54&40.74&$<$40.63&-&-&42.341&0.002&39.47&-\\
3C33&N&0.060&41.95&39.98&41.92&41.88&41.97&43.90&43.60&44.11&44.080&0.012&42.19&41.44\\
3C33.1&B&0.181&42.34&40.68&42.43&42.14&42.59&44.38&44.26&44.66&44.878&0.002&42.30&-\\
3C34&N&0.689&43.70&40.80&-&-&-&-&-&-&-&-&-&43.61\\
3C35&E&0.068&41.35&39.77&$<$40.85&-&-&$<$43.07&-&-&-&-&40.03&-\\
3C41&N&0.795&43.66&40.72&-&-&-&-&-&-&-&-&-&42.70\\
3C42&N&0.395&43.07&40.67&42.61&42.44&42.73&44.33&43.28&44.64&-&-&42.04&41.89\\
3C46&N&0.437&43.16&40.75&-&-&-&-&-&-&-&-&42.80&42.22\\
3C47&Q&0.425&43.52&42.23&45.01&44.97&45.04&45.05&44.77&45.21&45.805&0.004&43.28&42.63\\
3C48&Q&0.367&43.64&43.18&45.00&45.00&45.01&45.00&45.00&45.01&46.146&0.002&43.12&42.25\\
3C49&N&0.621&43.44&41.54&-&-&-&-&-&-&-&-&-&-\\
3C55&N&0.735&44.02&41.57&-&-&-&-&-&-&45.820&0.013&-&42.34\\
3C61.1&N&0.186&42.76&40.00&41.92&41.58&42.10&43.93&43.74&44.10&43.700&0.030&42.49&41.44\\
3C66B&E&0.022&40.69&39.97&41.04&41.00&41.08&$<$40.39&-&-&42.008&0.004&40.06&39.87\\
3C67&B&0.310&42.73&40.82&44.26&44.23&44.29&44.26&43.55&44.29&-&-&42.83&42.26\\
3C76.1&E&0.032&40.75&39.07&40.96&40.79&41.12&$<$41.28&-&-&41.966&0.017&$<$39.85&-\\
3C79&N&0.256&43.07&40.90&42.42&42.34&42.49&44.18&43.65&44.75&45.326&0.004&42.86&42.21\\
3C83.1B&E&0.026&40.88&39.46&40.91&40.15&41.54&$<$41.14&-&-&42.205&0.004&-&-\\
3C84&N&0.018&40.92&42.32&42.54&42.52&42.57&$<$42.37&-&-&44.217&-&41.62&41.09\\
3C98&N&0.031&41.29&38.97&40.65&40.51&40.76&42.71&42.67&42.74&-&-&41.02&40.24\\
3C109&B&0.306&43.08&42.48&45.23&45.18&45.29&45.23&44.60&45.29&45.975&0.001&43.32&42.09\\
4C14.11&E&0.206&42.41&41.18&43.01&42.94&43.07&$<$42.78&-&-&-&-&41.24&-\\
3C123&E&0.218&43.68&41.76&42.00&41.05&42.27&43.58&43.36&43.68&43.810&0.067&42.00&-\\
3C132&N&0.214&42.52&40.10&$<$41.99&-&-&43.25&43.04&43.40&-&-&-&-\\
3C138&Q&0.759&43.92&42.85&-&-&-&-&-&-&45.800&0.010&43.46&42.57\\
3C147&Q&0.545&44.04&43.98&-&-&-&-&-&-&45.500&0.010&43.79&43.45\\
3C153&N&0.277&42.82&$<$40.20&$<$41.99&-&-&$<$42.89&-&-&43.590&0.097&41.64&42.49\\
3C171&N&0.238&42.80&40.18&41.86&41.69&41.98&44.08&43.96&44.18&-&-&42.89&42.45\\
3C172&N&0.519&43.46&40.17&-&-&-&-&-&-&44.310&0.062&-&42.77\\
3C173.1&E&0.292&42.90&40.89&41.55&41.34&41.69&$<$43.13&-&-&43.400&0.079&40.85&-\\
3C175&Q&0.768&43.96&42.26&-&-&-&-&-&-&45.700&0.010&43.10&42.77\\
3C175.1&N&0.920&43.95&42.09&-&-&-&-&-&-&-&-&-&42.67\\
3C184&N&0.994&44.08&$<$40.41&43.48&42.48&43.95&44.76&44.57&44.90&45.300&0.010&-&42.89\\
3C184.1&N&0.119&41.95&39.99&41.73&41.45&41.89&43.91&43.70&44.22&-&-&42.23&41.48\\
DA240&E&0.036&41.08&40.17&40.90&40.78&41.01&$<$40.80&-&-&-&-&39.76&40.04\\
3C192&N&0.060&41.54&39.51&40.65&40.38&40.72&42.46&42.18&42.83&42.710&0.028&41.36&41.31\\
3C196&Q&0.871&44.63&41.84&-&-&-&-&-&-&46.000&0.010&-&-\\
3C200&E&0.458&43.21&41.97&43.58&43.52&43.64&$<$43.78&-&-&44.100&0.010&-&-\\
4C14.27&N&0.392&43.05&$<$39.68&42.34&42.17&42.48&$<$43.05&-&-&-&-&-&-\\
3C207&Q&0.684&43.71&43.49&45.14&45.06&45.19&45.14&45.06&45.19&45.500&0.010&43.05&$<$42.15\\
3C215&Q&0.411&43.13&41.54&44.84&44.81&44.87&44.84&44.46&44.87&-&-&42.59&42.22\\
3C217&N&0.898&43.88&$<$40.80&-&-&-&-&-&-&-&-&-&43.29\\
3C216&Q&0.668&43.84&43.79&-&-&-&-&-&-&45.700&0.010&$<$42.46&42.43\\
3C219&B&0.174&42.82&41.27&43.99&43.94&44.04&43.99&43.94&44.04&44.210&0.016&41.77&41.27\\
3C220.1&N&0.610&43.66&42.08&44.50&44.48&44.52&$<$44.04&-&-&44.700&0.010&42.79&42.46\\
3C220.3&N&0.685&43.74&$<$40.02&-&-&-&-&-&-&45.100&0.010&-&-\\
3C223&N&0.137&42.14&40.29&43.16&43.12&43.19&43.67&43.43&44.27&-&-&42.18&41.71\\
3C225B&N&0.580&43.74&40.68&-&-&-&-&-&-&44.483&0.129&-&42.62\\
3C226&N&0.820&43.94&41.82&-&-&-&-&-&-&46.261&0.006&-&42.74\\
4C73.08&N&0.058&41.35&39.62&41.48&41.36&41.57&43.59&43.45&43.90&-&-&40.94&40.57\\
3C228&N&0.552&43.71&41.72&43.86&42.81&43.91&43.65&42.81&43.94&44.574&0.087&-&42.15\\
3C234&N&0.185&42.76&41.56&42.89&42.87&42.91&44.36&44.26&44.60&45.590&0.006&43.13&42.12\\
3C236&E&0.099&41.82&40.98&42.84&41.80&43.18&$<$42.86&-&-&-&-&40.90&41.17\\
4C74.16&?&0.810&43.82&41.11&-&-&-&-&-&-&-&-&-&-\\
3C244.1&N&0.428&43.39&40.66&43.25&43.10&43.36&$<$42.92&-&-&45.130&0.009&43.03&-\\
3C247&N&0.749&43.63&41.41&-&-&-&-&-&-&-&-&-&43.01\\
3C249.1&Q&0.311&42.79&41.93&44.72&44.57&44.77&44.74&44.43&45.04&45.493&0.001&43.38&-\\
3C254&Q&0.734&43.96&42.13&45.32&45.25&45.40&45.32&45.25&45.40&45.600&0.010&43.71&43.13\\
3C263&Q&0.652&43.69&42.94&45.18&45.12&45.24&45.18&45.12&45.24&45.800&0.010&43.71&42.90\\
3C263.1&N&0.824&44.02&41.45&-&-&-&-&-&-&44.980&0.016&-&42.97\\
3C264&E&0.021&40.69&39.98&41.90&41.89&41.91&$<$40.60&-&-&42.315&0.003&39.16&40.10\\
3C265&N&0.811&44.06&41.40&43.45&43.33&43.54&44.49&44.28&44.63&45.860&0.010&43.80&43.85\\
3C268.1&N&0.973&44.19&41.39&-&-&-&-&-&-&45.300&0.010&-&42.27\\
3C268.3&B&0.371&42.92&40.24&-&-&-&-&-&-&-&-&42.49&-\\
3C272.1&E&0.003&38.84&38.22&39.69&39.63&39.75&$<$39.04&-&-&40.964&0.001&37.98&-\\
A1552&E&0.084&41.59&40.34&$<$40.92&-&-&$<$42.21&-&-&-&-&-&-\\
3C274&E&0.004&40.88&39.87&40.53&40.51&40.56&$<$39.28&-&-&41.506&0.003&38.95&-\\
3C274.1&N&0.422&43.29&40.83&43.27&43.21&43.33&$<$43.56&-&-&$<$43.910&-&41.36&-\\
3C275.1&Q&0.557&43.64&42.72&44.52&44.51&44.54&44.52&44.51&44.54&45.100&0.010&-&42.67\\
3C277.2&N&0.766&43.81&40.57&43.67&43.61&43.72&$<$43.81&-&-&-&-&-&43.21\\
3C280&N&0.996&44.32&41.11&42.85&42.55&43.03&45.00&44.81&45.13&45.800&0.010&-&43.68\\
3C284&N&0.239&42.57&40.35&42.22&42.19&42.26&43.98&42.80&44.63&-&-&41.60&-\\
3C285&N&0.079&41.53&39.64&40.54&40.26&40.76&43.38&43.30&43.46&-&-&40.56&40.46\\
3C286&Q&0.849&44.03&41.85&-&-&-&-&-&-&45.600&0.010&-&42.69\\
3C288&E&0.246&42.81&41.34&$<$41.41&-&-&$<$42.48&-&-&$<$43.250&-&-&-\\
3C289&N&0.967&43.99&42.11&-&-&-&-&-&-&45.400&0.010&-&42.57\\
3C292&N&0.710&43.60&40.82&43.62&43.32&43.80&44.40&44.26&44.51&44.800&0.010&-&-\\
3C293&E&0.045&41.06&40.36&40.97&40.79&41.15&42.88&42.85&42.91&43.300&0.001&39.81&41.56\\
3C295&N&0.461&44.05&40.91&42.50&42.18&42.68&44.48&44.43&44.97&45.004&0.005&41.99&42.33\\
3C296&E&0.024&40.51&39.68&41.38&41.15&41.62&$<$41.42&-&-&40.816&0.097&39.74&-\\
3C299&N&0.367&42.98&40.23&-&-&-&-&-&-&-&-&-&42.66\\
3C300&N&0.272&42.88&40.91&43.40&43.38&43.42&$<$42.49&-&-&43.400&0.146&42.02&42.48\\
3C303&B&0.141&42.05&41.54&43.91&43.85&43.97&43.91&43.85&43.97&-&-&41.74&41.90\\
3C305&N&0.042&41.09&39.75&40.56&40.30&40.72&$<$40.95&-&-&-&-&41.04&40.13\\
3C309.1&Q&0.904&44.12&44.40&45.78&45.76&45.79&45.78&45.76&45.79&46.000&0.010&43.70&42.94\\
3C310&E&0.054&41.87&40.42&40.26&39.00&40.58&$<$42.19&-&-&42.089&0.032&40.07&-\\
3C314.1&E&0.120&41.88&$<$39.22&41.38&41.12&41.54&$<$42.30&-&-&42.098&0.097&39.70&-\\
3C315&N&0.108&42.00&$<$41.31&$<$41.20&-&-&$<$42.36&-&-&43.010&0.048&40.88&-\\
3C319&E&0.192&42.49&$<$39.64&42.47&42.29&42.64&$<$42.80&-&-&$<$42.680&-&$<$40.18&39.98\\
3C321&N&0.096&41.77&40.50&41.54&41.45&41.62&42.80&42.40&43.94&44.916&0.001&40.91&41.32\\
3C326&E&0.090&41.89&40.08&42.20&42.15&42.23&$<$41.25&-&-&$<$42.160&-&40.40&41.25\\
3C325&Q&0.860&43.96&41.37&$<$43.16&-&-&44.56&44.43&44.70&45.600&0.010&-&42.79\\
3C330&N&0.549&43.76&40.46&43.08&42.99&43.15&43.90&43.60&44.00&45.000&0.010&-&43.19\\
NGC6109&E&0.030&40.62&39.44&40.04&39.60&40.26&$<$40.55&-&-&-&-&-&-\\
3C334&Q&0.555&43.39&42.64&45.08&44.99&45.15&45.08&44.99&45.15&45.700&0.010&43.37&42.54\\
3C336&Q&0.927&43.91&42.36&-&-&-&-&-&-&45.400&0.010&43.46&-\\
3C341&N&0.448&43.17&40.41&42.77&42.57&42.92&43.25&42.95&43.64&45.558&0.002&42.80&41.77\\
3C338&E&0.030&41.29&40.03&40.51&40.38&40.59&$<$40.76&-&-&42.018&0.007&39.54&40.79\\
3C340&N&0.775&43.67&40.94&-&-&-&-&-&-&44.900&0.010&-&42.67\\
3C337&N&0.635&43.52&40.18&-&-&-&-&-&-&44.300&0.010&-&41.63\\
3C343&Q&0.988&43.90&$<$43.58&-&-&-&-&-&-&45.900&0.010&42.68&41.99\\
3C343.1&N&0.750&43.59&$<$43.17&-&-&-&-&-&-&44.700&0.010&42.71&42.44\\
NGC6251&E&0.024&40.43&40.35&42.74&42.72&42.76&$<$41.08&-&-&42.873&0.001&-&-\\
3C346&N&0.162&42.15&41.83&43.40&43.38&43.41&$<$42.44&-&-&43.960&0.004&41.33&-\\
3C345&Q&0.594&43.34&44.59&45.64&45.58&45.71&45.64&45.58&45.71&-&-&-&-\\
3C349&N&0.205&42.48&41.10&41.82&41.52&41.92&43.87&43.82&43.91&-&-&41.56&-\\
3C351&Q&0.371&43.06&41.05&41.92&41.74&42.08&44.80&44.77&44.82&46.005&0.001&42.84&-\\
3C352&N&0.806&43.80&41.43&-&-&-&-&-&-&44.800&0.010&-&43.05\\
3C380&Q&0.691&44.32&44.67&45.81&45.72&45.89&45.81&45.72&45.89&45.900&0.010&43.76&42.99\\
3C381&B&0.161&42.34&40.18&42.11&42.00&42.20&44.31&44.18&44.44&44.650&0.010&42.38&40.92\\
3C382&B&0.058&41.48&40.85&44.58&44.57&44.59&44.58&44.57&44.59&44.240&0.008&41.78&40.73\\
3C386&E&0.018&40.51&39.62&39.49&38.45&39.83&$<$40.18&-&-&41.550&0.007&$<$40.25&-\\
3C388&E&0.091&41.98&40.77&41.74&41.65&41.81&$<$42.01&-&-&42.660&0.049&40.71&40.52\\
3C390.3&B&0.057&41.85&41.08&44.18&44.15&44.23&44.18&44.15&44.23&44.370&0.011&42.11&40.95\\
3C401&E&0.201&42.65&41.19&42.74&42.69&42.79&$<$43.05&-&-&43.170&0.125&41.06&-\\
3C427.1&E&0.572&43.83&40.53&$<$42.45&-&-&$<$43.24&-&-&$<$43.800&-&-&-\\
3C433&N&0.102&42.45&39.77&41.06&40.77&41.22&43.92&43.80&44.02&44.670&0.005&41.68&-\\
3C436&N&0.215&42.65&41.02&42.59&42.55&42.62&43.53&43.35&43.72&43.520&0.062&41.56&-\\
3C438&E&0.290&43.35&40.87&42.67&42.39&42.84&$<$43.14&-&-&$<$43.270&-&$<$41.47&-\\
3C441&N&0.708&43.70&41.36&-&-&-&-&-&-&44.800&0.010&-&42.42\\
3C442A&E&0.027&40.71&38.21&40.00&39.60&40.42&$<$40.95&-&-&-&-&-&40.56\\
3C449&E&0.017&40.16&39.08&40.49&40.43&40.54&$<$40.46&-&-&40.358&0.084&39.21&-\\
3C452&N&0.081&42.23&40.99&41.77&41.52&41.94&44.18&43.60&44.67&44.130&0.010&41.35&41.44\\
NGC7385&E&0.024&40.44&39.90&41.11&41.00&41.26&$<$41.10&-&-&-&-&-&-\\
3C454.3&Q&0.859&43.70&45.07&46.37&46.24&46.47&46.37&46.24&46.47&-&-&-&-\\
3C455&Q&0.543&43.41&40.72&-&-&-&-&-&-&-&-&43.07&42.81\\
3C457&N&0.428&43.23&40.69&43.35&43.30&43.40&44.56&44.52&44.88&-&-&42.49&-\\
3C465&E&0.029&41.16&40.41&40.91&40.57&41.42&$<$41.02&-&-&42.109&0.007&39.79&-\\\hline
\end{longtable}

\twocolumn

\end{document}